%% file: main.tex
\documentclass[11pt]{article}
\usepackage[utf8]{inputenc}

\usepackage{geometry} 
\usepackage{graphicx}

\usepackage{booktabs,array,paralist,verbatim,subcaption} 
\usepackage{amssymb,amsmath,amsthm,fourier,epstopdf,version}
\excludeversion{hide} 
\usepackage{algorithm2e}

\usepackage{fancyhdr} 
\usepackage{sectsty}
\allsectionsfont{\sffamily\mdseries\upshape}

\usepackage[nottoc,notlof,notlot]{tocbibind}
\usepackage[titles,subfigure]{tocloft}
\usepackage{mathrsfs,bm,color}

\usepackage{bbm}
\usepackage{tikz-qtree}
\usepackage{multicol}
\usepackage{soul}
\setstcolor{red}

\usepackage{hyperref}
\hypersetup{
    colorlinks=true,
    linkcolor=red,
    anchorcolor=black,
    citecolor=blue,
    filecolor=cyan,    
    menucolor=red,
    runcolor=cyan,
    urlcolor=magenta,
}
\newcommand{\footremember}[2]{%
    \footnote{#2}
    \newcounter{#1}
    \setcounter{#1}{\value{footnote}}%
}
\newcommand{\footrecall}[1]{%
    \footnotemark[\value{#1}]%
} 

\newcommand{\bdm}{\begin{displaymath}}
\newcommand{\edm}{\end{displaymath}}

\newcommand{\half}{\frac{1}{2}}

\newcommand{\R}{\mathbb{R}}
\newcommand{\Rf}{\mathbb{R}}

\newcommand{\N}{\mathbb{N}}
\newcommand{\SO}{\mathrm{SO}}
\newcommand{\abs}{\operatorname{abs}}
\newcommand{\distDAG}{d_{\mathrm{DAG}}}

\newcommand{\f}{\bm{f}}
\newcommand{\g}{\bm{g}}

\newcommand{\one}{\bm{1}}
\newcommand{\bdelta}{\bm{\delta}}

\newcommand{\bphi}{\bm{\phi}}
\newcommand{\bvphi}{\bm{\varphi}}
\newcommand{\bpsi}{\bm{\psi}}
\newcommand{\dd}{\mathrm{ d}}
\newcommand{\define}{\, := \,}

\newcommand{\diag}{\mathrm{diag}}
\newcommand{\transp}{{\scriptscriptstyle{\mathsf{T}}}}
\newcommand{\Lrw}{L_\mathrm{rw}}
\newcommand{\Lsym}{L_\mathrm{sym}}

\newcommand{\F}{\mathcal{F}_G}
\newcommand{\iF}{\mathcal{F}^{\scriptscriptstyle{-1}}_G}

\newcommand{\sspan}{\operatorname{span}}

\newcommand{\tol}{\mathrm{tol}}

\newcommand{\rank}{\operatorname{rank}}
\newcommand{\inner}[2]{\left\langle {#1}, {#2} \right\rangle}
\newcommand{\jmax}{{j_\mathrm{max}}}

\newcommand{\udots}{
  \mathinner {\mkern 1mu\raise 1pt \vbox {\kern 7pt \hbox {.}}\mkern 2mu
  \raise 4pt \hbox {.}\mkern 2mu\raise 7pt \hbox {.}\mkern 1mu}}
\newcommand{\score}{\operatorname{score}}

\newtheorem{theorem}{Theorem}[section]

\newtheorem{remark}[theorem]{Remark}
\newtheorem{conjecture*}{Conjecture}

\title{Natural Graph Wavelet Packet Dictionaries}
\author{%
  Alexander Cloninger\footnote{Department of Mathematics and Halicio{\u g}lu Data Science Institute, University of California, San Diego}%
  \and Haotian Li\footremember{ucd}{Department of Mathematics, University of California, Davis}%
  \and Naoki Saito\footrecall{ucd} \textsuperscript{ ,}\footnote{Corresponding author (Email: \texttt{saito@math.ucdavis.edu})}
}
\date{} 

\begin{document}

\maketitle

\begin{abstract}
\input{abs}

\end{abstract}

\input{intro_basis}

\input{background}

\input{vm_ngwp}

\input{pc_ngwp}

\input{applications}

\input{discussion}

\input{acknowledgments}

\appendix 
\input{varimax}

\newpage
\input{mgslp}


\end{document}

%% file: abs.tex
We introduce a set of novel multiscale basis transforms for signals on graphs that utilize their ``dual'' domains by incorporating the ``natural'' distances between graph Laplacian eigenvectors, rather than simply using the eigenvalue ordering. These basis dictionaries can be seen as generalizations of the classical Shannon wavelet packet dictionary to arbitrary graphs, and do not rely on the frequency interpretation of Laplacian eigenvalues.
We describe the algorithms (involving either vector rotations or orthogonalizations) to construct these basis dictionaries, use them to efficiently approximate graph signals through the best basis search, and demonstrate the strengths of these basis dictionaries for graph signals measured on sunflower graphs and street networks.

\vspace{0.5em}
\noindent
\textbf{Keywords:} Graph Laplacian eigenvectors, dual geometry of graph, wavelet packet dictionaries on graphs, best basis algorithm, graph signal approximation

\vspace{0.5em}
\noindent
\textbf{Mathematics Subject Classification:} 65T60, 68R10, 90C35, 94A08, 94A12, 94C15

%% file: intro_basis.tex

\section{Introduction and Motivation}
\label{sec:intro}
There is an explosion of interest and demand to analyze data sampled on graphs 
and networks. 
This has motivated development of more flexible yet mathematically sound 
\emph{dictionaries} (i.e., an overcomplete collection of atoms or basis
vectors) for data analysis and signal processing on graphs. 
Our main goal here is to build \emph{smooth} multiscale localized basis 
dictionaries on an input graph, with beneficial reconstruction and sparsity
properties, and to fill the ``gap'' left from our previous graph basis dictionary
constructions~\cite{IRION-SAITO-GHWT, IRION-SAITO-HGLETS, IRION-SAITO-SPIE, IRION-SAITO-TSIPN, SHAO-SAITO-SPIE} as we explain below. 
Our approach differs from the standard literature as we fully utilize both the 
similarities between the nodes (through the graph adjacency matrix) and the 
similarities between the eigenvectors of the graph Laplacian matrix (through new 
nontrivial eigenvector distances). 

Previous approaches to construct such graph basis dictionaries break down into
two main categories.  
The first category partitions the nodes through recursive graph cuts to generate
multiscale basis dictionaries. This includes: the Hierarchical Graph Laplacian
Eigen Transform (HGLET)~\cite{IRION-SAITO-HGLETS};
the Generalized Haar-Walsh Transform (GHWT)~\cite{IRION-SAITO-GHWT};
its extension, the eGHWT~\cite{SHAO-SAITO-SPIE};
and other Haar-like graph wavelets (see, e.g., \cite{COIF-GAVISH, GAVISH-NADLER-COIF, LEE-NADLER-WASSERMAN, MURTAGH-Haar, SZLAM-HAAR-GRAPH}).
But their basis vectors either are nonsmooth piecewise constants or have
non-overlapping supports.
The second category uses spectral filters on the Laplacian (or diffusion kernel)
eigenvalues to generate multiscale smooth dictionaries. This includes:
the Spectral Graph Wavelet Transform~\cite{hammond2011wavelets};
Diffusion Wavelets \cite{CoifmanRonaldR2006Dw};
extensions to spectral graph convolutional networks \cite{levie2018cayleynets}.
However, these dictionaries do not fully address the relationships
among eigenvectors \cite{CLONINGER-STEINERBERGER, LI-SAITO-SPIE, saito2018can},
which should be utilized for graph dictionary construction;
instead, they focus on the eigenvalue distributions to organize the
corresponding eigenvectors (although there are some works, e.g.,
\cite{PERRAUDIN-RICAUD-SHUMAN-VANDERGHEYNST, RICAUD-SHUMAN-VANDERGHEYNST, SHUMAN-RICAUD-VANDERGHEYNST}, which recognized the graph structures strongly influence
the eigenvector behaviors).
These relationships among eigenvectors can result from eigenvector
localization in different clusters, differing scales in multi-dimensional data,
etc. These notions of similarity and difference between eigenvectors,
while studied in the eigenfunction literature~\cite{CLONINGER-STEINERBERGER, LI-SAITO-SPIE, saito2018can}, have yet to be incorporated into building localized dictionaries on graphs.

We combine the benefits of both approaches to construct the graph 
equivalent of spatial-spectral filtering.
We have two approaches: one is to utilize the \emph{dual geometry}
of an input graph without partitioning the input graph, and
the other is to utilize clustering and partition information
in both the input graph and its dual domain.

Our first approach, detailed in Sect.~\ref{sec:varimax-ngwp}, fills the ``gap''
in the cycle of our development of the graph basis dictionaries, i.e.,
HGLET, GHWT, and eGHWT. This approach is a direct generalization of the
classical wavelet packet dictionary~\cite[Chap.~8]{MALLAT-BOOK3} to the graph
setting: we hierarchically partition the dual domain to generate a
tree-structured ``subbands'' each of which is an appropriate subset of the graph
Laplacian eigenvectors.
We also want to note the following correspondence:
The HGLET~\cite{IRION-SAITO-HGLETS} is a graph version of the Hierarchical Block
Discrete Cosine Transform (DCT) dictionary~\cite[Sect.~8.3]{MALLAT-BOOK3}
(i.e., the non-smooth non-overlapping version of the local cosine
dictionary~\cite{LCT}, \cite[Sect.~8.5]{MALLAT-BOOK3}),
and the former exactly reduces to the latter if the input graph is $P_N$,
a path graph with $N$ nodes. The former hierarchically partitions the input
graph while the latter does the same (with a non-adaptive manner) on the unit
interval $[0,1]$ in the \emph{time} domain.
On the other hand, the GHWT~\cite{IRION-SAITO-GHWT} is a graph version of
the Haar-Walsh wavelet packet dictionary~\cite{WPK},
\cite[Sect.~8.1]{MALLAT-BOOK3}, and the former exactly reduces to the latter
if the input graph is $P_N$. The latter hierarchically partitions the interval
$[0, N)$ in the \emph{sequency} domain while the former does the same by
the graph domain partitioning plus reordering;
see~\cite{IRION-SAITO-GHWT, IRION-SAITO-SPIE, IRION-SAITO-TSIPN} for the details.
Our graph basis dictionary using this first approach is a graph version of the
\emph{Shannon} wavelet packet dictionary~\cite[Sect.~8.1.2]{MALLAT-BOOK3},
which hierarchically partitions the interval $[0, 1/2)$ (or $[0, \pi]$ depending
on how one defines the Fourier transform) in the \emph{frequency} domain.
Again, the former essentially reduces to the latter if the input graph is $P_N$. 

Our second approach, detailed in Sect.~\ref{sec:pair-clustering},
partitions \emph{both} the input graph \emph{and} its dual domain;
more precisely, we first hierarchically partition the dual domain, and
then partition the input graph with constraints imposed by the dual domain
partition.
This approach parallels and generalizes classical time-frequency analysis,
where the time domain is replaced by a general \emph{node-domain} geometry
and the frequency domain is replaced by a general \emph{eigenvector-domain}
organization.
A version of this approach of node-eigenvector organization that embeds the 
eigenvectors to a one-dimensional Euclidean domain has also been considered as a 
visualization technique for low-frequency eigenvectors on clustered
graphs~\cite{girault2019s}.

We aim for the significance and impact of this research to be twofold.
First, these results will provide the first set of graph wavelet packet bases
that adaptively scale to the local structure of the graph.
This is especially important for graphs with complicated multiscale structure,
whose graph Laplacians have localized eigenvectors, for example.
This is an impactful direction, as most of graph wavelet packet bases previously
proposed only tile the node-eigenvector ``plane'' along the node ``axis,''
while Laplacian eigenvectors only tile that plane along the eigenvector ``axis''.
Our approach in Sect.~\ref{sec:pair-clustering} constructs filters in both
the node-domain and eigenvector-domain, which is related to the classical
\emph{time-frequency adapted} wavelet packets that tile both the time and
the frequency domains~\cite{herley1997joint, THIELE-VILLEMOES}.

Second, in the long term, this is a first method of systematically using the
novel concept of eigenvector dual geometry~\cite{CLONINGER-STEINERBERGER, LI-SAITO-SPIE, saito2018can}.  
This direction can set a path for future modification of spectral graph 
theory applications to incorporate dual geometry.

The structure of this article is organized as follows. 
Section~\ref{sec:back} reviews fundamentals: the basics of graphs, i.e.,
graph Laplacians and graph Fourier transform as well as graph wavelet
transforms and frames that were proposed previously.
It also reviews nontrivial metrics of graph Laplacian eigenvectors,
which are used to analyze the dual geometry/eigenvector-domain of an input graph.
Section~\ref{sec:varimax-ngwp} presents a natural graph wavelet packet dictionary
constructed through hierarchical partition of the eigenvector-domain.
Section~\ref{sec:pair-clustering} presents a second version of a natural graph
wavelet packet dictionary constructed through a pair of hierarchical partitions,
one on the input graph and one on its dual domain.
In Sect.~\ref{sec:applications}, we demonstrate the usefulness of our
proposed graph wavelet packet dictionaries in graph signal approximation using
numerical experiments. Code scripts to reproduce all the figures in this
article can be found at \cite{HAOTIAN-GITHUB}.
Finally, we discuss our findings gained through these numerical experiments and
near-future projects for further improvements of our dictionaries.

%% file: background.tex

\section{Background}
\label{sec:back}
\subsection{Graph Laplacians and Graph Fourier Transform}
\label{sec:graphlap}
Let $G=G(V,E)$ be an undirected connected graph.
Let ${V=V(G)=\{v_1,v_2,\ldots,v_N\}}$ denote the set of nodes (or vertices) of
the graph, where $N \define |V(G)|$.  For simplicity, we typically associate
each vertex with its index and write $i$ in place of $v_i$.
$E=E(G)=\{e_1,e_2,\ldots,e_M\}$ is the set of edges, where each $e_k$ connects
two vertices, say, $i$ and $j$, and $M \define |E(G)|$. In this article
we consider only finite graphs (i.e., $M, N < \infty$).  Moreover, we restrict
to the case of simple graphs; that is, graphs without loops (an edge connecting
a vertex to itself) and multiple edges (more than one edge connecting a pair of
vertices).  We use $\f = [ f(1), \ldots, f(N)]^\transp \in \R^N$ to denote
a graph signal on $G$, and we define $\one \define [1, \ldots, 1]^\transp \in \R^N$. 

We now discuss several matrices associated with graphs.  The information in both
$V$ and $E$ is captured by the \emph{edge weight matrix}
$W(G) \in \R^{N \times N}$, where $W_{ij} \geq 0$ is the edge weight between nodes
$i$ and $j$.  In an unweighted graph, this is restricted to be either $0$ or $1$,
depending on whether nodes $i$ and $j$ are adjacent, and we may refer to $W(G)$
as an \emph{adjacency matrix}.  In a weighted graph, $W_{ij}$ indicates the
\emph{affinity} between nodes $i$ and $j$.  In either case, since $G$ is
undirected, $W(G)$ is a symmetric matrix.  We then define the
\emph{degree matrix} $D(G)$ as the diagonal matrix with entries
$D_{ii} = \sum_j W_{ij}$.  With this in place, we are now able to define the
\emph{(unnormalized) Laplacian} matrix, \emph{random-walk normalized Laplacian}
matrix, and \emph{symmetric normalized Laplacian} matrix, respectively, as
\begin{align}
\label{eq:graphlaps}
	L(G)  &\define  D(G)-W(G) \nonumber \\
	\Lrw(G) &\define D(G)^{-1} L(G) \\
	\Lsym(G) &\define D(G)^{-1/2} L(G) D(G)^{-1/2} \nonumber .
\end{align}
We use $0=\lambda_0 \leq \lambda_1 \leq \ldots \leq \lambda_{N-1}$ to denote the
sorted Laplacian eigenvalues and $\bphi_0,\bphi_1,\ldots,\bphi_{N-1}$ to denote
their corresponding eigenvectors, where the specific Laplacian matrix to which
they refer will be clear from either context or subscripts.  
Denoting $\Phi \define [ \bphi_0, \ldots, \bphi_{N-1} ]$ and
$\Lambda \define \diag([\lambda_0, \ldots, \lambda_{N-1}])$,
the eigendecomposition of $L(G)$ can be written as
$L(G) = \Phi \Lambda \Phi^\transp$.
Since we only consider connected graphs here, we have
$0 = \lambda_0  \lneqq \lambda_1$, and $\bphi_0 = \one/\sqrt{N}$,
which is called the direct current component vector or the \emph{DC vector}
for short.
The second smallest eigenvalue $\lambda_1$
is called the \emph{algebraic connectivity} of $G$ and the corresponding
eigenvector $\bphi_1$ is called the \emph{Fiedler vector} of $G$.
The Fiedler vector plays an important role in graph partitioning and 
spectral clustering; see, e.g., \cite{vonLuxburg2007}, which suggests
the use of the Fiedler vector of $\Lrw(G)$ for spectral clustering
over that of the other Laplacian matrices.

\begin{remark}
  In this article, we use the Fiedler vectors of $\Lrw$ of an input graph
and its subgraphs as a tool to hierarchically bipartition the graph although
any other graph partition methods can be used in our proposed algorithms.
However, note that we use the unnormalized graph Laplacian eigenvectors of
$L(G)$ for simplicity to construct the dual domain of $G$ and consequently
our graph wavelet packet dictionaries. In other words, $\Lrw$ is only used
to compute its Fiedler vector for our graph partitioning purposes.
\end{remark}

The graph Laplacian eigenvectors are often viewed as generalized Fourier modes
on graphs.
Therefore, for any graph signal $\f \in \R^N$ and coefficient vector
$\g \in \R^N$, the \emph{graph Fourier transform} and
\emph{inverse graph Fourier transform}~\cite{SHUMAN-ETAL} are defined by
\begin{align}
  \F(\f) \define \Phi^\transp \cdot \f \in \R^N \quad \text{ and } \quad
  \iF(\g) \define \Phi \cdot \g \in \R^N .
\end{align}
Since $\Phi$ is an orthogonal matrix, it is not hard to see that 
$\iF \circ \F = I_{N}$. 
Thus, we can use $\F$ as an analysis operator and $\iF$ as a synthesis operator 
for graph harmonic analysis.

As an important example and for future reference,
let us consider the Laplacian eigenpairs of a path graph $P_N$, which
we also discussed earlier \cite{IRION-SAITO-SPIE, WHY4-LAA, SAITO-BJSIAM-ENGLISH, SAITO-WOEI-DENDRITES, SAITO-WOEI-KOKYUROKU}.
In this case, the eigenvectors of $L(P_N)$ are exactly the \emph{DCT Type II}
basis vectors (used in the JPEG standard) \cite{STRANG-DCT}:
\begin{equation}
\label{eqn:eigen1}
\lambda_k = \lambda_{k;N} \define 4 \sin^2\left(\frac{\pi k}{2N}\right) , \enskip
\phi_k(x) = \phi_{k;N}(x) \define a_{k; N} \cos\left(\frac{\pi k}{N}\left(x-\half \right)\right),
\end{equation}
where $k = 0:N-1$, $x = 1:N$, and $a_{k; N}$ is a normalization constant to have
$\| \bphi_{k;N} \|_2 = 1$. It is clear that the eigenvalue is a monotonically
increasing function of the \emph{frequency}, which is the eigenvalue index $k$
divided by $2$ in this case.

\subsection{Graph Wavelet Transforms and Frames}
\label{sec:graph-wavelet}
We now briefly review graph wavelet transforms and frames; see, e.g.,
\cite{ortega2018graph, SHUMAN-ETAL} for more information.
Translation and dilation are two important operators for
classical wavelet construction. 
However, unlike $\R^d$ ($d \in \N$) or its finite and discretized lattice
graph $P_{N_1} \times \cdots \times P_{N_d}$, we cannot assume the underlying
graph has self-symmetric structure in general, i.e., its interior nodes may
not always have the same neighborhood structure.
Therefore, it is difficult to construct graph wavelet bases or frames
by translating and dilating a single mother wavelet function of a fixed shape,
e.g., the Mexican hat mother wavelet in $\R$, because the graph structure
varies at different locations.
Instead, some researchers, e.g., Hammond et al.~\cite{hammond2011wavelets},
constructed wavelet frames by shifting smooth graph spectral filters to be
centered at different nodes.
A general framework of building wavelet frames can then be summarized as follows:
\begin{align}
  \bpsi_{j,n} \define \overbrace{\Phi F_j \Phi^\transp}^{\text{Filtering}}  \bdelta_n 
  \quad \text{for } j = 0,1,\cdots,J \text{ and } n = 1,2,\cdots, N, 
  \label{eq:wavelet-frame}
\end{align}
where the index $j$ stands for different scale of spectral filtering
(the greater $j$, the finer the scale, and $J \in \N$ represents
  the finest scale specified by the user),
the index $n$ represents the center location of the wavelet,
$\bdelta_n$ is the standard basis vector centered at node $n$,
and the diagonal matrices $F_j \in \R^{N\times N}$ satisfies 
$F_0(l,l) = h(\lambda_{l-1})$ and $F_j(l,l) = g(s_j\lambda_{l-1})$
for $l=1:N$, $j=1:J$. 
Here, $h$ is a scaling function (which mainly deals with the small eigenvalues),
while $g$ is a graph wavelet generating kernel.
For example, the kernel proposed in \cite{hammond2011wavelets} can be
approximated by the Chebyshev polynomial and lead to a fast algorithm.  
Note that $\{s_j\}_{j=1:J}$ are dilation parameters.
    
Furthermore, one can show that as long as the generalized partition of unity
\begin{equation}
  \label{eq:partitionofunity}
  A \cdot I_N \leq \sum_{j = 0}^J F_j \leq B \cdot I_N , \qquad 0 < A \leq B
\end{equation}
holds, $\{ \bpsi_{j,n} \}_{j=0:J; n=1:N}$ forms a
\emph{graph wavelet frame}, which can be used to decompose and recover any given
graph signals~\cite{hammond2011wavelets}.

\subsection{A Motivating Example}
\label{sec:spectral-filter-prob}
However, one important drawback of the above method is that the construction
of the spectral filters $F_j$ solely depends on the eigenvalue distribution
(except some flexibility in choosing the filter pair $(h, g)$, and the dilation
parameters $\{s_j\}$) and does \emph{not} reflect how the eigenvectors
\emph{behave}.
For simple graphs such as $P_N$ and $C_N$ (a cycle graph with $N$ nodes),
the graph Laplacian eigenvectors are global sinusoids whose frequencies can be
simply read off from the corresponding eigenvalues, as discussed in
Sect.~\ref{sec:graphlap}.
Hence, the usual \emph{Littlewood-Paley wavelet theory} (see, e.g.,
\cite[Sect.~4.2]{Daubechies1992}, \cite[Sect.~2.4]{jaffard2001wavelets}) applies
for those simple graphs.
Unfortunately, the graph Laplacian eigenvectors of a general graph --- even if
it is ever so slightly more complicated than $P_N$ and $C_N$ --- can behave in a
much more complicated or unexpected manner than those of $P_N$ or $C_N$, as
discussed in \cite{CLONINGER-STEINERBERGER, IRION-SAITO-SPIE, LI-SAITO-SPIE, WHY4-LAA, SAITO-BJSIAM-ENGLISH, saito2018can, SAITO-WOEI-DENDRITES, SAITO-WOEI-KOKYUROKU}. 

\begin{figure}
  \begin{subfigure}{0.45\textwidth}
    \centering{\includegraphics[width=\textwidth]{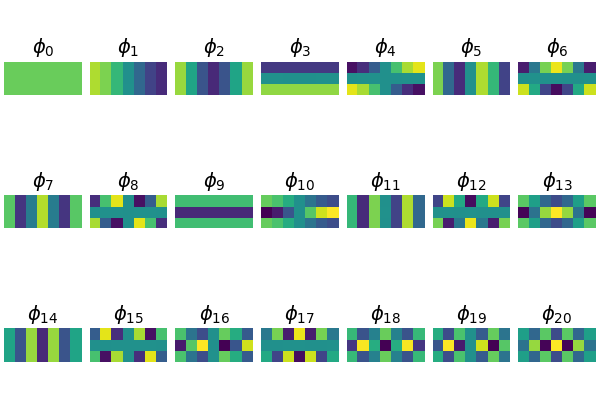}}
    \caption{Nondecreasing eigenvalue ordering}
    \label{fig:grid7x3evorder}
  \end{subfigure}
\hspace{3em}
  \begin{subfigure}{0.45\textwidth}
    \centering{\includegraphics[width=\textwidth]{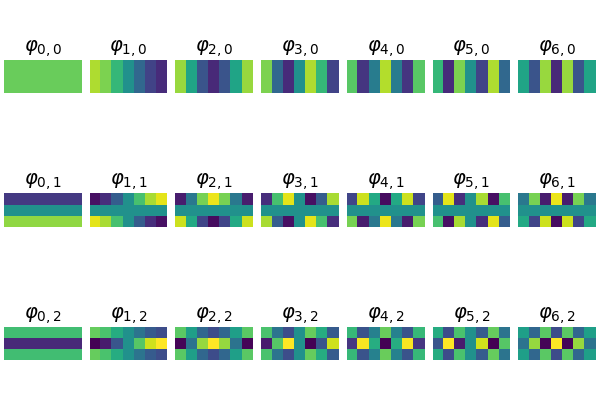}}
    \caption{Natural frequency ordering}
    \label{fig:grid7x3dctorder}
  \end{subfigure}
  \caption{Laplacian eigenvectors of $P_7 \times P_3$ ordered sequentially
  in terms of nondecreasing eigenvalues (a); those ordered in terms of
  their natural horizontal/vertical frequencies (b). The color scheme
  called \emph{viridis}~\cite{VIRIDIS} is used to represent the amplitude of eigenvectors
  ranging from deep violet (negative) to dark green (zero) to yellow (positive).}
  \label{fig:grid7x3order}
\end{figure}

In order to demonstrate this serious problem concretely and make this
article self-contained, let us examine the following example that was also
discussed in \cite{CLONINGER-STEINERBERGER, LI-SAITO-SPIE, saito2018can}.
Let us consider a thin rectangle in $\Rf^2$, and suppose that this
rectangle is discretized as $P_{N_x} \times P_{N_y}$ ($N_x > N_y > 1$).
The Laplacian eigenpairs of this lattice graph can be easily derived from
Eq.~\eqref{eqn:eigen1} as:
\begin{align*}
\lambda_k &= \lambda_{(k_x, k_y)} \define \lambda_{k_x; N_x} + \lambda_{k_y; N_y} \\
\phi_k(x,y) &= \varphi_{k_x,k_y}(x,y) \define \phi_{k_x; N_x}(x) \cdot \phi_{k_y; N_y}(y)
\end{align*}
where $k=0:N_xN_y-1$; $k_x=0:N_x-1$, $k_y=0:N_y-1$, $x=1:N_x$, and $y=1:N_y$.
As always, let $\{\lambda_k\}_{k=0:N_xN_y-1}$ be ordered in the nondecreasing
manner. 
Figure~\ref{fig:grid7x3evorder} shows the corresponding
eigenvectors ordered in this manner (with $N_x=7$, $N_y=3$).
Note that the layout of $3 \times 7$ grid of subplots is for the page saving
purpose: the layout of $1 \times 21$ grid of subplots would be more natural
if we use only the eigenvalue size for eigenvector ordering.
For such a 2D lattice graph, the smallest eigenvalue is still
$\lambda_0=\lambda_{(0,0)}=0$, and the corresponding eigenvector is constant.  
The second smallest eigenvalue $\lambda_1$ is
$\lambda_{(1,0)}=4\sin^2(\pi/2N_x)$, since $\pi/2N_x < \pi/2N_y$, and its
eigenvector has half oscillation (i.e., half period) in the $x$-direction.  
But, how about $\lambda_2$?
Even for such a simple situation
there are two possibilities for $\lambda_2$, depending on $N_x$ and $N_y$.
If $N_x > 2N_y$, then $\lambda_2=\lambda_{(2,0)} < \lambda_{(0,1)}$.
On the other hand, if $N_y < N_x < 2N_y$,
then $\lambda_2=\lambda_{(0,1)} < \lambda_{(2,0)}$.
More generally, if $KN_y < N_x < (K+1)N_y$ for some $K \in \N$, then 
$\lambda_k=\lambda_{(k,0)}=4\sin^2(k\pi/2N_x)$ for $k=0, \dots, K$.
Yet we have $\lambda_{K+1}=\lambda_{(0,1)}=4\sin^2(\pi/2N_y)$ and
$\lambda_{K+2}$ is equal to either $\lambda_{(K+1,0)}=4\sin^2((K+1)\pi/2N_x)$
or $\lambda_{(1,1)}=4[\sin^2(\pi/2N_x) + \sin^2(\pi/2N_y)]$ 
depending on $N_x$ and $N_y$.
Clearly, the mapping between $k$ and $(k_x,k_y)$ is quite nontrivial,
and moreover, the eigenpair computation does not tell us how to map
from $k$ to $(k_x, k_y)$.
In Fig.~\ref{fig:grid7x3evorder}, one can see this behavior with $K=2$, i.e.,
notice that $\bphi_{2} (\equiv \bvphi_{2,0})$ has one oscillation in the
$x$-direction and no oscillation in the $y$-direction whereas
$\bphi_{3} (\equiv \bvphi_{0,1})$ has no oscillation in the $x$-direction and
half oscillation in the $y$-direction.
In other words, all of a sudden the eigenvalue of a completely different
type of oscillation \emph{sneaks into} the eigenvalue sequence.
Hence, on a general graph, by simply looking at its
Laplacian eigenvalue sequence $\{\lambda_k\}_{k=0, 1, \dots}$,
it is \emph{almost impossible to organize the eigenvectors into physically
meaningful dyadic blocks and follow the Littlewood-Paley approach} unless
the underlying graph is of very simple nature, e.g., $P_N$ or $C_N$.
Therefore, it will be problematic to design graph wavelets by using spectral 
filters built solely upon eigenvalues and we need to find a way to distinguish 
eigenvector behaviors. 

What we really want to do is to \emph{organize} those eigenvectors based on 
their natural frequencies or their behaviors, as shown in
Fig.~\ref{fig:grid7x3dctorder} instead of Fig.~\ref{fig:grid7x3evorder},
without explicitly knowing the mapping from $k$ to $(k_x,k_y)$ in this example.
In order to do so for a general graph, we need to define and compute
quantitative similarity or difference between its eigenvectors. 
However, we cannot use the usual $\ell^2$-distances among them since they all
have the same value $\sqrt{2}$ due to their orthonormality.
Then a natural question is: how can we \emph{quantify the similarity/difference
between the eigenvectors} ?

\subsection{Nontrivial Eigenvector Distances}
\label{sec:nontrivial eigenvector distances}
  As a remedy to these issues, we measure the ``behavioral'' difference between
the eigenvectors using the so-called \emph{Difference of Absolute Gradient (DAG)
pseudometric} \cite{LI-SAITO-SPIE}, which is also used in all of our numerical
experiments in Sect.~\ref{sec:applications}.
Note that \cite{CLONINGER-STEINERBERGER, LI-SAITO-SPIE, saito2018can} proposed
several other affinities and distances between Laplacian eigenvectors.
The reasons why we decided to use the DAG pseudometric in this article are:
1) its computational efficiency compared to the other eigenvector metrics;
2) its superior performance for grid-like graphs; and
3) its close relationship with the Hadamard-product affinity proposed in
\cite{CLONINGER-STEINERBERGER}. See~\cite{LI-SAITO-SPIE, saito2018can}
for the details on the other eigenvector metrics and their performance
comparison. Below, we briefly summarize this DAG pseudometric.

Instead of the usual $\ell^2$-distance, we use the \emph{absolute gradient} of
each eigenvector as its feature vector describing its behavior. More precisely,
let $Q(G) \in \R^{N \times M}$ be the \emph{incidence matrix} of an input
graph $G(V,E,W)$ whose $k$th column indicates the head and tail of the
$k$th edge $e_k \in E$. However, we note that we need to orient each edge of $G$
in an arbitrary manner to form a directed graph temporarily in order to construct
its incidence matrix. For example, suppose $e_k$ joins nodes $i$ and $j$, then
we can set either $(Q_{ik}, Q_{jk})=(-\sqrt{W_{ij}}, \sqrt{W_{ij}})$ or
$(\sqrt{W_{ij}}, -\sqrt{W_{ij}})$. Of course, we set $Q_{lk}=0$ for $l \neq i, j$.
It is easy to see that $Q(G) \, Q(G)^\transp=L(G)$.

We now define the \emph{DAG pseudometric} between $\bphi_i$ and $\bphi_j$ by
\begin{equation}
  \label{eq:DAG}
  \distDAG(\bphi_i,\bphi_j) \define \| |\nabla_G| \bphi_i - |\nabla_G| \bphi_j \|_2 \quad \text{where $|\nabla_G| \bphi \define \abs.(Q(G)^\transp \bphi) \in
    \Rf^{M}_{\geq 0}$},
\end{equation}
where $\abs.(\cdot)$ applies the absolute value in the entrywise manner to
its argument.
We note that $|\nabla_G| \bphi$, the absolute gradient of an eigenvector
$\bphi$, is invariant with respect to:
1) sign flip, i.e., $|\nabla_G| \bphi \equiv |\nabla_G|(-\bphi)$ and
2) choice of sign of each column (i.e., edge orientation) of the incidence
matrix $Q(G)$.
We also note that this quantity is not a metric but a \emph{pseudometric}
because the identity of discernible of the axioms of metric is not satisfied.
In order to see the meaning of this quantity, let us analyze its square
as follows.
\begin{align*}
  \distDAG(\bphi_i,\bphi_j)^2 & = \inner{|\nabla_G| \bphi_i - |\nabla_G| \bphi_j}{|\nabla_G| \bphi_i - |\nabla_G| \bphi_j}_E \\
  &= \inner{|\nabla_G| \bphi_i}{|\nabla_G | \bphi_i}_E + 
  \inner{|\nabla_G| \bphi_j}{|\nabla_G| \bphi_j}_E
  - 2 \inner{|\nabla_G| \bphi_i}{|\nabla_G| \bphi_j}_E\\
  &= \lambda_i + \lambda_j - \sum_{x\in V} \sum_{y \sim x}|\bphi_i(x) - 
  \bphi_i(y)| \cdot |\bphi_j(x) - \bphi_j(y)| \quad \text{thanks to
    $Q(G) \, Q(G)^\transp = L(G)$}
\end{align*}
where $\langle \cdot,\cdot \rangle_E$ is the inner product over edges.
The last term of the formula can be viewed as \emph{a global average 
of absolute local correlation} between eigenvectors. In this sense,
this quantity is related to the \emph{Hadamard-product affinity} between
eigenvectors proposed by Cloninger and Steinerberger~\cite{CLONINGER-STEINERBERGER}. Note that the computational cost is $O(M)$ for each $\distDAG(\cdot,\cdot)$
evaluation provided that the eigenvectors have already been computed.

\begin{figure}
\centering\includegraphics[width=0.8\textwidth]{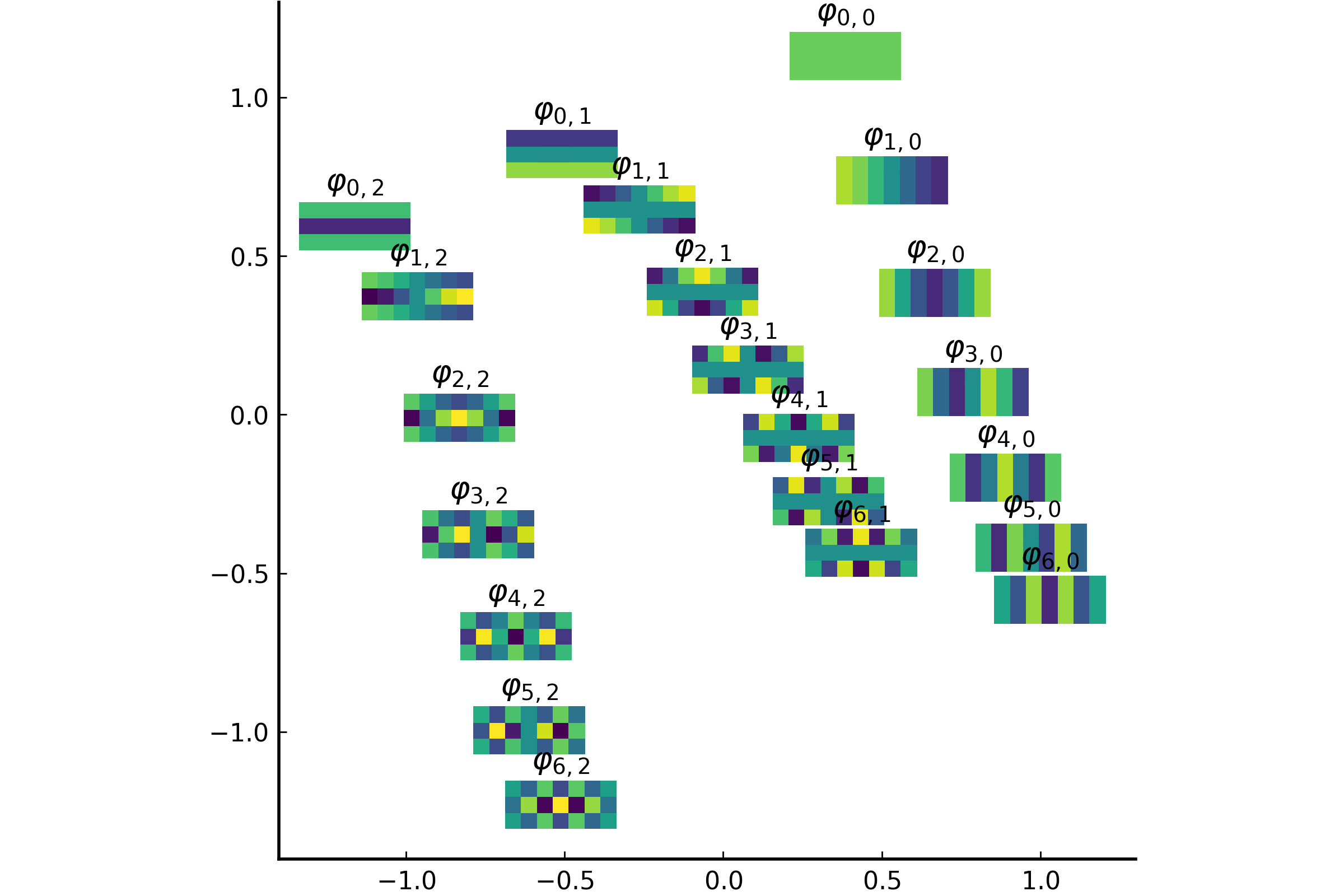}
\caption{Embedding of the Laplacian eigenvectors of $P_7 \times P_3$ into
  $\Rf^2$ via $\distDAG$ and the classical MDS.}
\label{fig:grid7x3map}
\end{figure}

Let us demonstrate the power of the DAG pseudometric using the 2D lattice
graph $P_7 \times P_3$ used in the previous subsection.
Figure~\ref{fig:grid7x3map} displays the embedding of the 21 eigenvectors shown
in Fig.~\ref{fig:grid7x3evorder} into $\Rf^2$ by computing the distances
among all the eigenvectors via Eq.~\eqref{eq:DAG} followed by applying the
classical Multidimensional Scaling (MDS) \cite[Chap.~12]{BORG-GROENEN}.
Of course, in general, when a graph is given, we cannot assume the best
embedding dimension a priori. Here we simply embedded into $\Rf^2$ because the
top two eigenvalues of the Gram matrix of the configurations (i.e., the outputs
of the MDS) were about twice the third eigenvalue.
Figure~\ref{fig:grid7x3map} clearly reveals the natural two-dimensional
organization of the eigenvectors, and is similar to a rotated version
of Fig.~\ref{fig:grid7x3dctorder}.

\subsection{Graph Wavelet Packets}
Instead of building the graph wavelet packet dictionary by graph wavelet frames
using spectral filters as summarized in Sect.~\ref{sec:graph-wavelet},
one could also accomplish it by generalizing the classical wavelet packets to
graphs.
The classical wavelet packet decomposition (or dictionary construction)
of a 1D discrete signal is obtained by passing it through a full binary tree of
filters (each node of the tree represents either low-pass filtered or high-pass
filtered versions of the coefficients entering that node followed by the
subsampling operation) to get a set of binary-tree-structured coefficients~\cite{Coifman-Wickerhauser}, \cite[Sect.~8.1]{MALLAT-BOOK3}.
This basis dictionary for an input signal of length $N$ has up to $N(1+\log_2N)$
basis vectors (hence clearly redundant), yet contains more than $1.5^N$
searchable orthonormal bases (ONBs)~\cite{Coifman-Wickerhauser, THIELE-VILLEMOES}.
For the purpose of efficient signal approximation,
the \emph{best-basis algorithm} originally proposed by Coifman and
Wickerhauser~\cite{Coifman-Wickerhauser} can find the most desirable ONB
(and the expansion coefficients of the input signal) for a given task
among such an immense number of ONBs.
The best-basis algorithm requires a user-specified cost function, e.g., the
$\ell^p$-norm ($0 < p \leq 1$) of the expansion coefficients for sparse signal
approximation, and the basis search starts at the bottom level of the dictionary
and proceeds upwards, comparing the cost of the coefficients at the children
nodes to the cost of the coefficients at their parents nodes. This best-basis
search procedure only costs $O(N)$ operations provided that the expansion
coefficients of the input signal have already been computed.

In order to generalize the classical wavelet packets to the graph setting,
however, there are two main difficulties: 1) the concept of the frequency domain
of a given graph is not well-defined (as discussed in Sect.~\ref{sec:spectral-filter-prob});
and 2) the relation between the Laplacian eigenvectors and sample locations are
much more subtle on general graphs.
For 1), we propose to construct a \emph{dual graph}\footnote{Our definition
of a dual graph is different from the standard definition in the graph theory;
see Remark~\ref{rem:dualgraph} for the details.} $G^\star$ of the input
graph $G$ and view it as the natural spectral domain of $G$, and use any graph
partition method to hierarchically bipartition $G^\star$ instead of building
low and high pass filters like the classical case.
This can be viewed as the \emph{generalized Littlewood-Paley theory}.
For 2), we propose a node-eigenvector organization algorithm called
the \emph{pair-clustering algorithm}, which implicitly provides a downsampling
process on graphs; see Sect.~\ref{sec:pair-clustering} for the details.

%% file: vm_ngwp.tex

\section{Natural Graph Wavelet Packets using \em{Varimax} Rotations}
\label{sec:varimax-ngwp}
Given a graph $G = G(V,E,W)$ with $|V| = N$ and the nontrivial distance
$d$ between its eigenvectors (e.g., $\distDAG$ of Eq.~\eqref{eq:DAG}), we build a
\emph{dual graph} $G^\star = G^\star(V^\star, E^\star, W^\star)$ by viewing the
eigenvectors as its nodes, $V^\star = \{ \bphi_0, \ldots, \bphi_{N-1}\}$,
and the nontrivial affinity between eigenvector pairs as its edge weights,
$W^\star_{ij} = 1 / d(\bphi_{i-1}, \bphi_{j-1})$, $i,j = 1,2,\cdots,N$. 
We note that one can use the alternative and popular Gaussian affinity,
i.e., $\exp(-d(\bphi_{i-1}, \bphi_{j-1})^2/\epsilon)$. This affinity, however,
requires a user to select an appropriate scale parameter $\epsilon > 0$, which
is not a trivial task as explained in \cite{LINDENBAUM-ETAL}, for example.
Moreover, our edge weights using the inverse distances tend to connect
the eigenvectors more globally compared to the Gaussian affinity with
a fixed bandwidth.
Using $G^\star$, which is a complete graph, for representing the graph
spectral domain and studying relations between the eigenvectors is clearly more
\emph{natural} and effective than simply using the eigenvalue magnitudes,
as \cite{CLONINGER-STEINERBERGER, LI-SAITO-SPIE, saito2018can} hinted at.
In this section, we will propose one of our graph wavelet packet dictionary
constructions solely based on hierarchical bipartitioning of $G^\star$.
Basic Steps to generate such a graph wavelet packet dictionary for $G$ are
quite straightforward:
\begin{description}
\item[Step 1:] \emph{Bipartition the dual graph $G^\star$ recursively} via any method,
  e.g., spectral graph bipartition using the \emph{Fiedler vectors};
\item[Step 2:] \emph{Generate wavelet packet vectors} using the eigenvectors
 belonging to each subgraph of $G^\star$ that are \emph{well localized on $G$}.
\end{description}
Note that Step~1 corresponds to bipartitioning the frequency band of an input
signal using the \emph{characteristic functions} in the classical setting.
Hence, our graph wavelet packet dictionary constructed as above can be viewed as
a graph version of the \emph{Shannon} wavelet packet dictionary~\cite[Sect.~8.1.2]{MALLAT-BOOK3}.

\begin{remark}
\label{rem:dualgraph}
  Our definition of the ``dual graph'' $G^\star$ of a given (primal) graph $G$
  is a graph representing the dual geometry/eigenvector domain of the primal
  graph; in particular, it is not related to the graph-theoretic notion of dual
  graph (see, e.g., \cite[Sect.~1.8]{GODSIL-ROYLE}), and does not satisfy the
  equivalence of the double dual graph and the primal graph.
  Our definition is also different from that of Leus et al.~\cite{LEUS-ETAL}
  who defined the dual graph of a primal graph by first assuming that the
  eigenbasis of a graph shift operator (e.g., the adjacency matrix)
  of the dual graph is the transpose of the eigenbasis of the graph shift
  operator of the primal graph. Furthermore, their definition does not work
  for graph Laplacian matrices.
\end{remark}

\begin{remark}
  Our dual domain using the DAG pseudometric among eigenvectors is a finite
  pseudometric space. In order to hierarchically partition the points in that
  space, our strategy --- constructing a complete graph connecting all these
  points using an appropriate affinity measure as its edge weights followed by
  the recursive applications of the spectral graph partitioning, which we will
  discuss in detail in the next subsection, and which has been used in all of
  our numerical examples in this article --- is the most convenient and efficient
  approach as far as we know.
\end{remark}

We now describe the details of each step of our graph wavelet packet
dictionary construction below.

\subsection{Hierarchical Bipartitioning of \texorpdfstring{$G^\star$}{G*}}
\label{sec:gstar-partition}
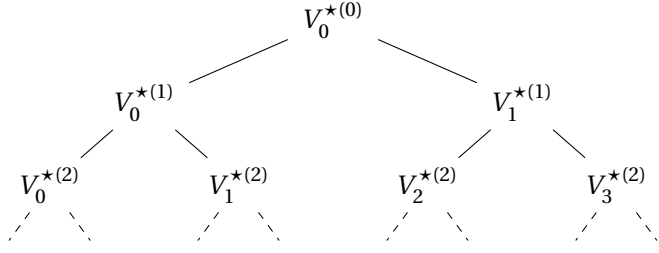
\begin{figure}
  \begin{center}
    \begin{tikzpicture}[scale = 1]
      \tikzset{every tree node/.style={minimum width=3.5em,align=center,anchor=north}, blank/.style={draw=none}, edge from parent/.style={draw, edge from parent path={(\tikzparentnode) -- (\tikzchildnode)}}, level distance=1.1cm}
      \Tree [.$V^{\star(0)}_0$ [.$V^{\star(1)}_0$ [.$V^{\star(2)}_0$ \edge[dashed]; $\,$ \edge[dashed]; $\,$ ] [.$V^{\star(2)}_1$ \edge[dashed]; $\,$ \edge[dashed]; $\,$ ] ] [.$V^{\star(1)}_1$ [.$V^{\star(2)}_2$ \edge[dashed]; $\,$ \edge[dashed]; $\,$ ] [.$V^{\star(2)}_3$ \edge[dashed]; $\,$ \edge[dashed]; $\,$ ] ] ]
    \end{tikzpicture}
  \end{center}
  \caption{The hierarchical bipartition tree of the dual graph nodes
    $V^\star \equiv V^{\star(0)}_0$, which corresponds to the frequency domain
    bipartitioning used in the classical wavelet packet dictionary.}
  \label{fig:binary-tree-dual}
\end{figure}

Let $V^{\star(0)}_0 \define V^\star$ be the node set of the dual graph $G^\star$,
which is simply the set of the eigenvectors of the unnormalized graph Laplacian
matrix $L(G)$.
Suppose we get the \emph{hierarchical bipartition tree} of $V^{\star(0)}_0$
as shown in Fig.~\ref{fig:binary-tree-dual}.
Hence, each $V^{\star(j)}_k$ contains an appropriate subset of the eigenvectors
of $L(G)$.
As we mentioned earlier, any graph bipartitioning method can be used
to generate this hierarchical bipartition tree of $G^\star$. Typically,
we use the Fiedler vector of the random-walk normalized graph Laplacian
matrix $\Lrw$ (see Eq.~\eqref{eq:graphlaps}) of each subgraph of $G^\star$,
whose use is preferred over that of $L$ or $\Lsym$ as von Luxburg
discussed~\cite{vonLuxburg2007}.

  \begin{remark}
    We recursively apply the above bipartition algorithm until
    we reach $j=\jmax > 0$, where each $V_k^{\star(\jmax)}$, $k=0:N-1$,
    contains a single eigenvector. Note that our previous graph basis
    dictionaries, i.e., HGLET \cite{IRION-SAITO-HGLETS},
    GHWT \cite{IRION-SAITO-GHWT}, and eGHWT \cite{SHAO-SAITO-SPIE},
    also constructed such ``full'' hierarchical bipartition trees
    in the primal (input graph) domain, not in the dual domain.
    Note also that during the hierarchical bipartition procedure,
    some $V^{\star(j)}_k$ may become a singleton before reaching $j=\jmax$.
    If this happens, such a subset is copied to the next lower level $j+1$.
    See \cite{IRION-SAITO-TSIPN} for the detailed explanation of such situations.
    We can also stop the recursion at some level $J (< \jmax)$, of course.
    Below, we denote $\jmax$ as the deepest possible level at which
    every subset becomes a singleton for the first time whereas we denote
    $J (\leq \jmax)$ as a more general deepest level specified by a user.
  \end{remark}

\begin{figure}
\centering\includegraphics[width=0.8\textwidth]{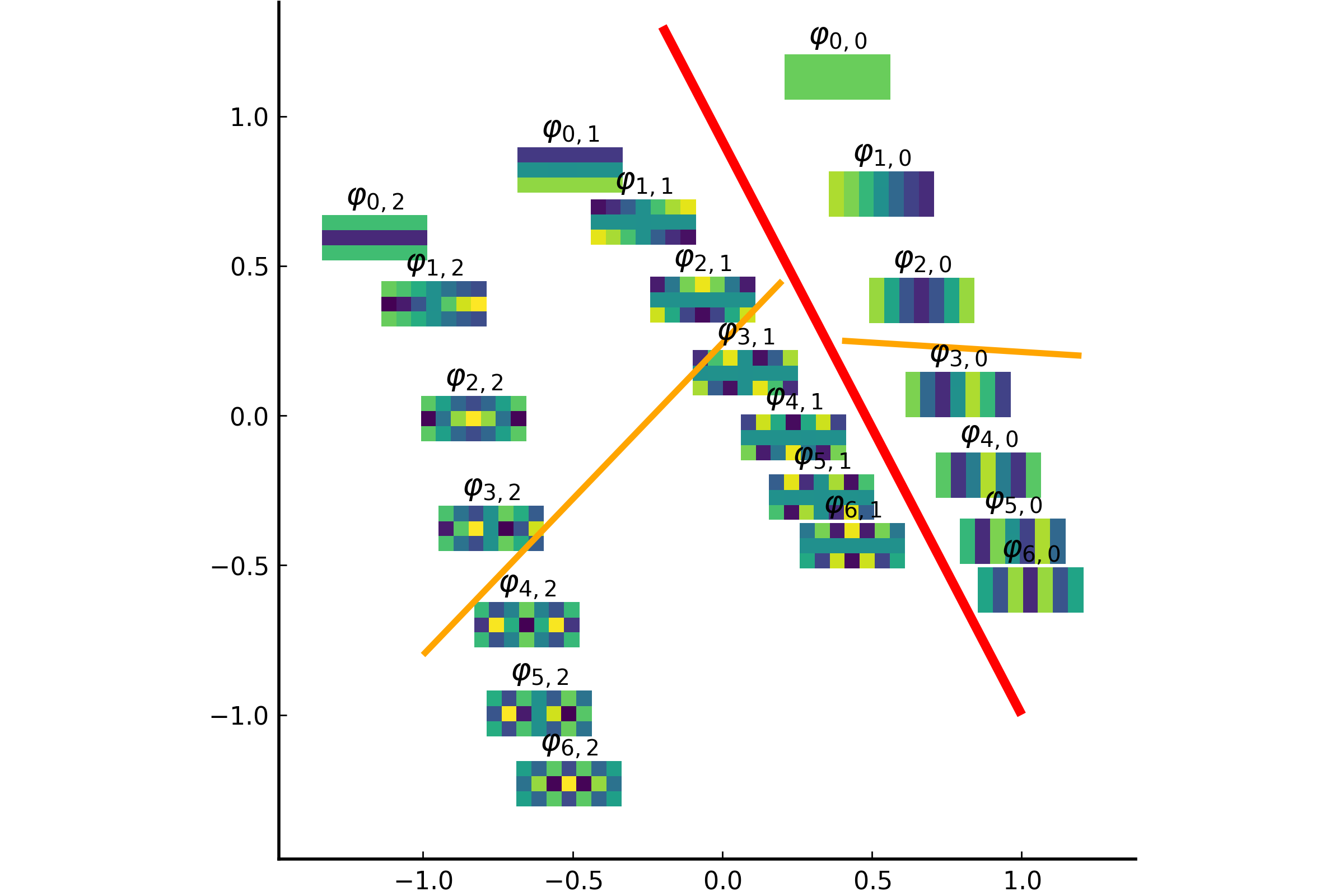}
\caption{The result of the hierarchical bipartition algorithm applied to
  the dual geometry of the 2D lattice graph $P_7 \times P_3$ shown in
  Fig.~\ref{fig:grid7x3map} with $J=2$. The thick red line indicates the
  bipartition at $j=1$ while the orange lines indicate those at $j=2$.}
\label{fig:grid7x3bipartree}
\end{figure}

Figure~\ref{fig:grid7x3bipartree} demonstrates the above strategy
for the 2D lattice graph discussed in Sect.~\ref{sec:back}, whose
dual domain geometry together with the graph Laplacian eigenvectors belonging to
$V^{\star(0)}_0$ was displayed in Fig.~\ref{fig:grid7x3map}.
The thick red line indicates the first split of $V^{\star(0)}_0$, i.e.,
all the eigenvectors above this red line belong to $V^{\star(1)}_0$ while
those below it belong to $V^{\star(1)}_1$. Then, our hierarchical bipartition
algorithm further splits them into $\left\{V^{\star(2)}_0, V^{\star(2)}_1\right\}$
and $\left\{V^{\star(2)}_2, V^{\star(2)}_3\right\}$, respectively.
This two-level bipartition pattern is quite reasonable and natural considering
the fact that the size of the original rectangle is $7 \times 3$, both of
which are odd integers.

\subsection{Localization on \texorpdfstring{$G$}{G} via Varimax Rotation}
For realizing Step~2 of the above basic algorithm, 
we propose to use the \emph{varimax rotation} on the eigenvectors in
$V^{\star(j)}_k$ for each $j$ and $k$. 
Let $\Phi^{(j)}_k \in \R^{N \times N^j_k}$ be a matrix whose columns are the
eigenvectors belonging to $V^{\star(j)}_k$.
A varimax rotation is an orthogonal rotation, originally proposed by
Kaiser~\cite{kaiser1958varimax} and often used
in \emph{factor analysis} (see, e.g., \cite[Chap.~11]{MULAIK}),
to maximize the variances of energy distribution (or a scaled version of
the \emph{kurtosis}) of the input column vectors, which can also be
interpreted as the \emph{approximate entropy minimization of the distribution
of the eigenvector components}~\cite[Sect.~3.2]{SAITO-GENSPIKE-FINAL}.
For the implementation of the varimax rotation algorithm,
see Appendix~\ref{app:varimax}, which is based on the Basic Singular Value (BSV)
Varimax Algorithm of \cite{JENNRICH}.
Thanks to the orthonormality of columns of $\Phi^{(j)}_k$, this is equivalent to
finding an orthogonal rotation that maximizes the overall
\emph{4th order moments}, i.e.,
\begin{equation}
  \label{eq:varimax}
\Psi^{(j)}_k \define \Phi^{(j)}_k \cdot R^{(j)}_k, \quad \text{where $R^{(j)}_k = \arg \max_{R \in \SO(N^j_k)} \sum_{p=1}^N \sum_{q=1}^{N^j_k} \left[ \left( \Phi^{(j)}_k \cdot R \right)^4\right]_{p,q}$}.
\end{equation}
The column vectors of $\Psi^{(j)}_k$ are \emph{more ``localized'' in the
primal domain $G$} than those of $\Phi^{(j)}_k$.
This type of localization is important since the graph Laplacian eigenvectors in
$\Phi^{(j)}_k$ are of \emph{global} nature in general.
We also note that the column vectors of $\Psi^{(j)}_k$ are orthogonal to those of
$\Psi^{(j')}_{k'}$ as long as the latter is neither a direct ancestor nor a direct
descendant of the former.
Hence, Steps~1 and 2 of the above basic algorithm truly generate the graph
wavelet packet dictionary for an input graph signal. We refer to this graph
wavelet packet dictionary $\left\{\Psi^{(j)}_k\right\}_{j=0:J; \, k=0:2^j-1}$
generated by this algorithm as the \emph{Varimax Natural Graph Wavelet Packet}
(VM-NGWP) dictionary. One can run the \emph{best-basis algorithm} of
Coifman-Wickerhauser~\cite{Coifman-Wickerhauser} on this dictionary to
extract the ONB most suitable for a task at hand (e.g., an efficient graph
signal approximation) once an appropriate cost function is specified
(e.g., the $\ell^p$-norm minimization, $0 < p \leq 1$).
Note also that it is easy to extract a graph Shannon wavelet basis from this
dictionary by specifying the appropriate dual graph nodes, i.e.,
$\Psi^{(1)}_1, \Psi^{(2)}_1, \ldots, \Psi^{(J)}_1$, and the father wavelet
vectors $\Psi^{(J)}_0$ where $J (\leq \jmax)$ is the user-specified
deepest level of the hierarchical bipartition tree.
We point out that the meaning of the level index $j$ in our NGWP
dictionaries is different from that in the general graph wavelet frames
\eqref{eq:wavelet-frame} discussed in Sect.~\ref{sec:graph-wavelet}:
in our NGWP dictionaries, a smaller $j$ corresponds to a finer and more localized
(in the primal graph domain) basis vector in $V^{\star(j)}_k$. 

Let us now demonstrate that our algorithm actually generates the classical
\emph{Shannon} wavelet packets dictionary~\cite[Sect.~8.1.2]{MALLAT-BOOK3}
when an input graph is the simple path $P_N$.
\begin{figure}
  \begin{subfigure}{0.33\textwidth}
    \centering\includegraphics[width=\textwidth]{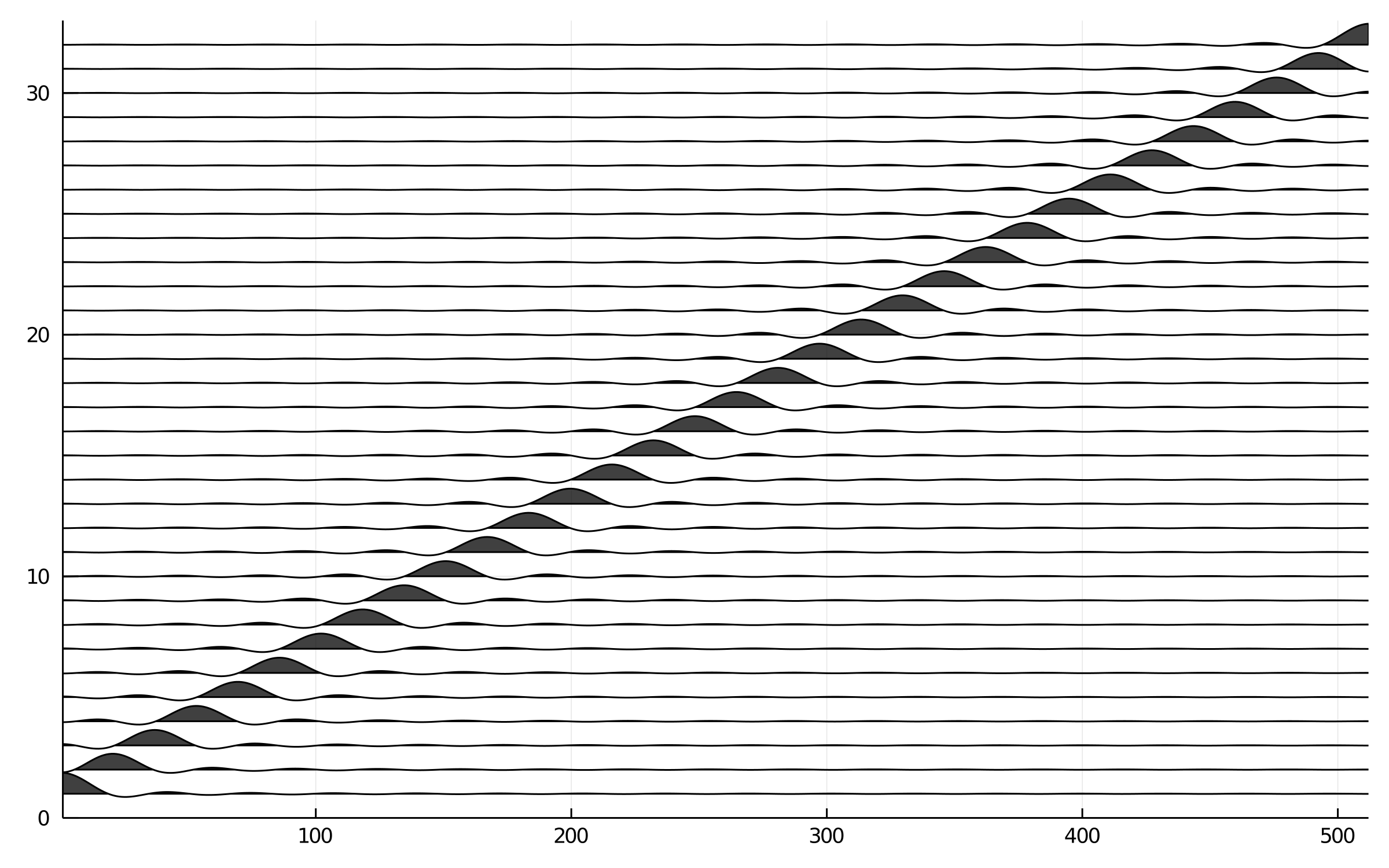}
    \caption{Father wavelet vectors $\Psi^{(4)}_0$}
  \end{subfigure}
  \begin{subfigure}{0.33\textwidth}
    \centering\includegraphics[width=\textwidth]{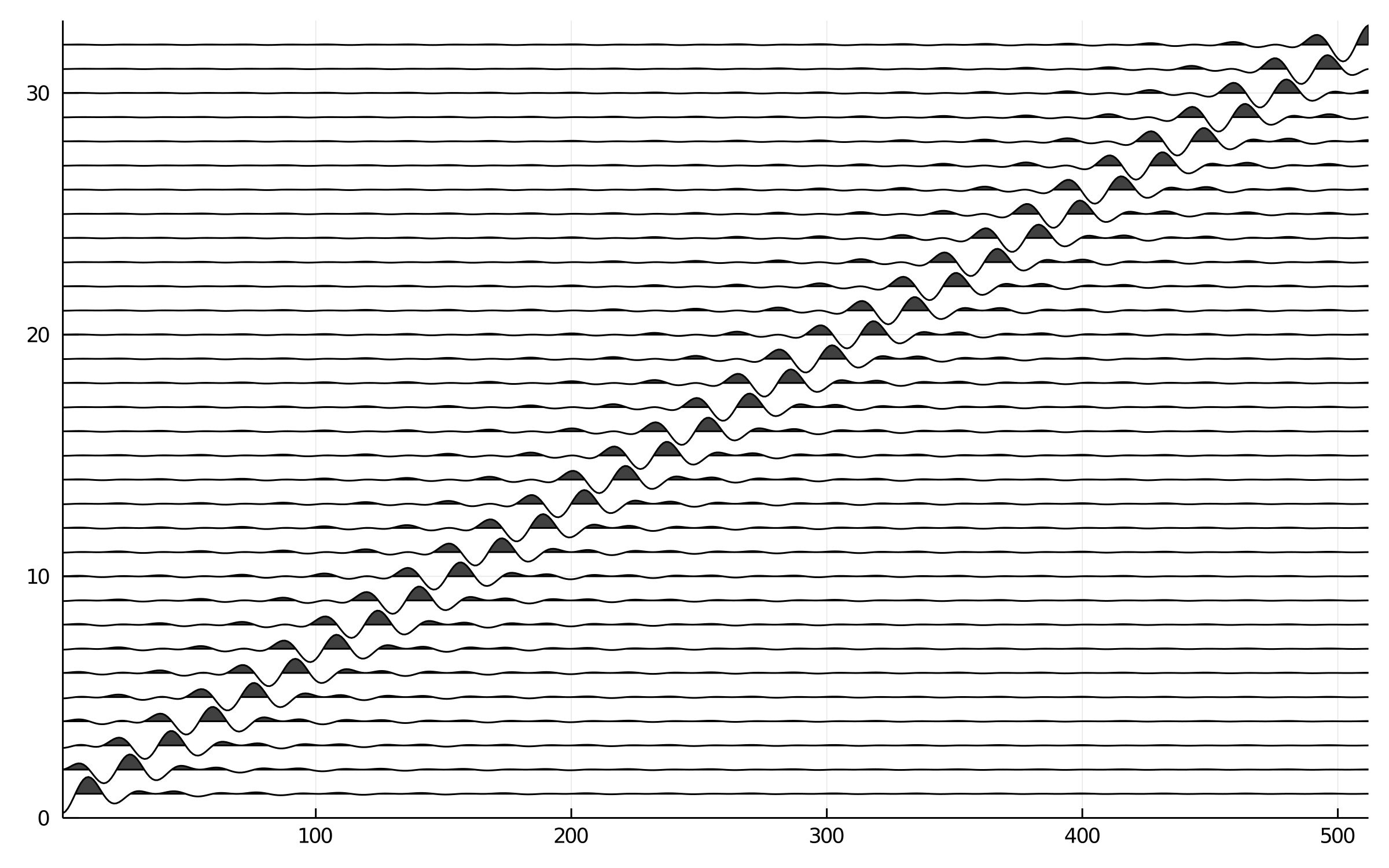}
    \caption{Mother wavelet vectors $\Psi^{(4)}_1$}
  \end{subfigure}
  \begin{subfigure}{0.33\textwidth}
    \centering\includegraphics[width=\textwidth]{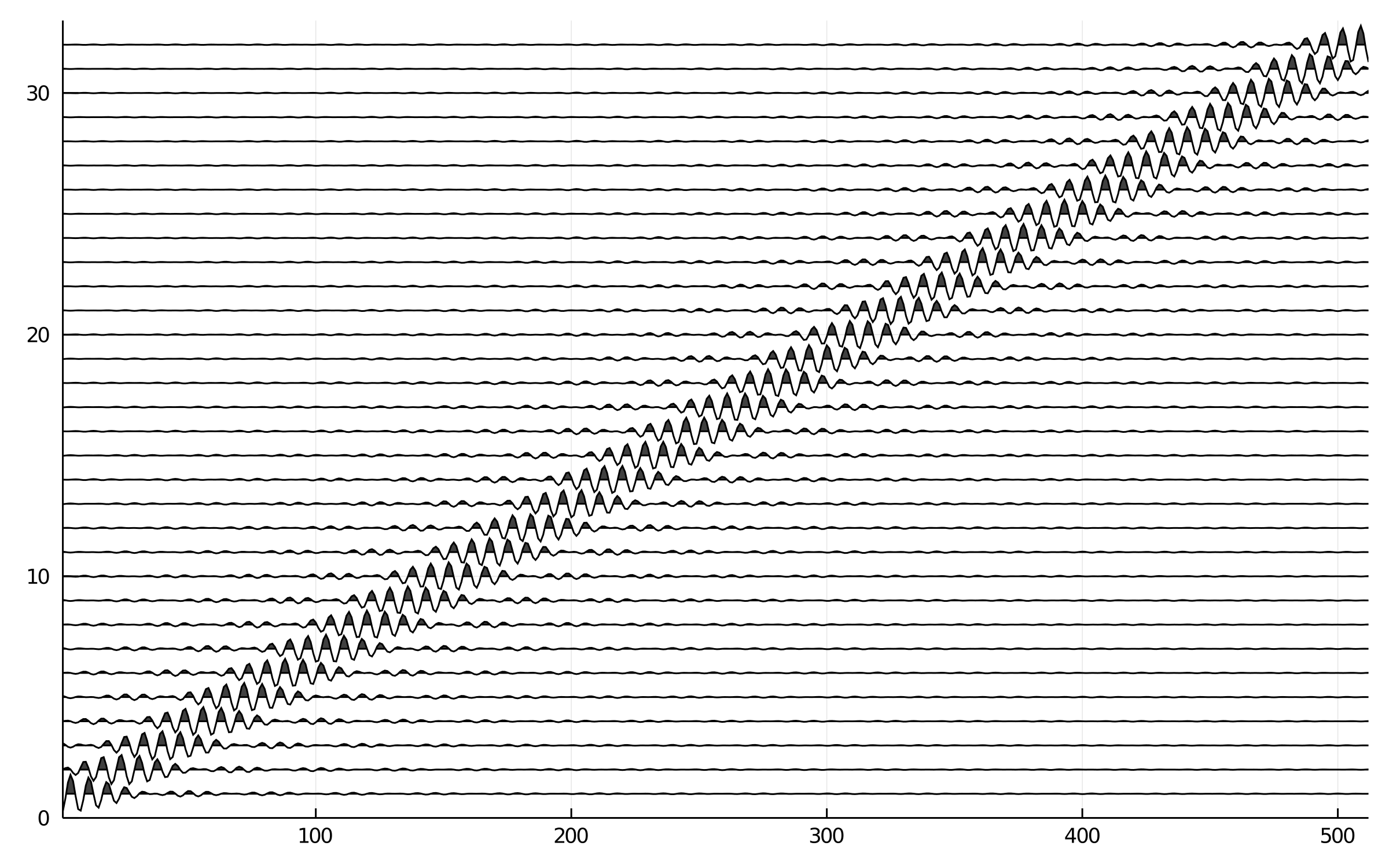}
    \caption{Wavelet packet vectors $\Psi^{(4)}_4$}
  \end{subfigure}
  \caption{Some of the Shannon wavelet packet vectors on $P_{512}$}
  \label{fig:shannon512}
\end{figure}
Note that the varimax rotation algorithm does not necessarily
sort the vectors as shown in Fig.~\ref{fig:shannon512} because
the minimization in Eq.~\eqref{eq:varimax} is the same modulo to
any permutation of the columns and any sign flip of each column.
In other words, to produce Fig.~\ref{fig:shannon512}, we carefully applied
sign flip to some of the columns, and sorted the whole columns so that each
subfigure simply shows translations of the corresponding wavelet packet vectors.

\begin{figure}
\centering\includegraphics[width=0.8\textwidth]{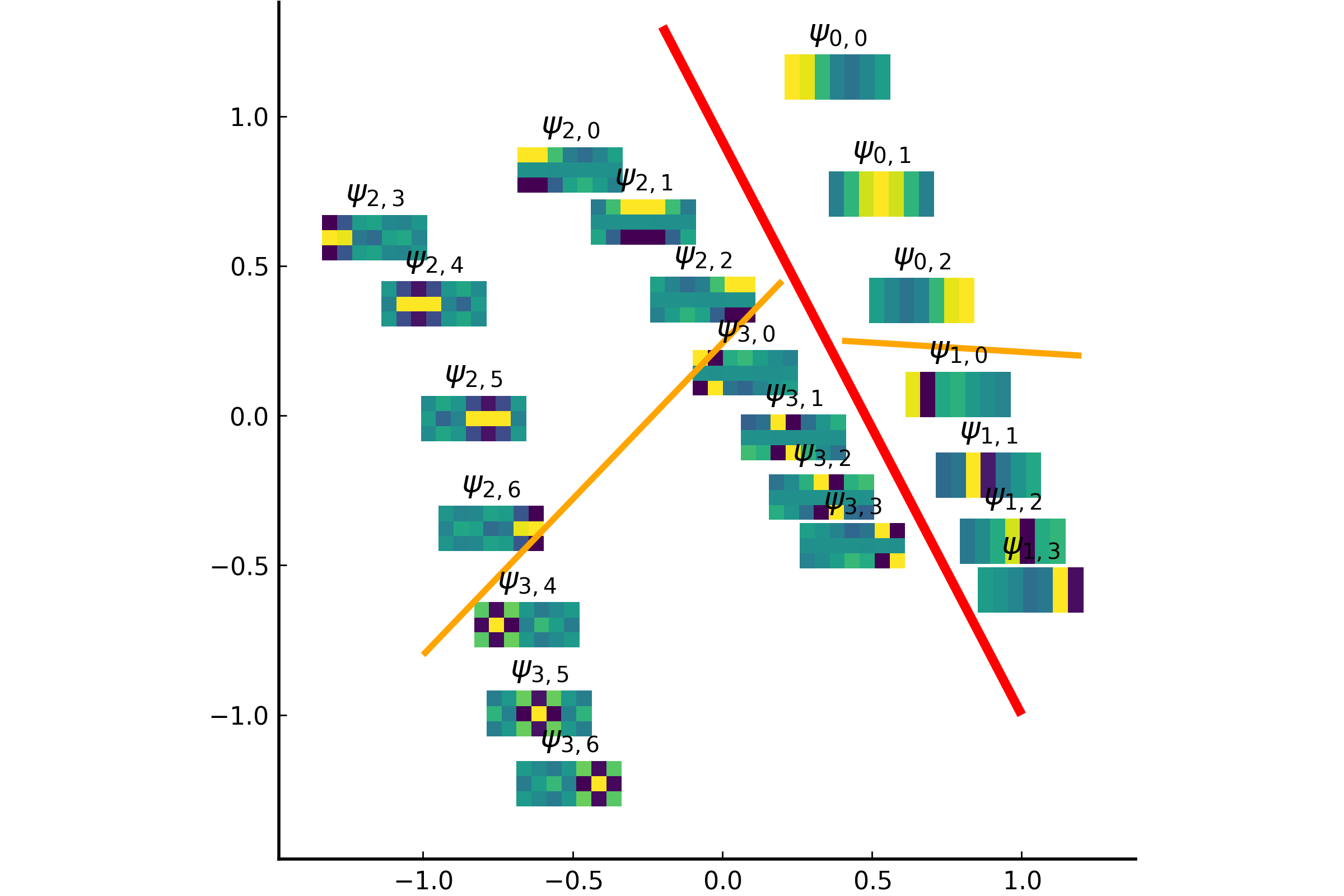}
\caption{The VM-NGWP basis vectors of the 2D lattice graph $P_7 \times P_3$
    computed by the varimax rotations in the hierarchically partitioned dual
    domain shown in Fig.~\ref{fig:grid7x3bipartree}.
    Note that the column vectors of the basis matrix $\Psi^{(2)}_{k}$ are
    denoted as $\psi_{k,l}$, $l=0, 1, \ldots$, in this figure instead of
    $\psi^{(2)}_{k,l}$ for simplicity.}
\label{fig:grid7x3varimax}
\end{figure}

Let us also demonstrate how some VM-NGWP basis vectors of the 2D lattice graph
$P_7 \times P_3$ look like. Figure~\ref{fig:grid7x3varimax} shows
such VM-NGWP basis vectors with $J=2$. Those basis vectors are placed at
the same locations as the graph Laplacian eigenvectors in the dual domain
shown in Fig.~\ref{fig:grid7x3bipartree} for the demonstration purpose.
It is quite clear that those VM-NGWP basis vectors are more localized in
the primal graph domain than those graph Laplacian eigenvectors shown in
Fig.~\ref{fig:grid7x3bipartree}.
We note that we determined the index $l$ in $\psi_{k,l}$ for each $k$
in such a way that the main features of the VM-NGWP basis vectors translates
nicely in the horizontal and vertical directions, and some sign flips were
applied as in the case of the 1D Shannon wavelet packets shown in
Fig.~\ref{fig:shannon512}.
As one can see, like the classical wavelet packet vectors on a rectangle,
$\{\psi_{0,l}\}_{l=0:2}$, are the father wavelets and clearly function
as local averaging operators along the horizontal direction
while $\{\psi_{1,l}\}_{l=0:3}$, work as localized first order differential
operators along the horizontal direction.
On the other hand, $\{\psi_{2,l}\}_{l=0:2}$ work as localized first order
differential operators along the vertical direction; $\{\psi_{2,l}\}_{l=3:6}$ work
as localized second order differential operators along the vertical direction;
$\{\psi_{3,l}\}_{l=0:3}$ work as localized mixed differential operators;
and finally, $\{\psi_{3,l}\}_{l=4:6}$ work as localized Laplacian operators.

\subsection{Computational Complexity}
\label{sec:varimax-cost}
The varimax rotation algorithm of Appendix~\ref{app:varimax} is of iterative
nature and is an example of the BSV algorithms~\cite{JENNRICH}:
for each iteration at the dual node set $V^{\star(j)}_k$, it requires computing
the full Singular Value Decomposition (SVD) of a matrix of size
$N^j_k \times N^j_k$ representing a gradient of the objective function, which
itself is computed by multiplying matrices of sizes $N^j_k \times N$ and
$N \times N^j_k$. The convergence is checked with respect to the relative error
between the current and previous gradient estimates measured in the nuclear norm
(i.e., the sum of the singular values).
For our numerical experiments in Sect.~\ref{sec:applications}, we set the
maximum iteration as 1000 and the error tolerance as $10^{-12}$.
Therefore, to generate $\Psi^{(j)}_k$ for each $(j,k)$, the computational cost in
the worst case scenario is $O\left(c \cdot (N^j_k)^3 + N \cdot (N^j_k)^2\right)$
where $c=1000$ and the first term accounts for the SVD computation and
the second does for the matrix multiplication.
For a perfectly balanced and fully developed bipartition tree with $N=2^\jmax$,
we have $N^j_k = 2^{\jmax-j}$, $j=0:\jmax$, $k=0:2^j-1$.
Hence we have:
\begin{equation}
\label{eq:Njk2}
\sum_{k=0}^{2^j-1} (N^j_k)^2 = \sum_{k=0}^{2^j-1} 2^{2(\jmax-j)} = 2^{2\jmax-2j} \cdot 2^j = N^2 \cdot 2^{-j} ,
\end{equation}
and
\bdm
\sum_{k=0}^{2^j-1} (N^j_k)^3 = \sum_{k=0}^{2^j-1} 2^{3(\jmax-j)} = 2^{3\jmax-3j} \cdot 2^j = N^3 \cdot 2^{-2j} .
\edm
Note that at the bottom level $j=\jmax$, each node is a leaf containing only one
eigenvector, and there is no need to do any rotation estimation and computation.
Note also that at the root level $j=0$, the columns of $\Phi^{(0)}_0$
span the whole $\Rf^N$, and we know that the varimax rotation turns
$\Phi^{(0)}_0$ into the identity matrix (or its permuted version).
Hence, we do not need to run the varimax rotation algorithm on the root node.
Finally, summing the cost $O\left(c \cdot (N^j_k)^3 + N \cdot (N^j_k)^2\right)$
from $j=1$ to $\jmax-1$, the total worst case computational cost becomes
$O((1+c/3)N^3 - 2N^2 - 4c/3 N)$. So after all, it is an $O(N^3)$ algorithm.
In practice, the convergence is often achieved with less than 1000 iterations
at each node except possibly for the nodes with small $j$ where $N^j_k$ 
is large. For example, when computing the VM-NGWP dictionary for the path graph
$P_{512}$ ($\jmax=9$) shown in Fig.~\ref{fig:shannon512}, the average number of
iterations over all the dual graph nodes
$\left\{ V^{\star(j)}_k\right\}_{j=0:9; \, k=0:2^j-1}$
was $68.42$ with the standard deviation $98.09$.

%% file: pc_ngwp.tex

\section{Natural Graph Wavelet Packets using \emph{Pair-Clustering}}
\label{sec:pair-clustering}
Another way to construct a natural graph wavelet packet dictionary is to mimic
the convolution and subsampling strategy of the classical wavelet packet
dictionary construction: form a binary tree of spectral filters in the
dual domain via $\left\{ V^{\star (j)}_k \right\}_{j=0:J; k=0:2^j-1}$
and then perform the filtering/downsampling process based on the relations
between the sampling points (primal nodes) and the eigenvectors of $L(G)$.
In order to fully utilize such relations, we look for a coordinated pair of
partitions on $G$ and $G^\star$, which is realized by our \emph{pair-clustering}
algorithm described below.
We will first describe the one-level pair-clustering algorithm and then proceed
to the hierarchical version.

\subsection{One-Level Pair-Clustering}
\label{sec:1levpc}
Suppose we partition the dual graph $G^\star$ into $K \geq 2$ clusters
using any method including the spectral clustering~\cite{vonLuxburg2007}
as we used in the previous section.
Let $V^\star_1, \ldots, V^\star_K$ be those mutually disjoint $K$ clusters of the nodes $V^\star$, i.e., $\displaystyle V^\star = \bigsqcup_{k=1}^K V^\star_k$,
which is also often written as $\displaystyle \bigoplus_{k=1}^K V^\star_k$.
Denote the cardinality of each cluster as $N_k \define |V^\star_k|$,
$k = 1:K$, and we clearly have $\displaystyle \sum_{k = 1}^K N_k = N$. 
Then, we also partition the primal graph  nodes $V$
into mutually disjoint $K$ clusters, $V_1, \ldots, V_K$ with the constraint that
$|V_k|=|V^\star_k|=N_k$, $k = 1:K$, and the members of $V_k$ and
$V^\star_k$ are as ``closely related'' as possible. The purpose of partitioning
$V$ is to select appropriate primal graph nodes as sampling points around
which the graph wavelet packet vectors using the information on $V^\star_k$ are
localized.
With a slight abuse of notation, let $V$ also represent a collection of
the standard basis vectors in $\R^N$, i.e.,
$V \define \{\bdelta_1, \ldots, \bdelta_N\}$,
where $\bdelta_k(k)=1$ and $0$ otherwise.
In order to formalize this constrained clustering of $V$,
we define the \emph{affinity measure} $\alpha$ between $V_k$ and $V^\star_k$
as follows:
\begin{align}
  \alpha(V_k,V^\star_k) \define \sum_{\bdelta \in V_k, \bphi \in V^\star_k}
  | \inner{\bdelta}{\bphi} |^2,
\end{align}
where $\inner{\cdot}{\cdot}$ is the standard inner product in $\Rf^N$.
Note that $\displaystyle \alpha(V, V^\star) = \sum_{\bdelta \in V, \bphi \in V^\star} | \inner{\bdelta}{\bphi} |^2 = \sum_{\bphi \in V^\star} \| \bphi \|^2 = N$.
Denote the feasible partition set as
\begin{displaymath}
  U(V; N_1, \ldots, N_K) \define \left\{ (V_1,\ldots,V_K) \, \biggl| \,
  \bigsqcup_{k=1}^K V_k = V; \, |V_k|=N_k, k=1:K \right\}.
\end{displaymath}
Now we need to solve the following optimization problem for a given partition of 
$\displaystyle V^\star= \bigsqcup_{k=1}^K V^\star_k$:
\begin{equation}
  \label{eq:objective}
  (V_1, \ldots, V_K) = \arg \max_{(V_1,\ldots,V_K) \in U(V; N_1,\ldots,N_K)} 
  \sum_{k = 1}^K \alpha(V_k,V^\star_k)
\end{equation}
This is a discrete optimization problem. 
In general, it is not easy to find the global optimal solution except for 
the case $K = 2$.  For $K=2$, we can find the desired partition of $V$ by
the following greedy algorithm:
1) compute $\score(\bdelta) \define \alpha(\{\bdelta\}, V^\star_1) - \alpha(\{\bdelta\},V^\star_2)$ for each $\bdelta \in V$; 2) select $N_1$ $\bdelta$'s in $V$
that give the largest $N_1$ values of $\score(\cdot)$, set them as $V_1$,
and set $V_2 = V \setminus V_1$.

When $K>2$, we can find a \emph{local optimum} by the similar strategy as above:
1) compute the values $\alpha(\{\bdelta\},V^\star_1)$ for each $\bdelta \in V$;
2) select $N_1$ $\bdelta$'s giving the largest $N_1$ values, and set them
as $V_1$; 
3) compute the values $\alpha(\{\bdelta\},V^\star_2)$ for each
$\bdelta \in  V \setminus V_1$, select $N_2$ $\bdelta$'s giving the largest
$N_2$ values, and set them as $V_2$;
4) repeat the above process to produce $V_3, \ldots, V_K$.
While this greedy strategy does not reach the global optimum of
Eq.~\eqref{eq:objective}, we find that empirically the algorithm attains
a reasonably large value of the objective function.
We note that our one-level pair-clustering problem is a particular example of
the so-called \emph{submodular welfare problem}~\cite{VONDRAK} with cardinality
constraints; however, we will not pursue this direction for a general $K >2$
with the one-level pair clustering. Rather, we will apply it with $K=2$ in
a hierarchical manner, which will be discussed next.

\subsection{Hierarchical Pair-Clustering}
\label{sec:hierarchical-clustering}
In order to build a multiscale graph wavelet packet dictionary,
we develop a hierarchical (i.e., multilevel) version of the
pair-clustering algorithm. First, let us assume that the hierarchical
bipartition tree of $V^\star$ is already computed using the same
algorithm discussed in Sect.~\ref{sec:gstar-partition}.
We now begin with level $j=0$ where $V^{(0)}_0$ is simply
$V = \{\bdelta_1,\bdelta_2,\cdots,\bdelta_N\}$ and $V^{\star(0)}_0$ is
$V^\star = \{\bphi_0,\bphi_1,\cdots,\bphi_{N-1}\}$.
Then, we perform one-level pair-clustering algorithm ($K=2$) to get 
$\left(V^{\star(1)}_0, V^{(1)}_0\right)$ and then $\left(V^{\star(1)}_1, V^{(1)}_1\right)$.
We iterate the above process to generate paired clusters
$\left(V^{\star(j)}_k, V^{(j)}_k\right)$, $j=0:J$, $k=0:2^j-1$.
Note that the hierarchical pair-clustering algorithm ensures nestedness in 
both the primal node domain $V$ and the dual/eigenvector domain $V^\star$.

\subsection{Generating the NGWP Dictionary}
Once we generate two hierarchical bipartition trees $\left\{V^{(j)}_k\right\}$ and
$\left\{V^{\star(j)}_k\right\}$, we can proceed to generate the NGWP vectors
$\left\{\Psi^{(j)}_k\right\}$ that are necessary to form an NGWP dictionary.
For each $\bdelta_l \in V^{(j)}_k$, we first compute the orthogonal projection of
$\bdelta_l$ onto the span of $V^{\star(j)}_k$,
i.e., $\sspan\left(\Phi^{(j)}_k\right)$ where $\Phi^{(j)}_k$ are those
eigenvectors of $L(G)$ belonging to $V^{\star(j)}_k$. Unfortunately,
$\Phi^{(j)}_k \left(\Phi^{(j)}_k\right)^\transp \bdelta_l$ and
$\Phi^{(j)}_k \left(\Phi^{(j)}_k\right)^\transp \bdelta_{l'}$ are not mutually
orthogonal for $\bdelta_l, \bdelta_{l'} \in V^{(j)}_k$ in general.
Hence, we need to perform orthogonalization of the vectors
$\left\{\Phi^{(j)}_k \left(\Phi^{(j)}_k\right)^\transp \bdelta_l\right\}_l$.
We use the \emph{modified Gram-Schmidt with $\ell^p (0 < p < 2)$ pivoting
orthogonalization} (MGSLp)~\cite{CoifmanRonaldR2006Dw} to generate the
orthonormal graph wavelet packet vectors associated with $V^{\star(j)}_k$
(and hence also $V^{(j)}_k$). This MGSLp algorithm listed in
Appendix~\ref{app:mgslp} tends to generate localized orthonormal vectors
because the $\ell^p$-norm\footnote{We typically set $p=1$ here, and in fact,
that setting was used in all the numerical experiments with the PC-NGWP
dictionary in this article.} pivoting promotes sparsity.
We refer to the graph wavelet packet dictionary
$\left\{\Psi^{(j)}_k\right\}_{j=0:J; \, k=0:2^j-1}$ generated by this algorithm
as the \emph{Pair-Clustering Natural Graph Wavelet Packet} (PC-NGWP) dictionary.

Let us now briefly discuss the performance of the PC-NGWP dictionary
on the same examples in Sect.~\ref{sec:varimax-ngwp}, i.e., $P_{512}$
and $P_7 \times P_3$, without displaying figures to save pages.
We essentially obtained the similar wavelet packet vectors in both cases as
those shown in Figs.~\ref{fig:shannon512} and \ref{fig:grid7x3varimax} using
the VM-NGWP dictionaries; yet they are not exactly the same: the localization
of those PC-NGWP vectors in the primal node domain is worse (e.g., with
larger sidelobes) than that of the VM-NGWP vectors mainly due to the
MGSLp orthogonalization procedure (even if it promoted sparsity).

\subsection{Computational Complexity}
\label{sec:pc-cost}
At each $V^{\star(j)}_k$ of the hierarchical bipartition tree of the dual
graph $G^\star$, the orthogonal projection of the standard basis vectors
in $V^{(j)}_k$ onto $\sspan\left(\Phi^{(j)}_k\right)$ and the MGSLp procedure
are the two main computational burden for our PC-NGWP dictionary construction.
The orthogonal projection costs $O\left(N \cdot (N^j_k)^2 + N \cdot N^j_k\right)$
while the MGSLp costs $O\left(2 N \cdot (N^j_k)^2\right)$. Hence, the dominating cost for
this procedure is $O\left(3 N \cdot (N^j_k)^2\right)$ for each $(j,k)$.
And we need to sum up this cost on all the tree nodes.
Let us analyze the special case of the perfectly balanced and fully developed
bipartition tree with $N=2^\jmax$ as we did for the VM-NGWP in
Sect.~\ref{sec:varimax-cost}.
In this case, the bipartition tree has $1+\jmax$ levels, and $N^j_k = 2^{\jmax-j}$,
$k=0:2^j-1$.
So, for the $j$th level, using Eq.~\eqref{eq:Njk2}, we have
$O(3 N^3 \cdot 2^{-j})$. 
Finally, by summing this from $j=1$ to $\jmax-1$ (again, no computation is
needed at the root and the bottom levels), the total cost for
PC-NGWP dictionary construction in this ideal case is:
$O(3N^3 \cdot (1-2/N)) \approx O(3N^3)$.
So, it still requires $O(N^3)$ operations; the difference from that of
the VM-NGWP is the constants, i.e.,
$3$ (PC-NGWP) vs $1+1000/3 \approx 334$ (the worst case VM-NGWP).

%% file: applications.tex

\section{Applications in Graph Signal Approximation}
\label{sec:applications}
In this section, we demonstrate the usefulness of our proposed NGWP dictionaries
in efficient approximation of graph signals on two graphs, and compare the
performance with the other previously proposed methods:
the global graph Laplacian eigenbasis;
the HGLET best basis~\cite{IRION-SAITO-HGLETS};
the graph Haar basis; the graph Walsh basis;
the GHWT coarse-to-fine (c2f) best basis~\cite{IRION-SAITO-GHWT};
the GHWT fine-to-coarse (f2c) best basis~\cite{IRION-SAITO-GHWT}; and
the eGHWT best basis~\cite{SHAO-SAITO-SPIE}.
Recall that the HGLET is a graph version of the Hierarchical Block
DCT dictionary, which is based on the hierarchical bipartition tree of a
primal graph, as briefly discussed in Introduction. Although the HGLET
can choose three different types of graph Laplacian eigenvectors
of $L$, $\Lrw$, and $\Lsym$ at each subgraph (see Eq.~\eqref{eq:graphlaps}),
we only use those of the unnormalized $L$ at each subgraph in order to
compare its performance in a fair manner with the NGWP dictionaries that
are based on the eigenvectors of $L(G)$.
Note also that the graph Haar basis is a particular basis choosable from
the GHWT f2c dictionary and the eGHWT dictionary while the graph Walsh basis
is choosable from both versions of the GHWT dictionaries as well as the eGHWT;
see~\cite{IRION-SAITO-TSIPN, SHAO-SAITO-SPIE} for the details.
We use the $\ell^1$-norm minimization as the best-basis selection criterion
for all the best bases in our experiments.
The edge weights of the dual graph $G^\star$ are the reciprocals of the DAG
pseudometric between the corresponding eigenvectors of $L(G)$ as defined in
Eq.~\eqref{eq:DAG}.
For a given graph $G$ and a graph signal $\f$ defined on it,
we decompose $\f$ into those dictionaries and select those bases first.
Then, to measure the approximation performance, we sort the expansion
coefficients in the nonincreasing order of their magnitude, and use
the top $k$ most significant terms to approximate $\f$ where $k$ starts from $0$
up to about 50\% of the total number of terms, or more precisely,
$\lfloor 0.5 N \rfloor + 1$.
All of the approximation performance is measured by the relative $\ell^2$
approximation error with respect to the fraction of coefficients retained,
which we denote $FCR$ for simplicity.

\subsection{Sunflower Graph Signals Sampled on Images} 
\begin{figure}
  \begin{subfigure}{0.45\textwidth}
    \centering\includegraphics[width = \textwidth]{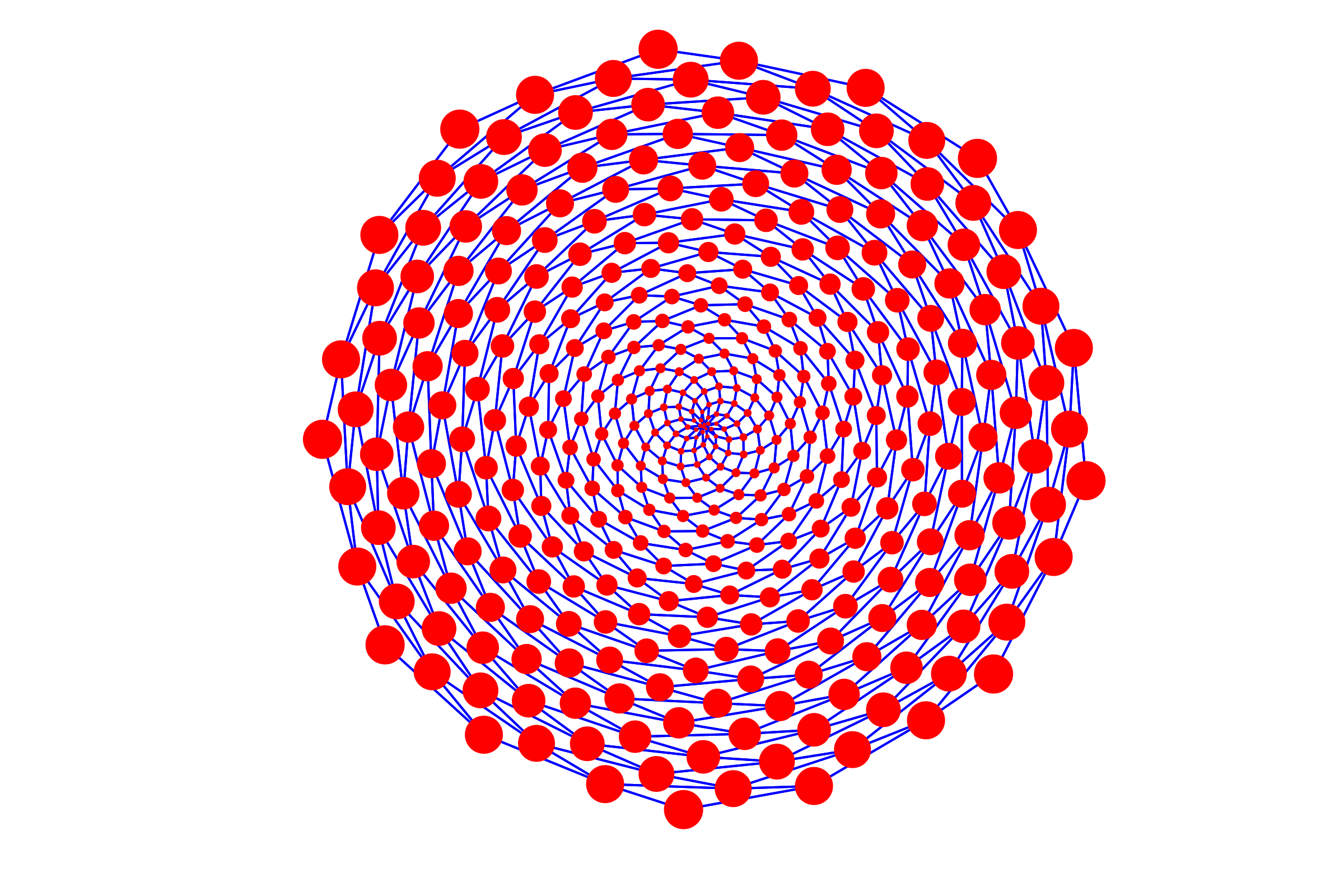}
    \caption{Sunflower graph}
    \label{fig:sunflower-graph}
  \end{subfigure}
  \begin{subfigure}{0.45\textwidth}
    \centering\includegraphics[width = \textwidth]{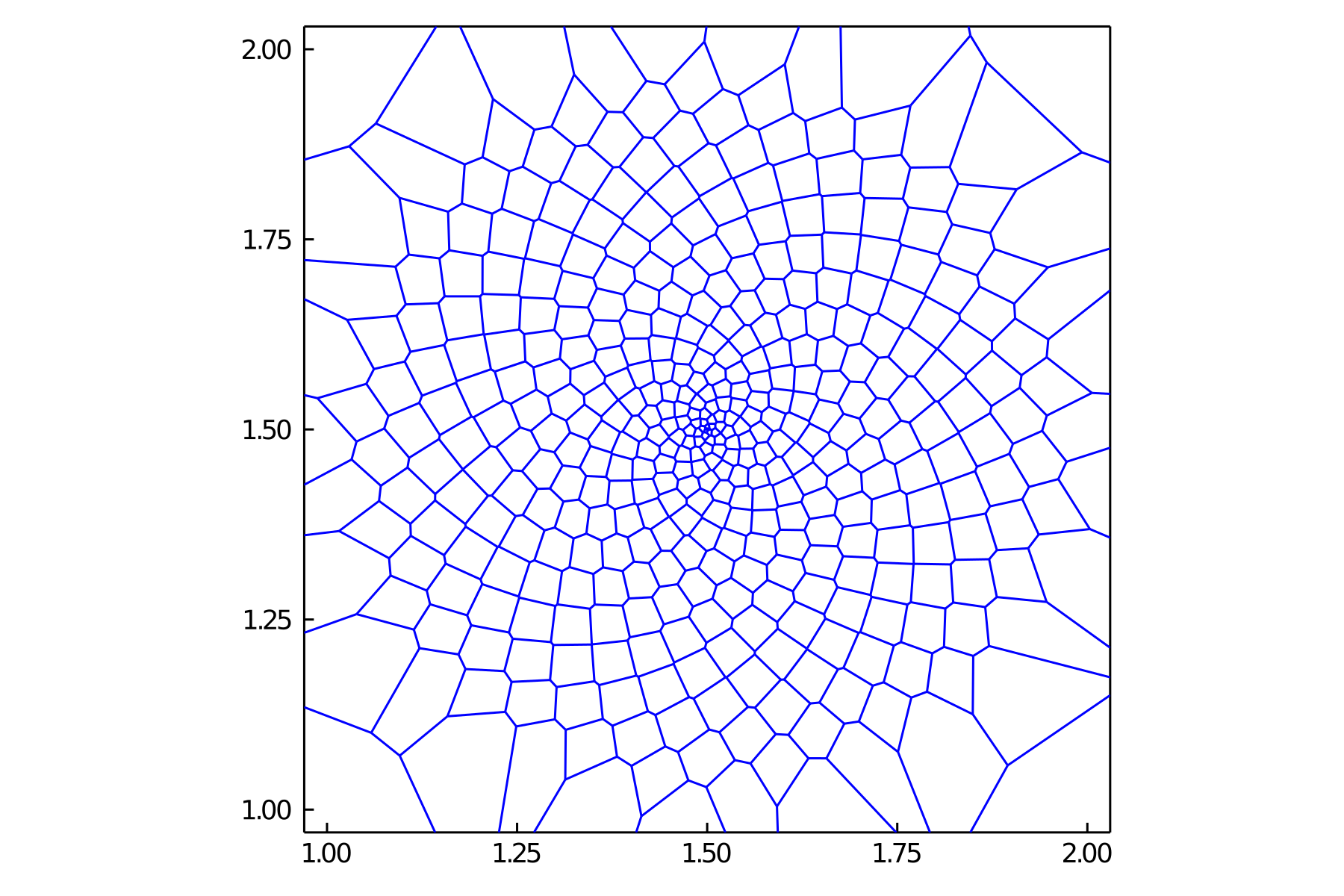}
    \caption{Voronoi tessellation}
    \label{fig:sunflower-voronoi}
  \end{subfigure}
  \caption{Sunflower graph ($N=400$) (a); node radii vary for visualization
    purpose; its Voronoi tessellation (b)}
\end{figure}

We consider the so-called ``sunflower'' graph shown in
Fig.~\ref{fig:sunflower-graph}. This particular graph has $400$ nodes and
each edge weight is set as the reciprocal of the Euclidean distance between
the endpoints of that edge.
Consistently counting the number of spirals in such a sunflower graph gives rise
to the \emph{Fibonacci numbers}: $0, 1, 1, 2, 3, 5, 8, 13, 21, 34, 55, \dots$;
see Fig.~\ref{fig:sunflower-graph}.
We also note that the majority of nodes ($374$ among $400$) have degree $4$
while there are eight nodes with degree $2$, $17$ nodes with degree $3$, and
the central node has the greatest degree $9$.
See, e.g., \cite{MATHAI-DAVIS, VOGEL} and our code \texttt{SunFlowerGraph.jl} in
\cite{HAOTIAN-GITHUB}, for algorithms to construct such sunflower grids and
graphs.
We can also view such a distribution of nodes as a simple model of the
distribution of \emph{photoreceptors in mammalian visual systems} due to cell
generation and growth; see, e.g., \cite[Chap.~9]{RODIECK}.
Such a viewpoint motivates us the following sampling scheme:
1) overlay the sunflower graph on several parts of the standard Barbara image;
2) construct the Voronoi tessellation of the bounding square region
with the nodes of the sunflower graph as its seeds as shown in
Fig.~\ref{fig:sunflower-voronoi}; 
3) compute the average pixel value within each Voronoi cell; and
4) assign that average pixel value to the corresponding seed/node\footnote{If a Voronoi cell does not contain any original image pixels (which occurs
at some tiny cells around the center), we bilinearly interpolate
the pixel value at the node location using the nearest image pixel values.}.  
See \cite{YAMAGISHI-SUSHIDA} for more about the relationship between
the Voronoi tessellation and the sunflower graph.
We also note that for generating the Voronoi tessellation,
we used the following open source Julia packages developed by
the JuliaGeometry team~\cite{JuliaGeometry}: 
\texttt{VoronoiDelaunay.jl}; \texttt{VoronoiCells.jl}; and
\texttt{GeometricalPredicates.jl}.
For our numerical experiments, we sampled
two different regions: her left eye and pants, where
quite different image features are represented, i.e., a piecewise-smooth image
containing oriented edges and a textured image with directional oscillatory
patterns, respectively.

\begin{figure}
  \begin{subfigure}{0.325\textwidth}
    \centering\includegraphics[width = \textwidth]{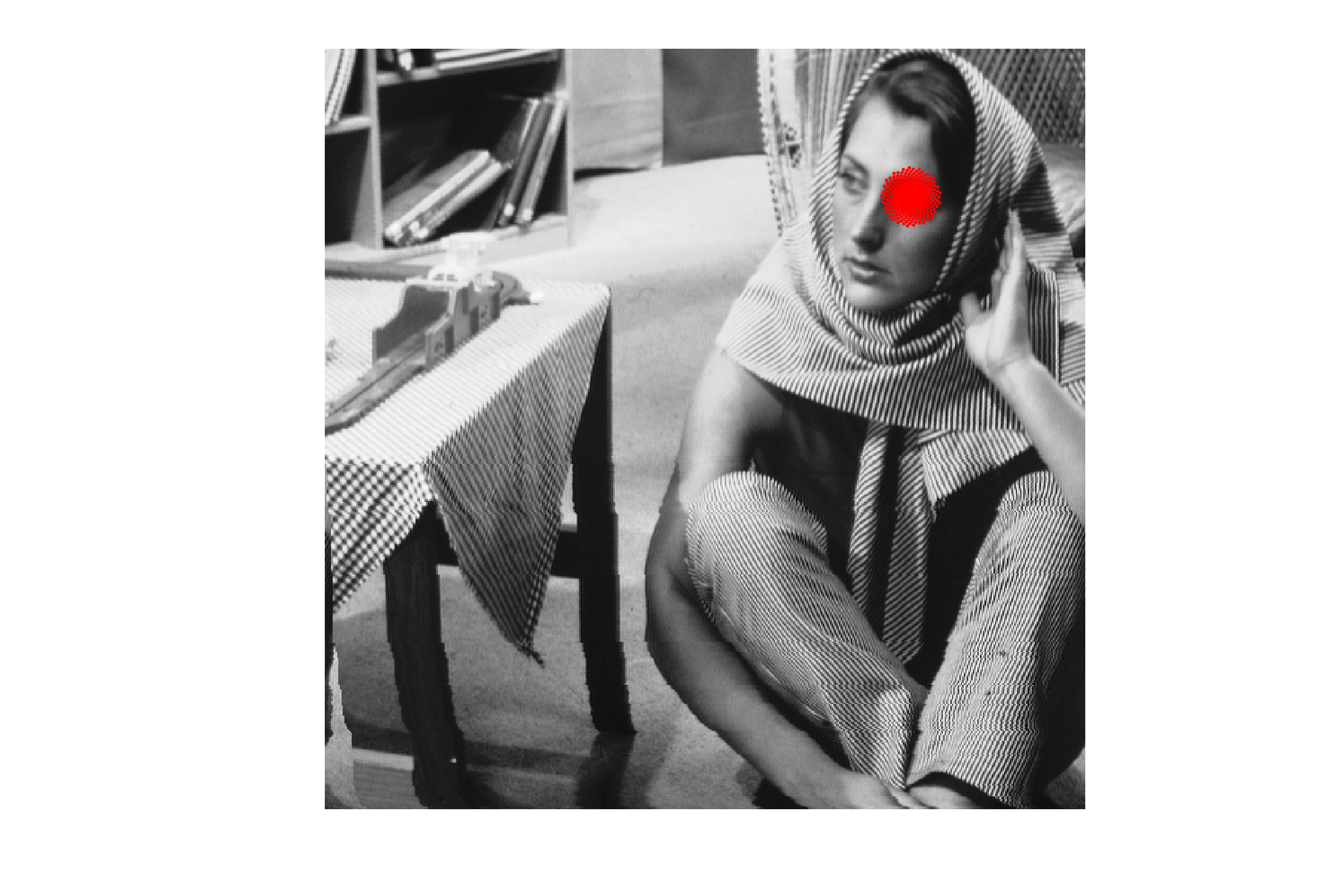}
    \caption{The sunflower graph overlaid on Barbara's left eye}
  \end{subfigure}
  \begin{subfigure}{0.325\textwidth}
    \centering\includegraphics[width = \textwidth]{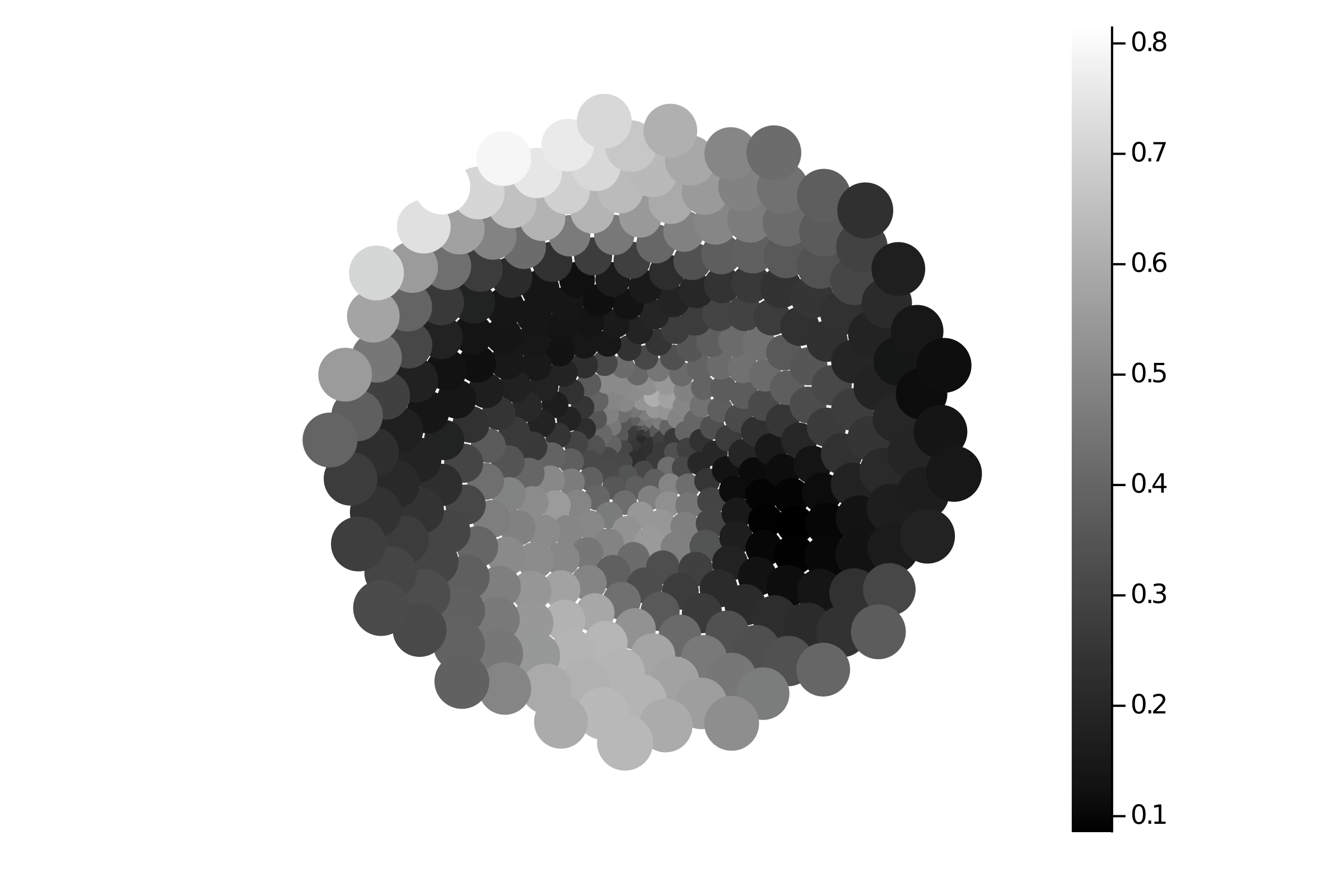}
      \caption{Barbara's left eye as an input graph signal}
  \end{subfigure}
  \begin{subfigure}{0.325\textwidth}
    \includegraphics[width = \textwidth]{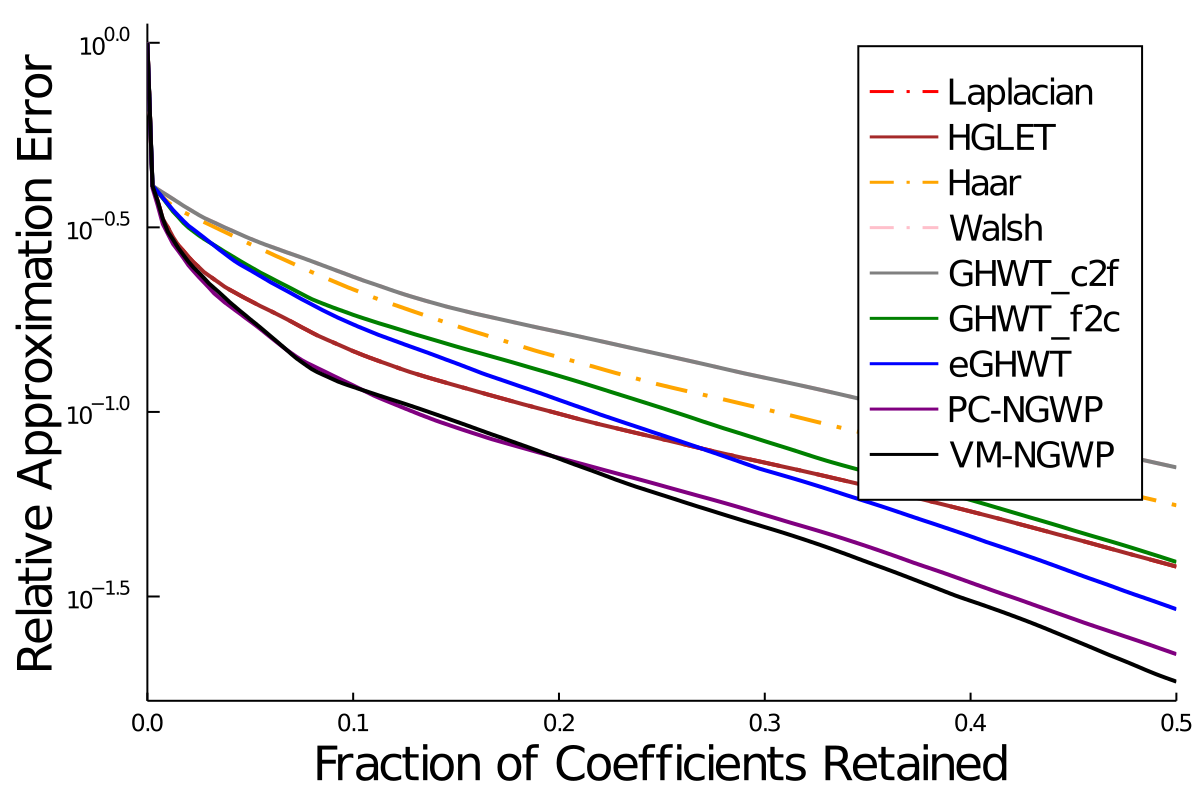}
    \caption{Approximation performance of various methods}
    \label{fig:barb-eye-approx}
  \end{subfigure}
  \caption{Barbara's left eye region sampled on the sunflower graph nodes (a)
    as a graph signal (b); the relative $\ell^2$ approximation errors by
    various methods (c)}
  \label{fig:barb-eye}
\end{figure}

First, let us discuss our approximation experiments on Barbara's eye graph
signal, which are shown in Fig.~\ref{fig:barb-eye}.
From Fig.~\ref{fig:barb-eye-approx}, we observe the following:
1) the VM-NGWP best basis performed best closely followed by the PC-NGWP
best basis;
2) the HGLET best basis chose the global graph Laplacian eigenbasis,
which worked relatively well particularly up to $FCR \approx 0.27$;
and 3) those bases chosen from the Haar-Walsh wavelet packet dictionaries did
not perform well; among them, the eGHWT best basis performed well in the range
$FCR \gtrapprox 0.27$. Note also that the GHWT c2f best basis turned out to be
the graph Walsh basis for this graph signal.
These observations can be attributed to the fact that this Barbara's eye graph
signal is not of piecewise-\emph{constant} nature; rather, it is a
\emph{locally smooth} graph signal.
Hence, the NGWP dictionaries containing smooth \emph{and} localized basis vectors
made a difference in performance compared to the global graph Laplacian
eigenbasis and the eGHWT best basis.

\begin{figure}
  \begin{center}
    \begin{subfigure}{0.245\textwidth}
      \includegraphics[width = \textwidth]{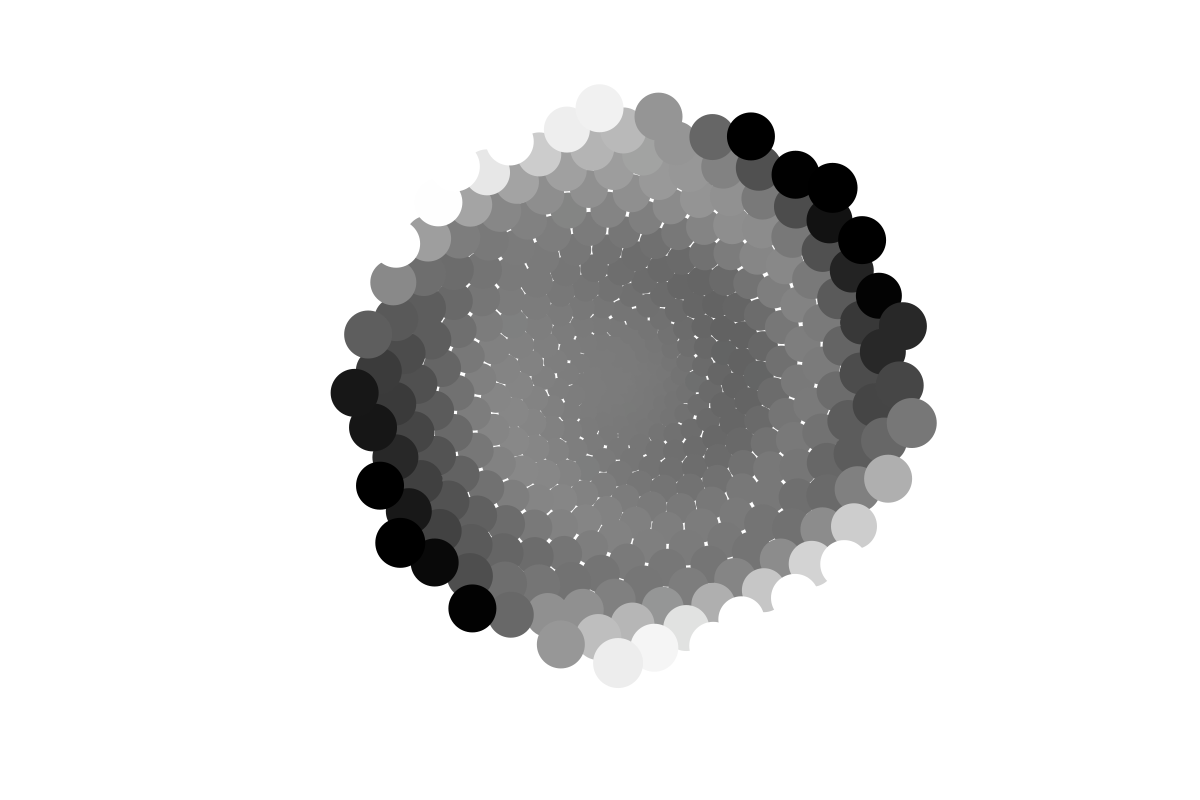}
      \caption{$\bpsi^{(11)}_{13,0} \equiv \bphi_{11}$}
    \end{subfigure}
    \begin{subfigure}{0.245\textwidth}
      \includegraphics[width = \textwidth]{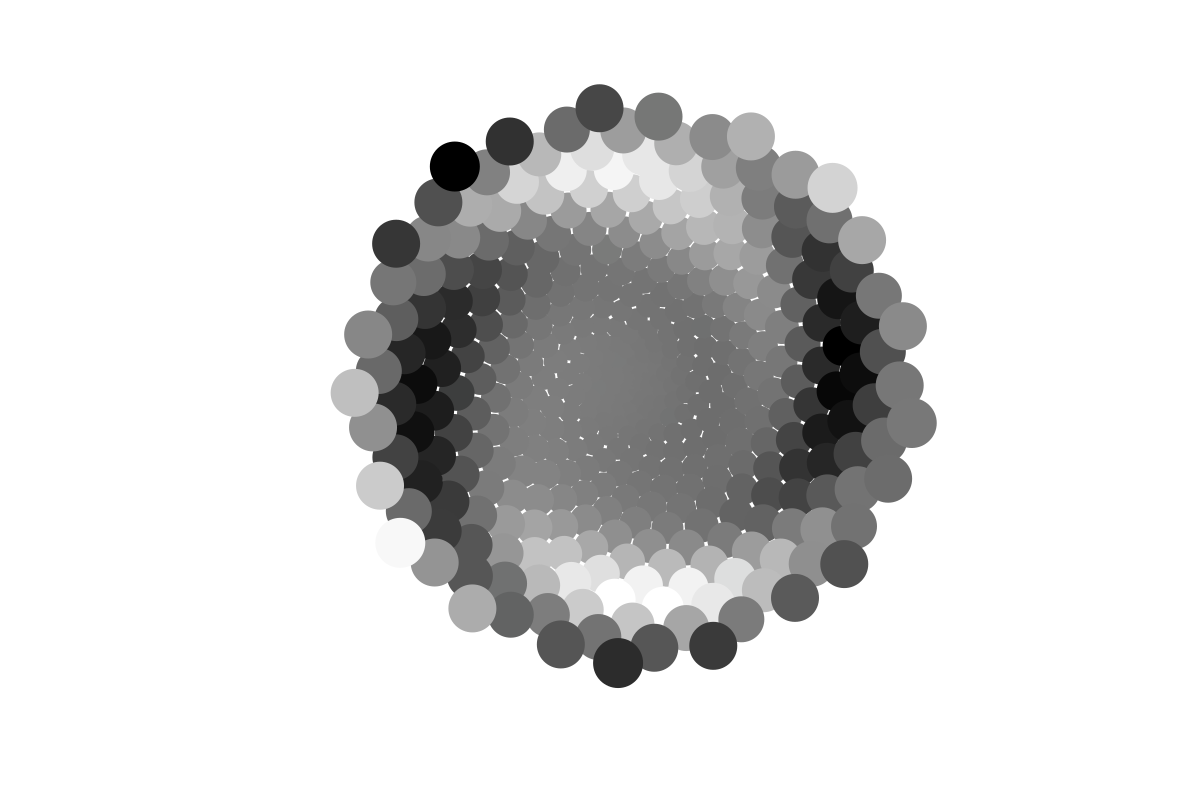}
      \caption{$\bpsi^{(11)}_{26,0} \equiv \bphi_{16}$}
    \end{subfigure}
    \begin{subfigure}{0.245\textwidth}
      \includegraphics[width = \textwidth]{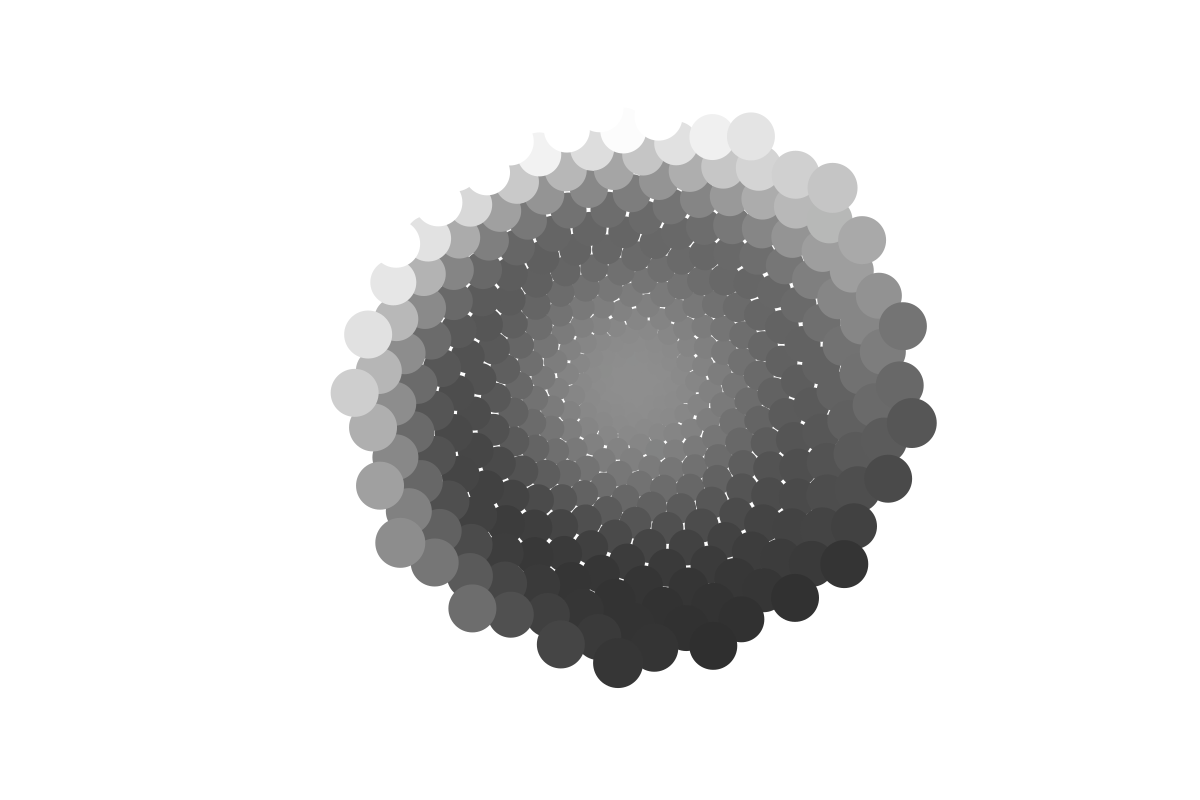}
      \caption{$\bpsi^{(5)}_{1,2}$}
    \end{subfigure}
    \begin{subfigure}{0.245\textwidth}
      \includegraphics[width = \textwidth]{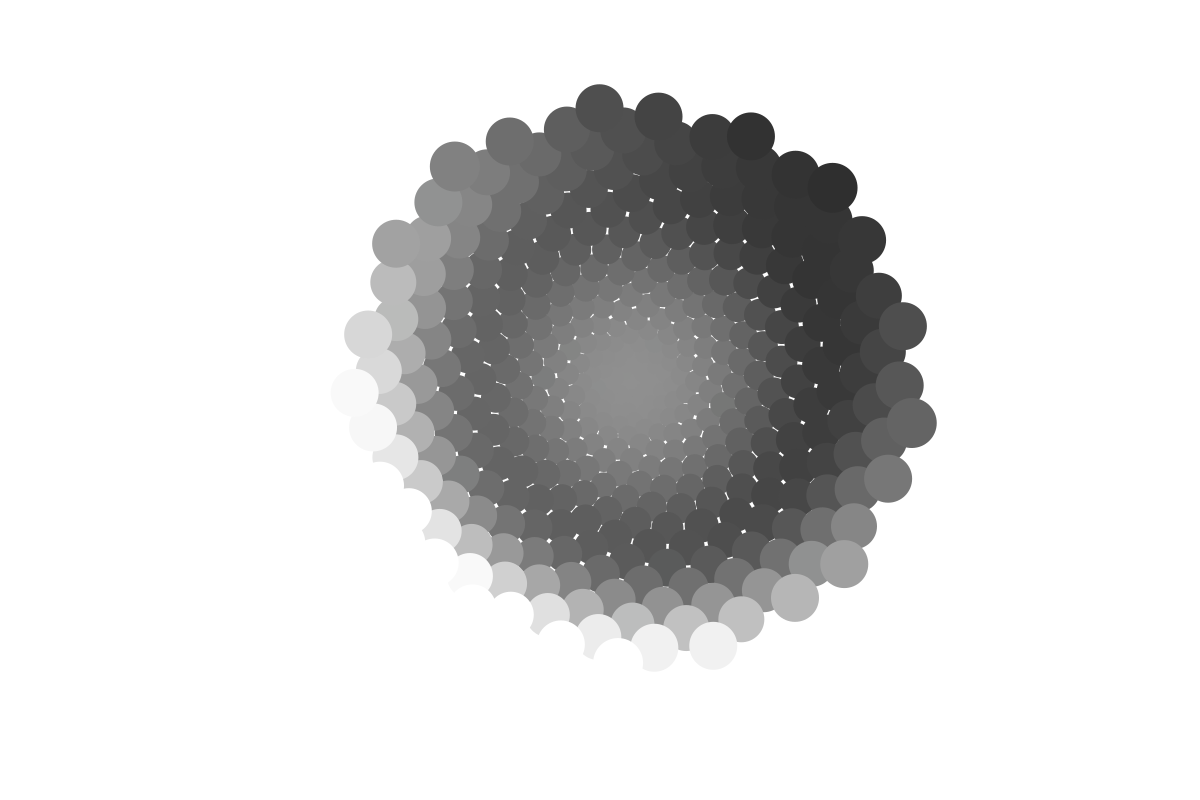}
      \caption{$\bpsi^{(5)}_{1,0}$}
    \end{subfigure}
    \\
    \begin{subfigure}{0.245\textwidth}
      \includegraphics[width = \textwidth]{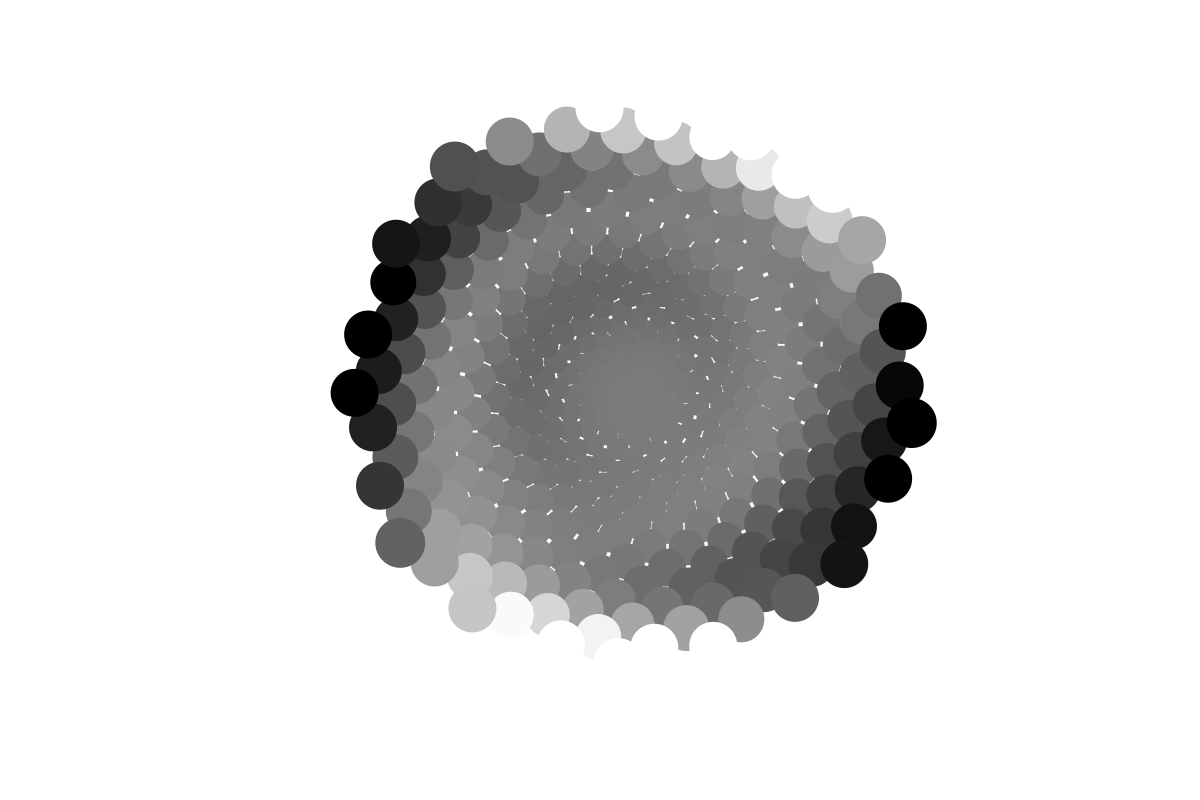}
      \caption{$\bpsi^{(11)}_{14,0} \equiv \bphi_{12}$}
    \end{subfigure}
    \begin{subfigure}{0.245\textwidth}
      \includegraphics[width = \textwidth]{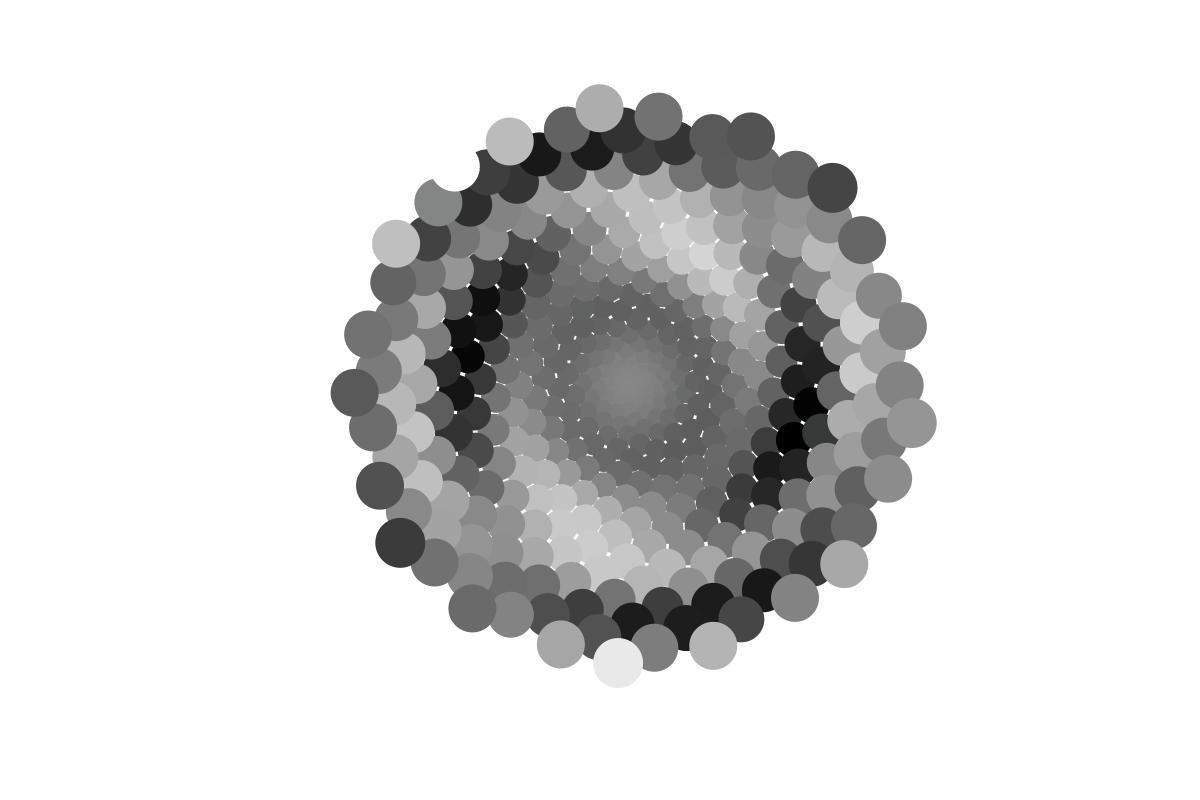}
      \caption{$\bpsi^{(7)}_{21,0}$}
    \end{subfigure}
    \begin{subfigure}{0.245\textwidth}
      \includegraphics[width = \textwidth]{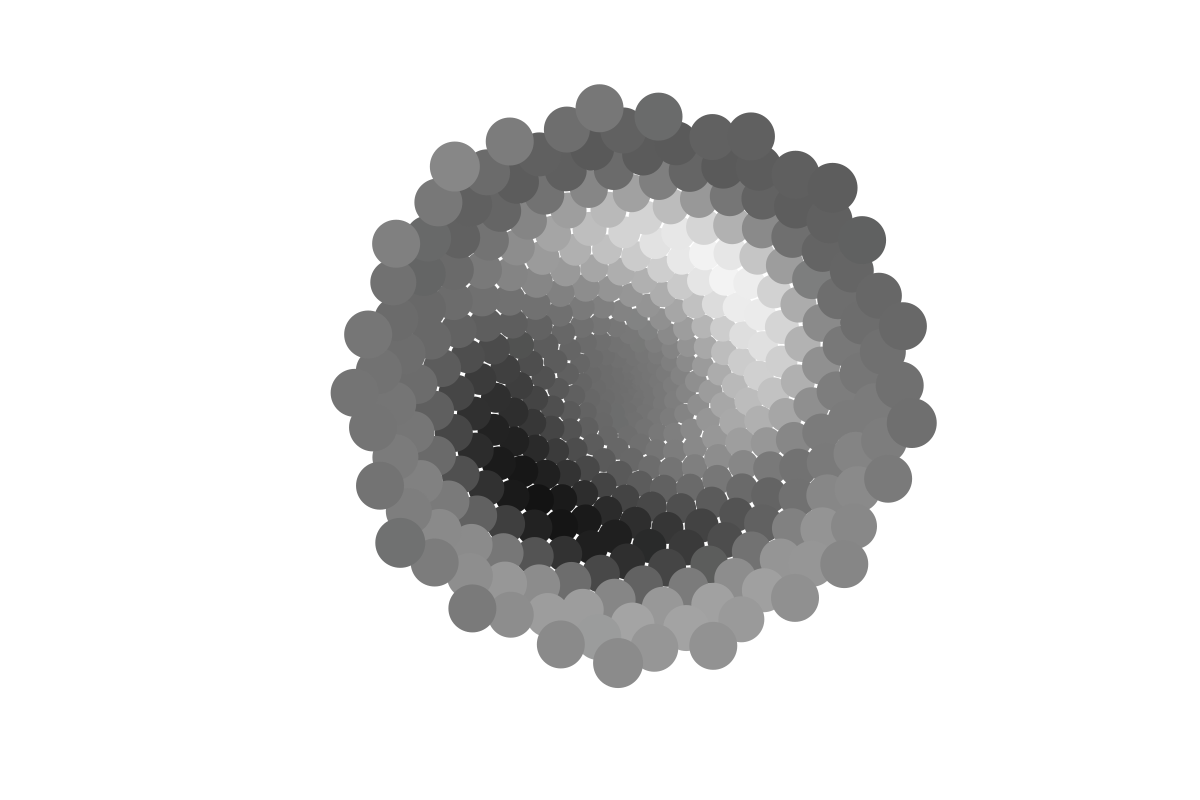}
      \caption{$\bpsi^{(7)}_{5,2}$}
    \end{subfigure}
    \begin{subfigure}{0.245\textwidth}
      \includegraphics[width = \textwidth]{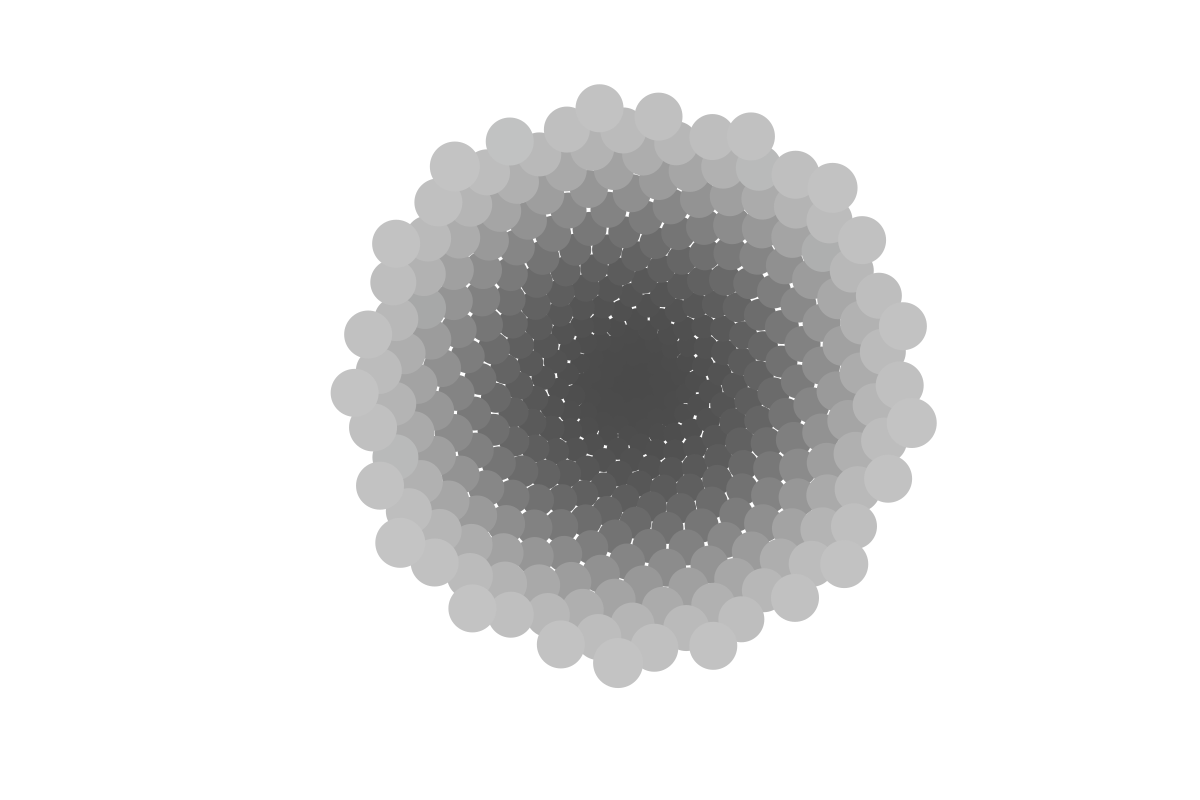}
      \caption{$\bpsi^{(11)}_{1,0} \equiv \bphi_1$}
    \end{subfigure}
    \\
    \begin{subfigure}{0.245\textwidth}
      \includegraphics[width = \textwidth]{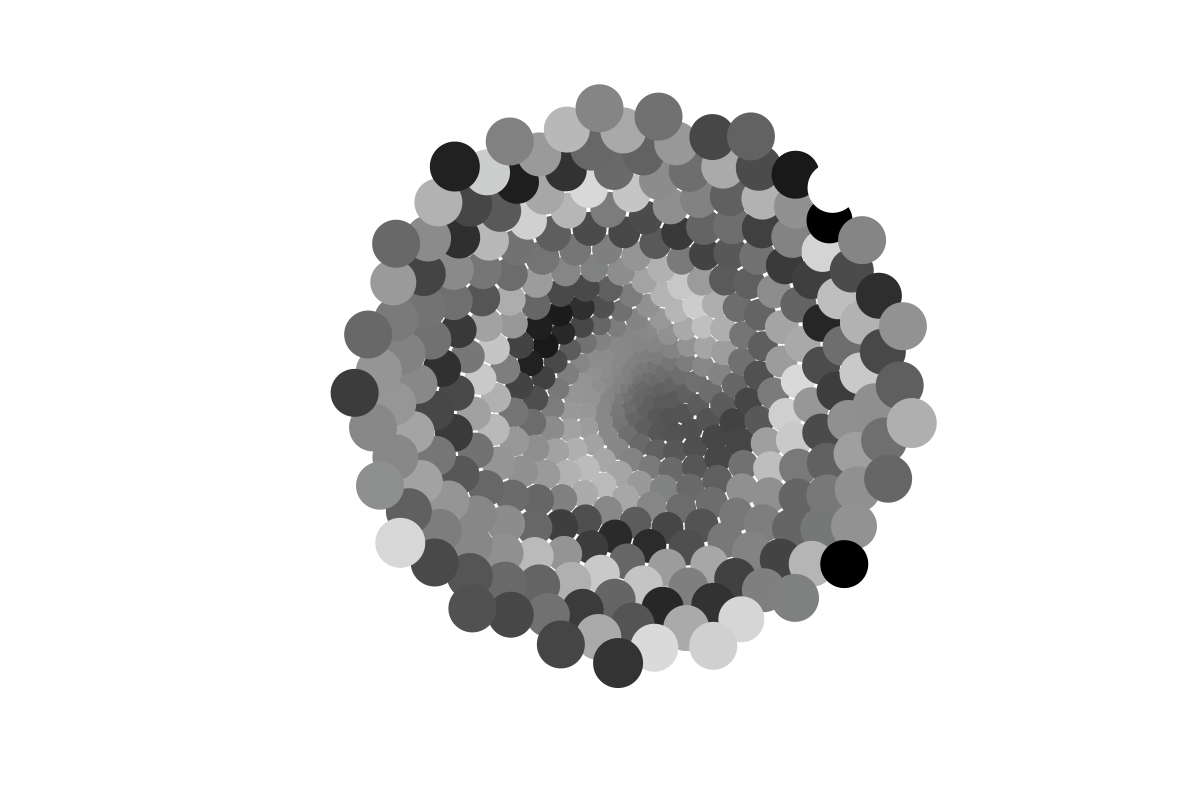}
      \caption{$\bpsi^{(11)}_{40,0} \equiv \bphi_{42}$}
    \end{subfigure}
    \begin{subfigure}{0.245\textwidth}
      \includegraphics[width = \textwidth]{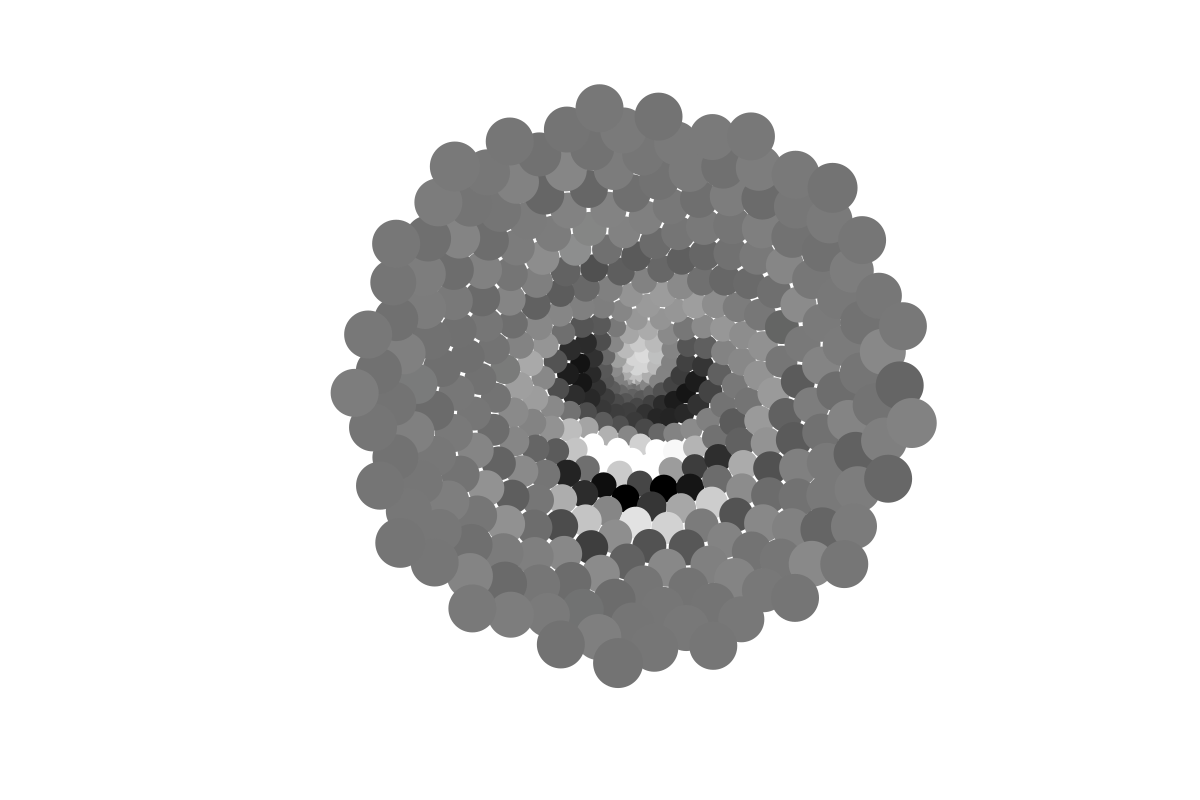}
      \caption{$\bpsi^{(3)}_{2,18}$}
    \end{subfigure}
    \begin{subfigure}{0.245\textwidth}
      \includegraphics[width = \textwidth]{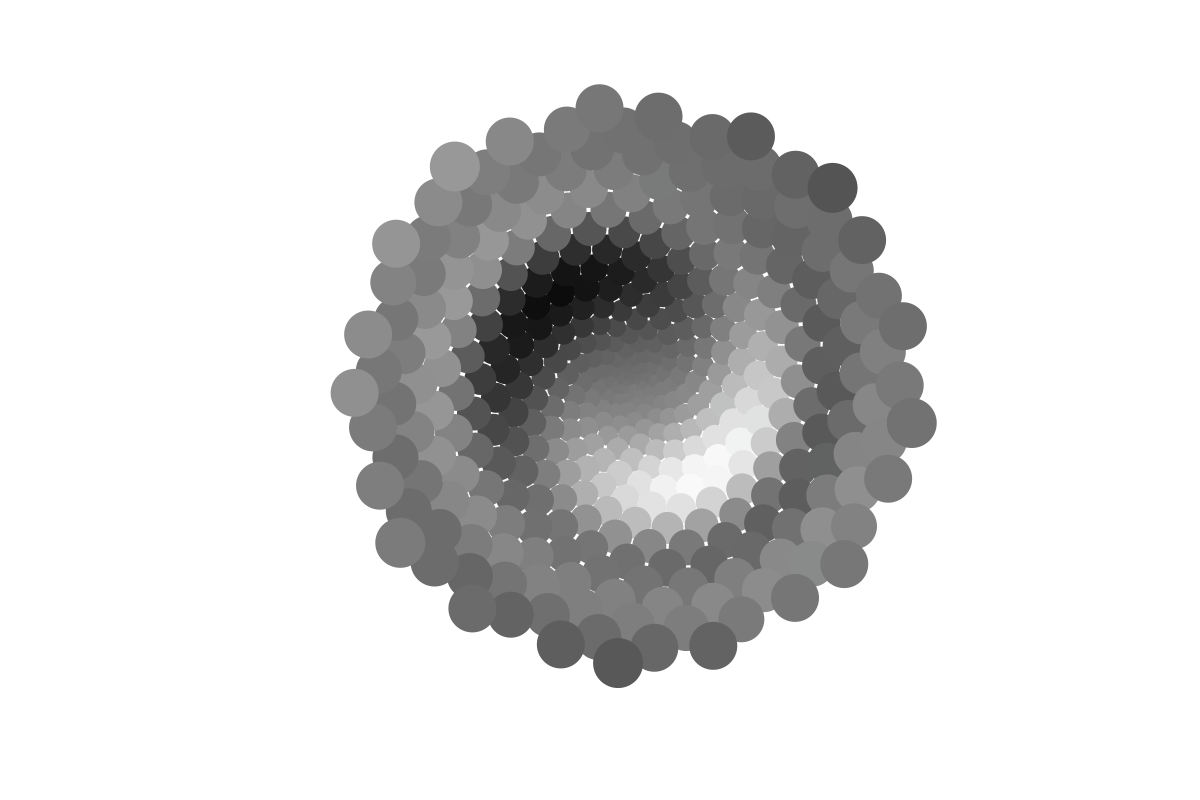}
      \caption{$\bpsi^{(6)}_{5,1}$}
    \end{subfigure}
    \begin{subfigure}{0.245\textwidth}
      \includegraphics[width = \textwidth]{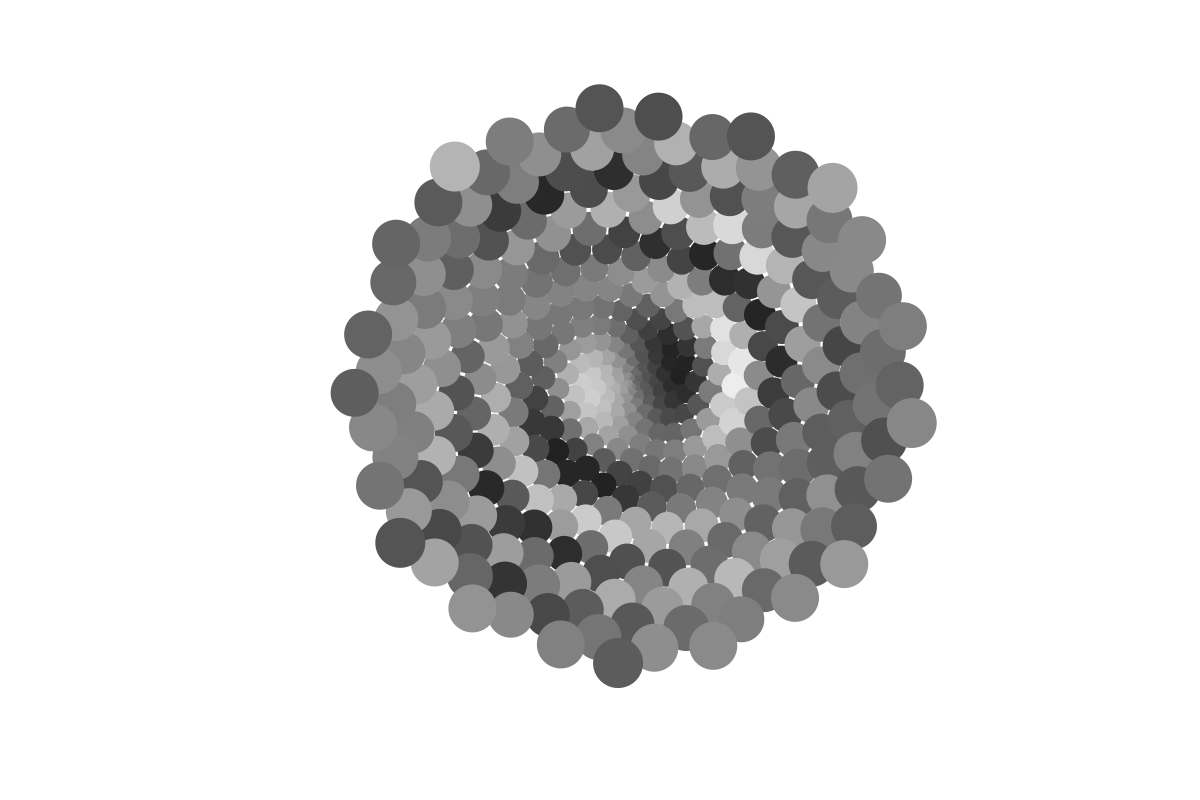}
      \caption{$\bpsi^{(6)}_{11,1}$}
    \end{subfigure}
    \\
    \begin{subfigure}{0.245\textwidth}
      \includegraphics[width = \textwidth]{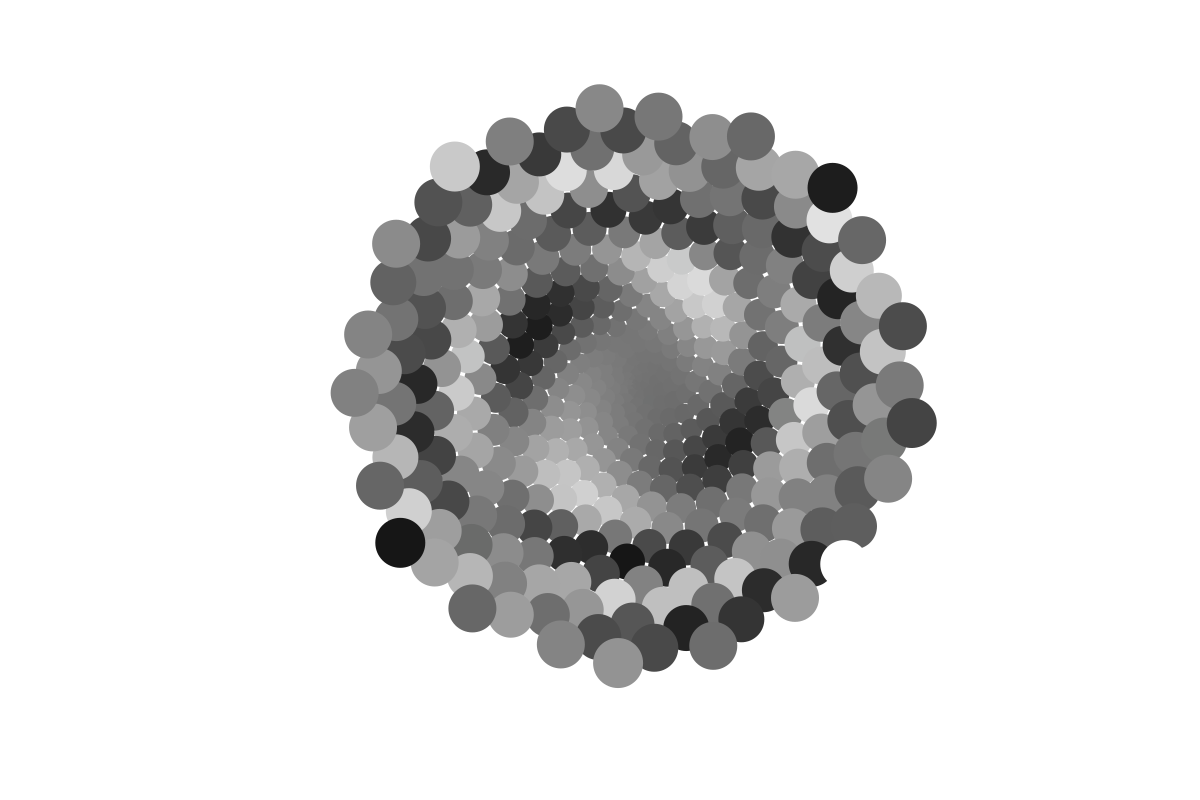}
      \caption{$\bpsi^{(11)}_{21,0} \equiv \bphi_{35}$}
    \end{subfigure}
    \begin{subfigure}{0.245\textwidth}
      \includegraphics[width = \textwidth]{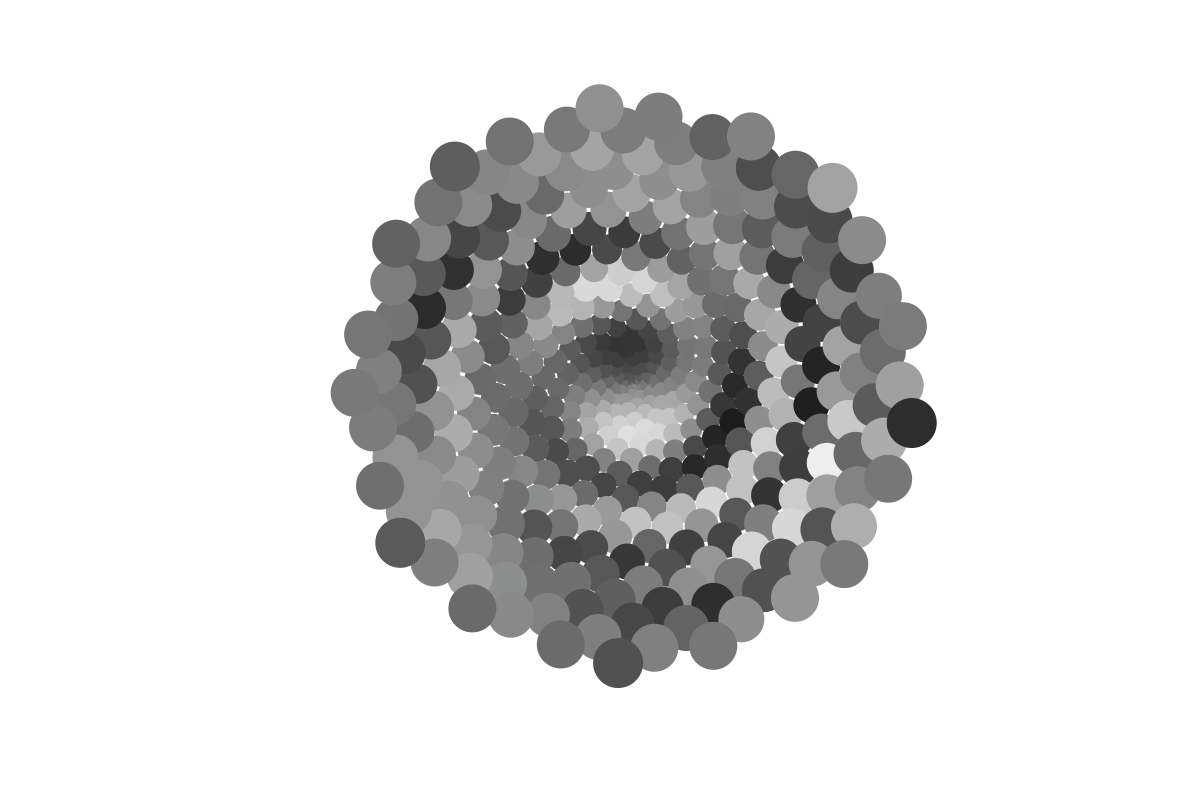}
      \caption{$\bpsi^{(6)}_{15,3}$}
    \end{subfigure}
    \begin{subfigure}{0.245\textwidth}
      \includegraphics[width = \textwidth]{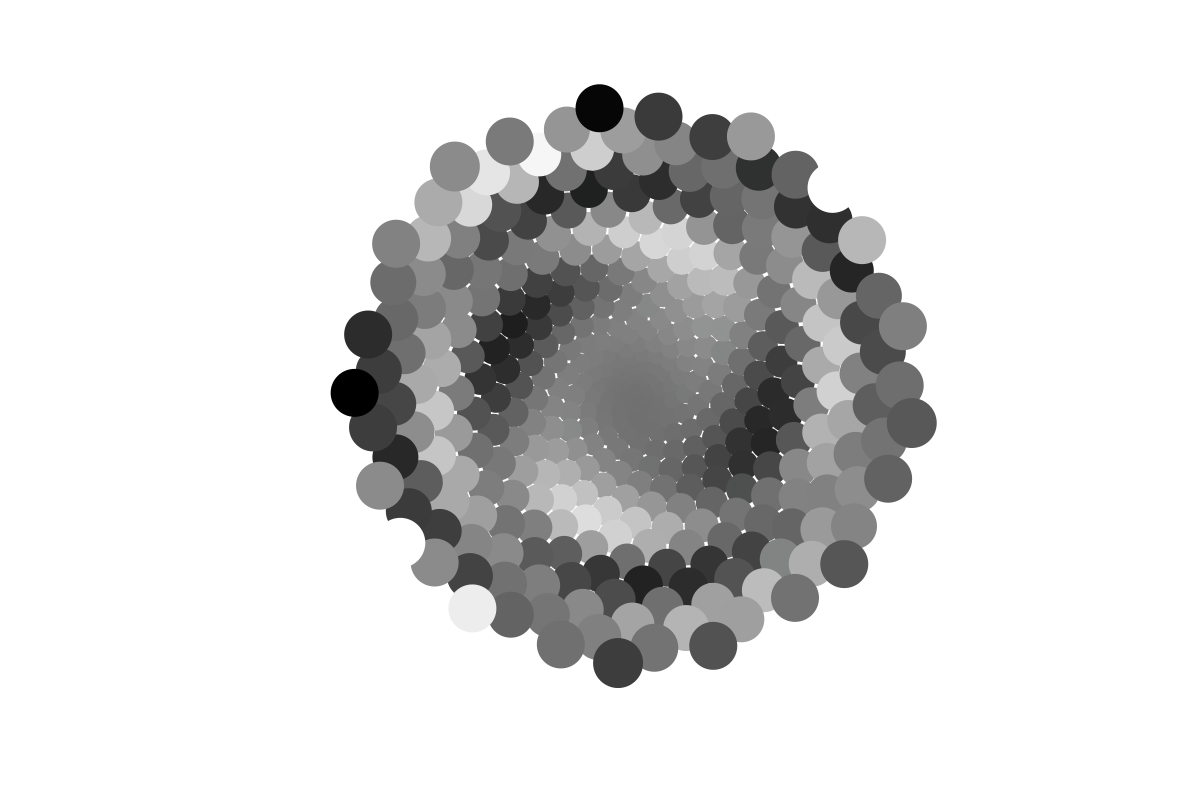}
      \caption{$\bpsi^{(11)}_{30,0} \equiv \bphi_{27}$}
    \end{subfigure}
    \begin{subfigure}{0.245\textwidth}
      \includegraphics[width = \textwidth]{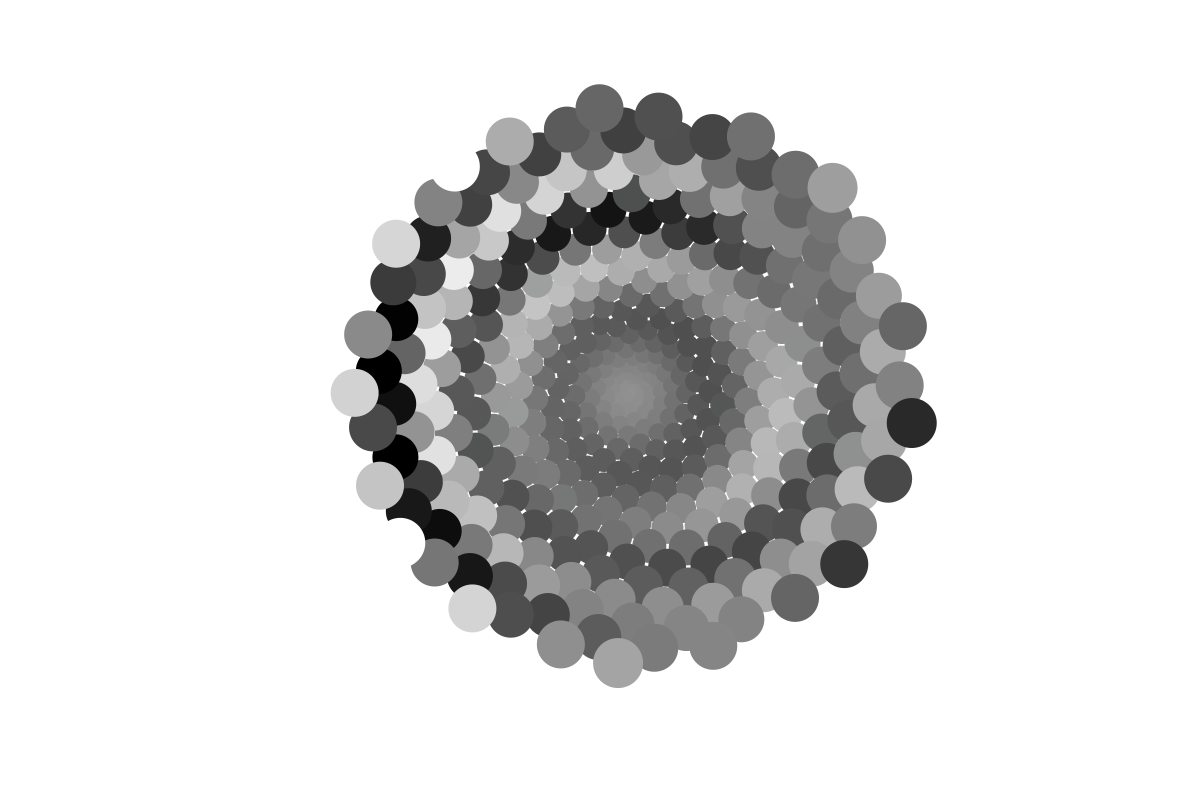}
      \caption{$\bpsi^{(6)}_{9,2}$}
    \end{subfigure}
    \caption{Sixteen most significant VM-NGWP best basis vectors (the DC vector not shown)
      for Barbara's eye. The basis vector amplitudes within $(-0.15, 0.15)$
        are mapped to the grayscale colormap.}
    \label{fig:barb-eye-top16}
  \end{center}
\end{figure}

In order to examine what kind of basis vectors were chosen as the best basis
to approximate this Barbara's eye signal, we display the 16 most
significant VM-NGWP best basis vectors in Fig.~\ref{fig:barb-eye-top16}.
The corresponding PC-NGWP best basis vectors are relatively similar;
hence they are not shown here.
We note that many of these top basis vectors essentially work as oriented
edge detectors for Barbara's eye. For example, $\bpsi^{(7)}_{5,2}$
(Fig.~\ref{fig:barb-eye-top16}g) and
$\bpsi^{(6)}_{5,1}$ (Fig.~\ref{fig:barb-eye-top16}k) try to capture her eyelid
while $\bpsi^{(3)}_{2,18}$ (Fig.~\ref{fig:barb-eye-top16}j),
$\bpsi^{(6)}_{11,1}$ (Fig.~\ref{fig:barb-eye-top16}l),
and $\bpsi^{(6)}_{15,3}$ (Fig.~\ref{fig:barb-eye-top16}n)
do the same for her iris and sclera.
The other basis vectors take care of shading and peripheral features of her
eye region.
We also note that seven among these top 16 best basis vectors are the global
graph Laplacian eigenvectors; see Fig.~\ref{fig:barb-eye-top16}a, b, e, h, i, m, o.

\begin{figure}
  \begin{subfigure}{0.325\textwidth}
    \centering\includegraphics[width = \textwidth]{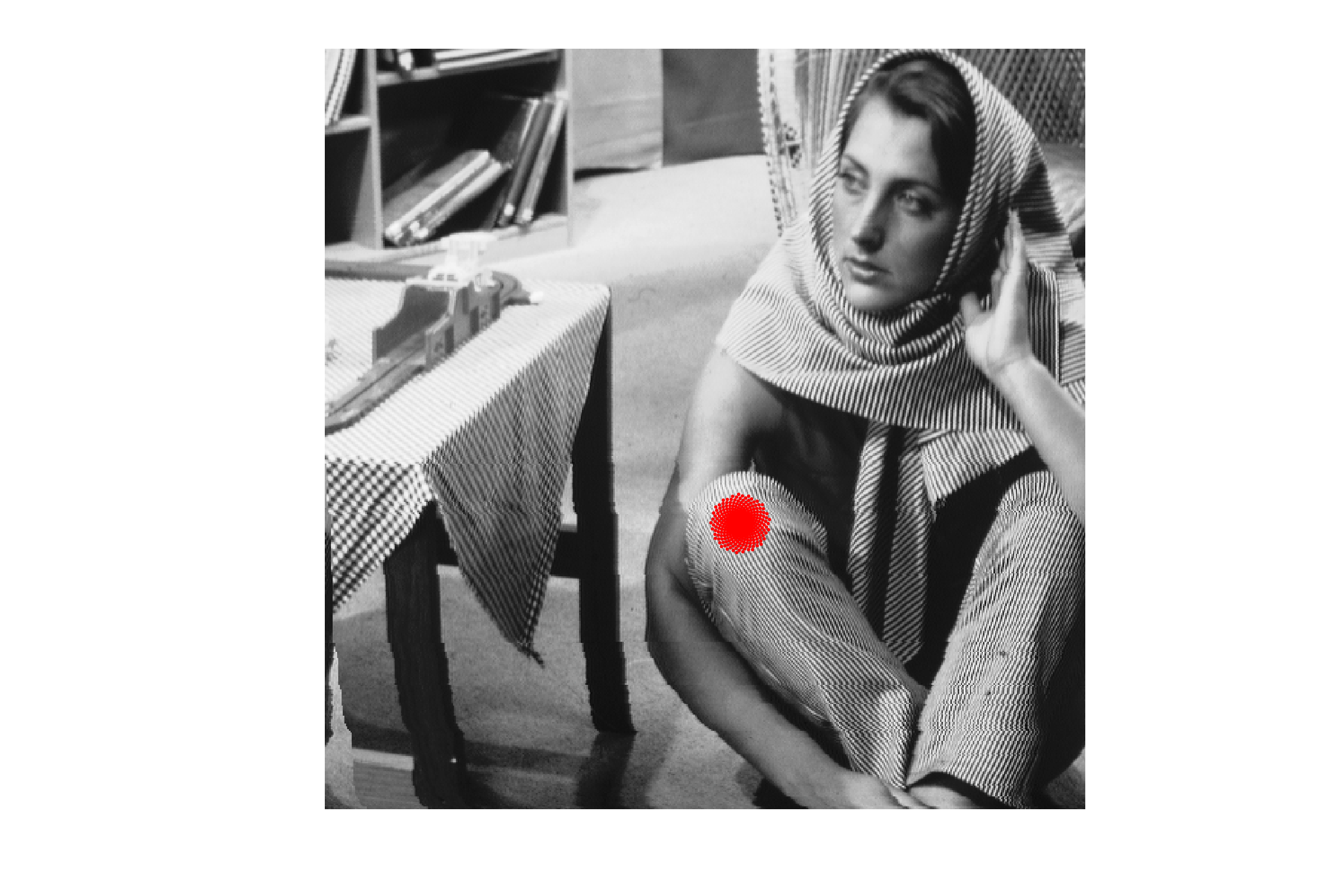}
    \caption{The sunflower graph overlaid on Barbara's pants/knee region}
  \end{subfigure}
  \begin{subfigure}{0.325\textwidth}
    \centering\includegraphics[width = \textwidth]{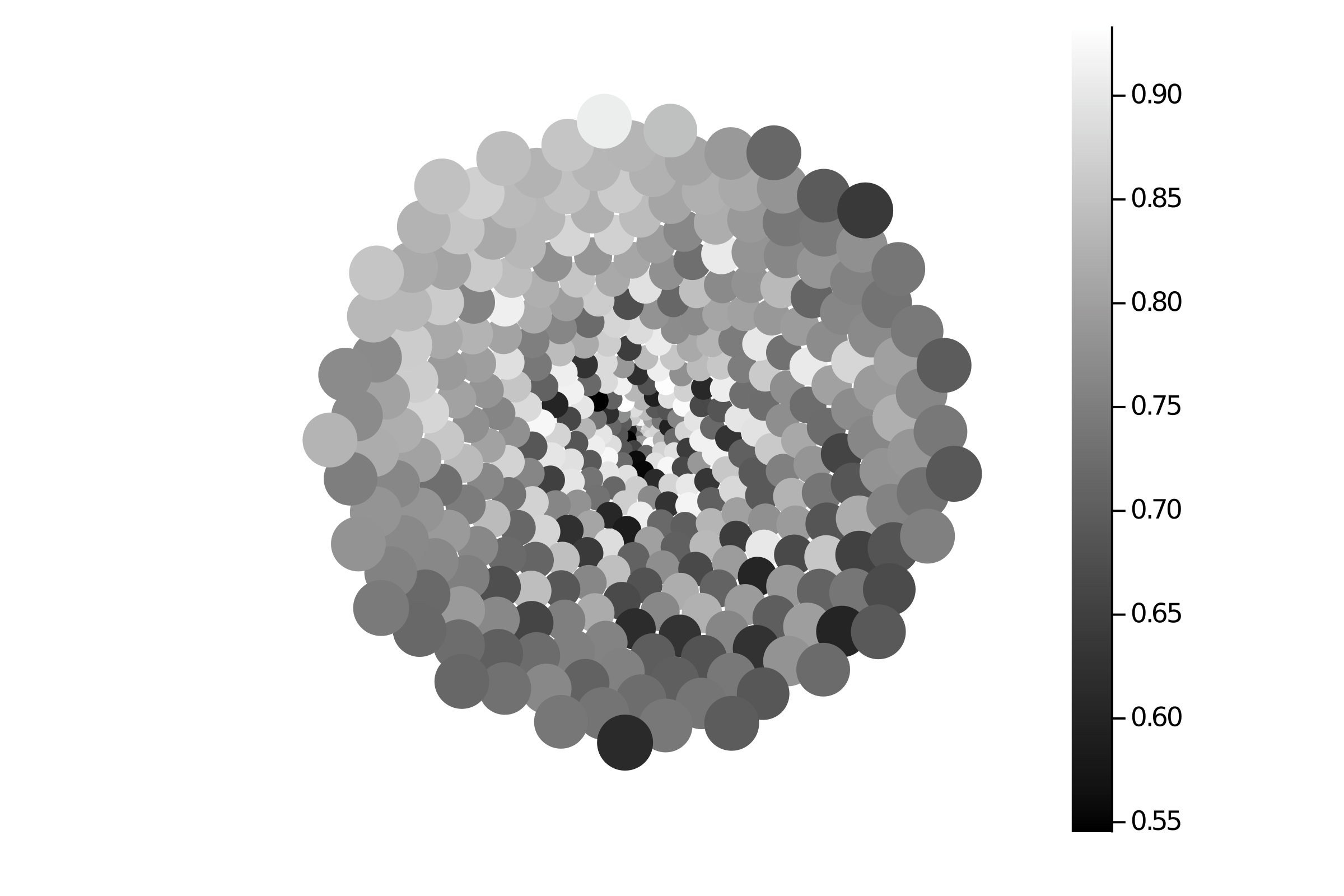}
    \caption{Barbara's pants region as an input graph signal}
    \label{fig:barb-pants-sunflower}
  \end{subfigure}
  \begin{subfigure}{0.325\textwidth}
    \includegraphics[width = \textwidth]{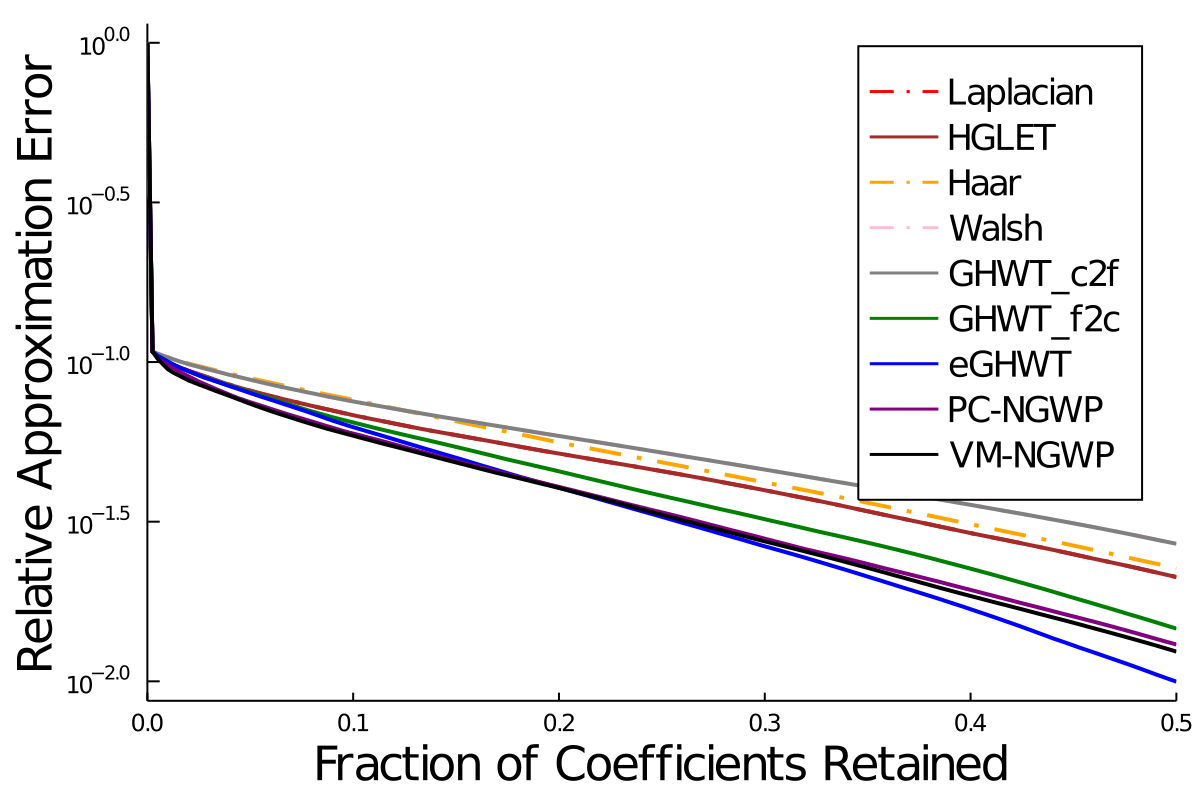}
    \caption{Approximation performance of various methods}
    \label{fig:barb-pants-approx}
  \end{subfigure}
  \caption{Barbara's pants region sampled on the sunflower graph nodes (a)
    as a graph signal (b); the relative $\ell^2$ approximation errors by
    various methods (c)}
  \label{fig:barb-pants}
\end{figure}

Now, let us discuss our second approximation experiments: Barbara's pants
region as an input graph signal as shown in Fig.~\ref{fig:barb-pants}.
The nature of this graph signal is completely different from the eye region:
it is dominated by directional oscillatory patterns of her pants.
From Fig.~\ref{fig:barb-pants-approx}, we observe the following:
1) the NGWP best bases and the eGHWT best basis performed very well and
competitively; the NGWP best bases performed better than the eGHWT best basis
up to $FCR \approx 0.2$ while the latter outperformed all the others for
$FCR \gtrapprox 0.2$;
2) the GHWT f2c best basis performed relatively well behind those three bases;
3) there is a substantial gap in performance between those four bases and
the rest: the graph Haar basis; the GHWT c2f best basis; and the HGLET best basis.
Note that similarly to the case of Barbara's eye, the latter two best bases
coincide with the graph Walsh basis and the global graph Laplacian eigenbasis,
respectively.
We knew that the eGHWT is known to be quite efficient in capturing oscillating
patterns as shown by Shao and Saito for the graph setting~\cite{SHAO-SAITO-SPIE}
and by Lindberg and Villemoes for the classical non-graph
setting~\cite{LINDBERG-VILLEMOES}.
Hence, it is a good thing to observe that our NGWPs are competitive with the
eGHWT for this type of textured signal.

\begin{figure}
  \begin{center}
    \begin{subfigure}{0.245\textwidth}
      \includegraphics[width = \textwidth]{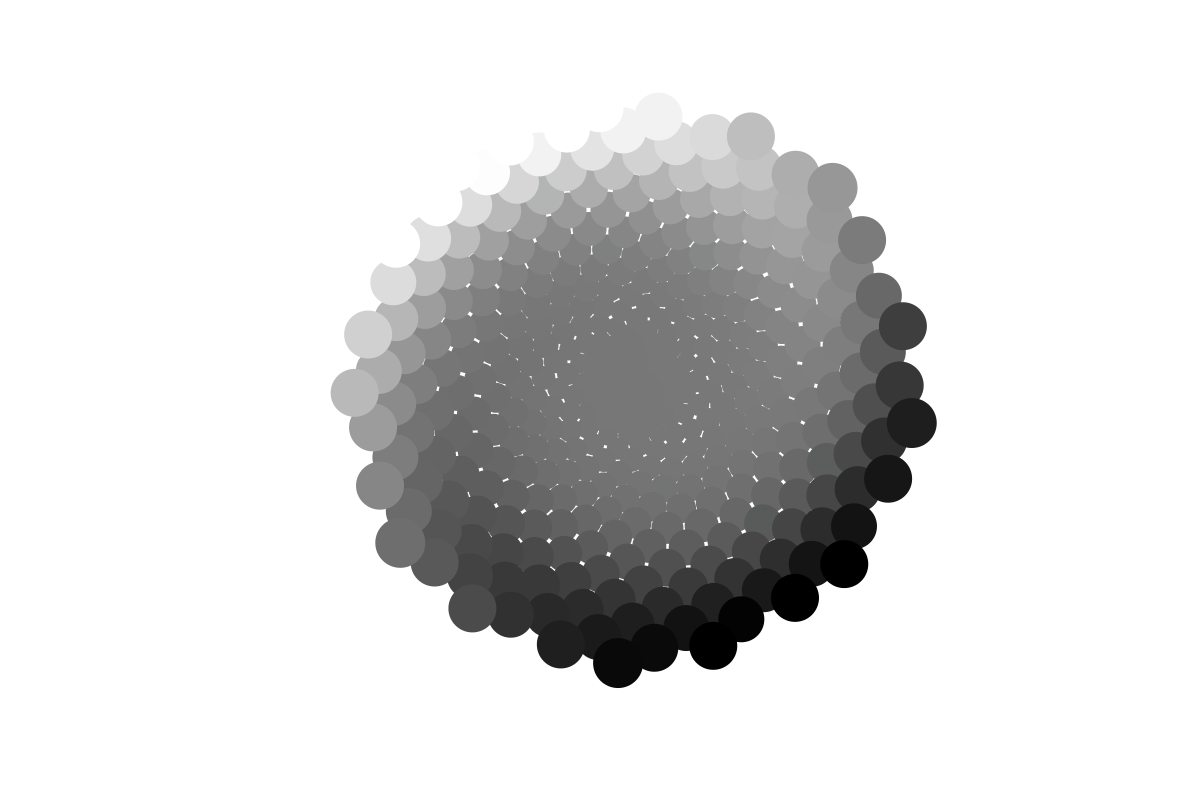}
      \caption{$\bpsi^{(6)}_{3,1}$}
    \end{subfigure}
    \begin{subfigure}{0.245\textwidth}
      \includegraphics[width = \textwidth]{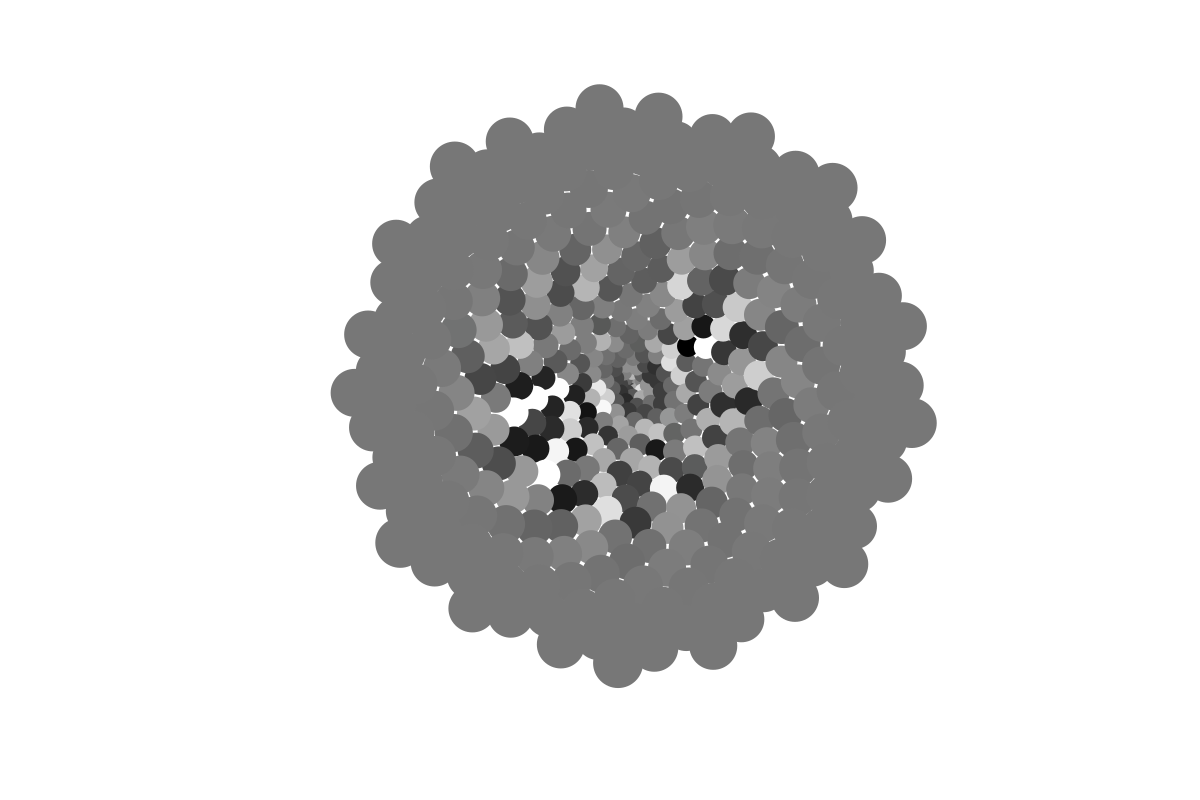}
      \caption{$\bpsi^{(6)}_{49,4}$}
    \end{subfigure}
    \begin{subfigure}{0.245\textwidth}
      \includegraphics[width = \textwidth]{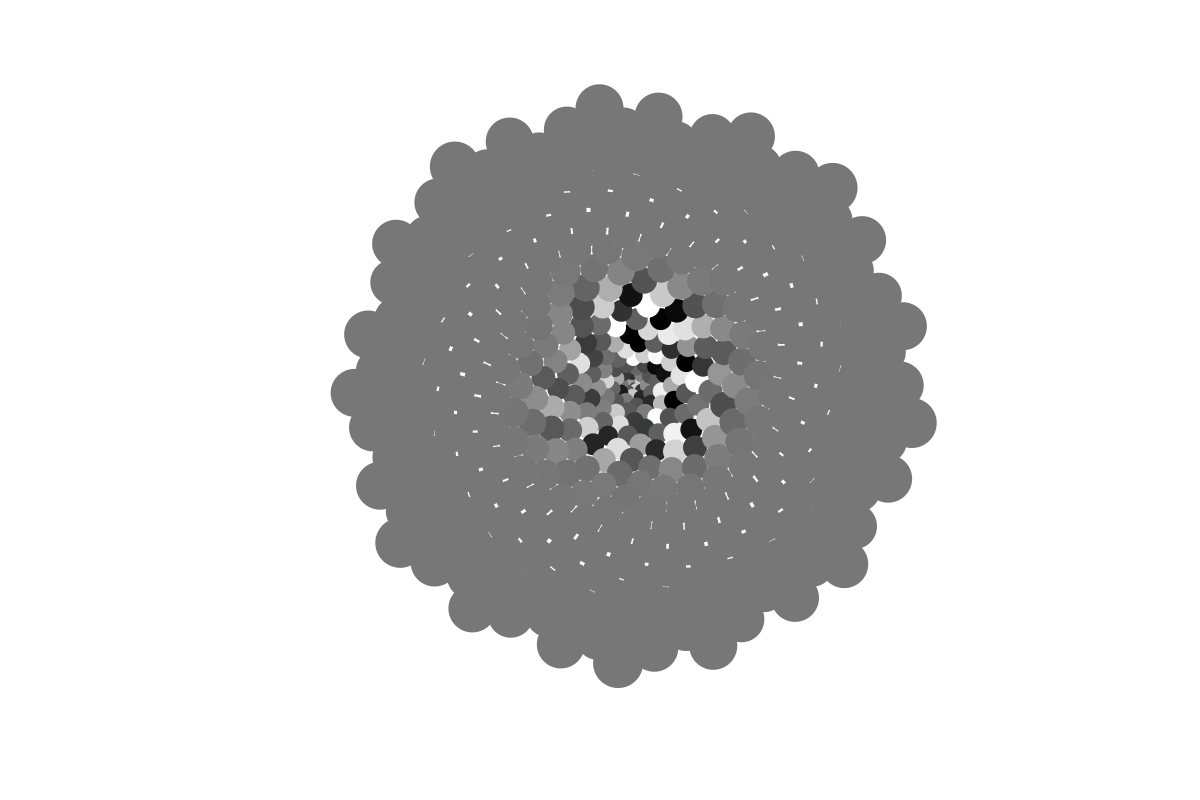}
      \caption{$\bpsi^{(9)}_{325,0}$}
    \end{subfigure}
    \begin{subfigure}{0.245\textwidth}
      \includegraphics[width = \textwidth]{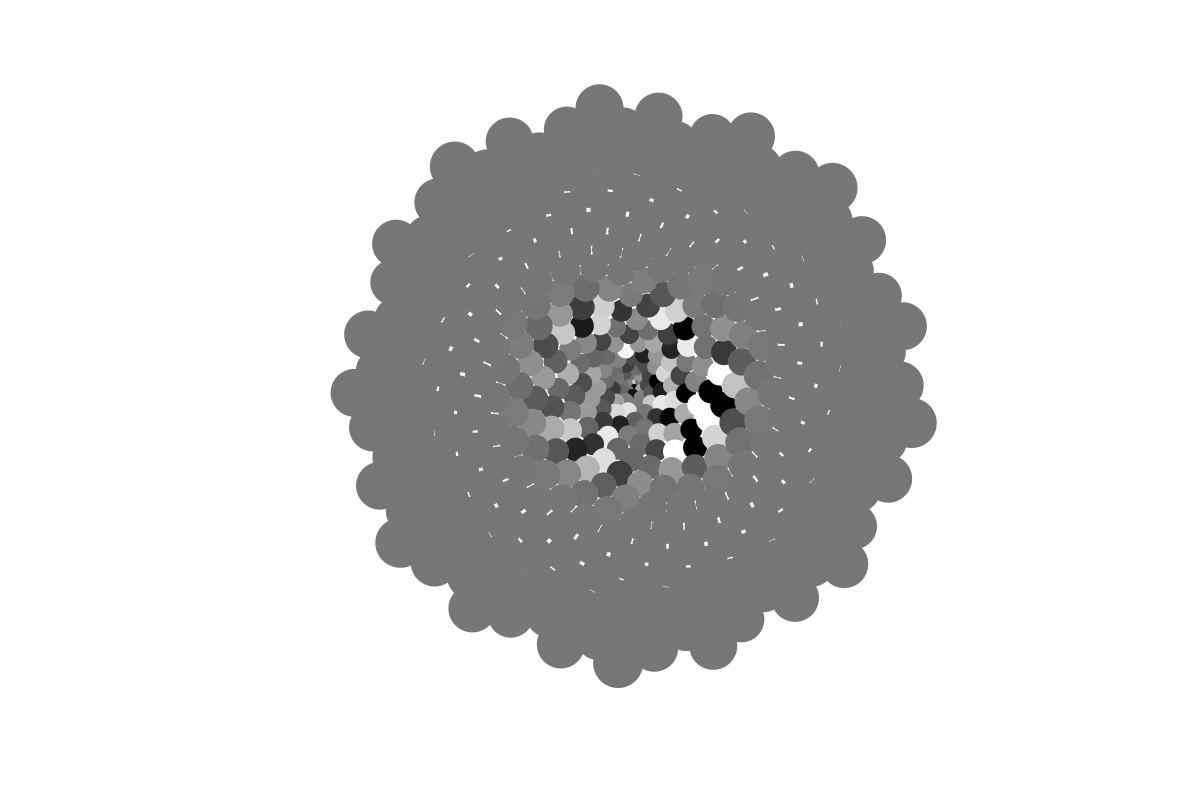}
      \caption{$\bpsi^{(6)}_{56,0}$}
    \end{subfigure}
    \\
    \begin{subfigure}{0.245\textwidth}
      \includegraphics[width = \textwidth]{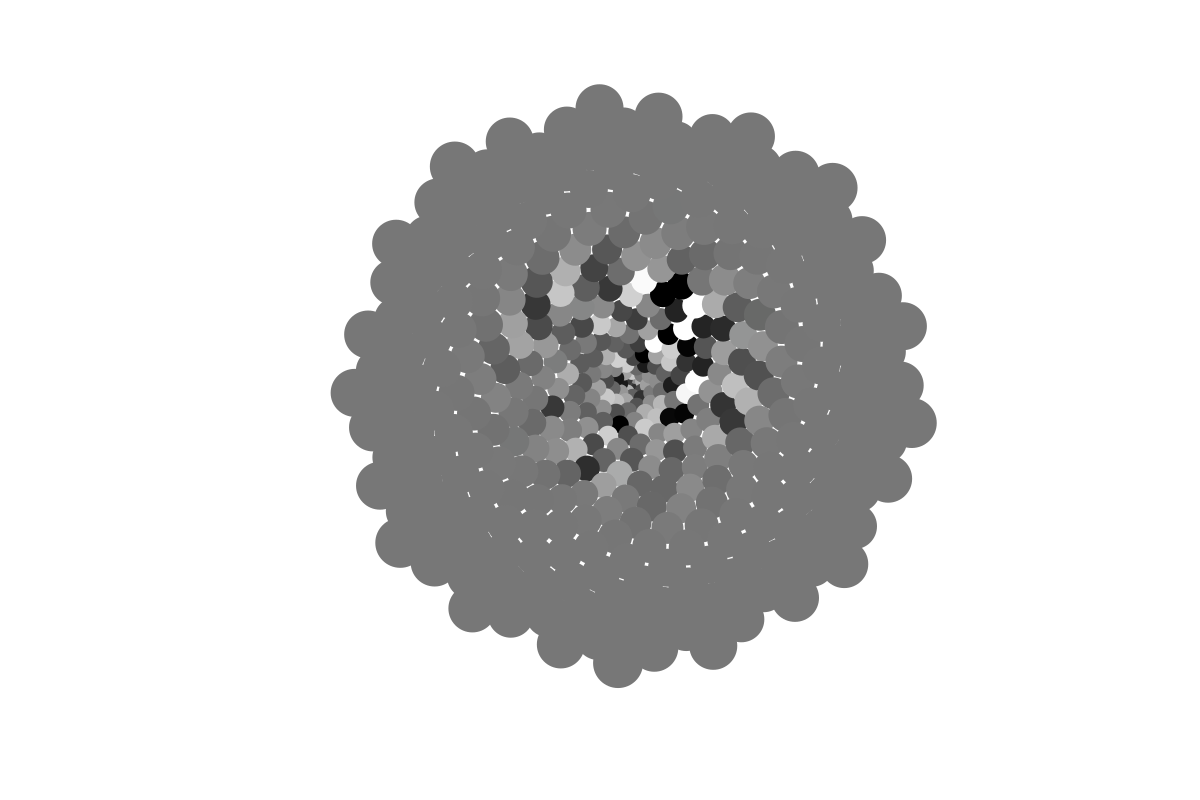}
      \caption{$\bpsi^{(5)}_{26,13}$}
    \end{subfigure}
    \begin{subfigure}{0.245\textwidth}
      \includegraphics[width = \textwidth]{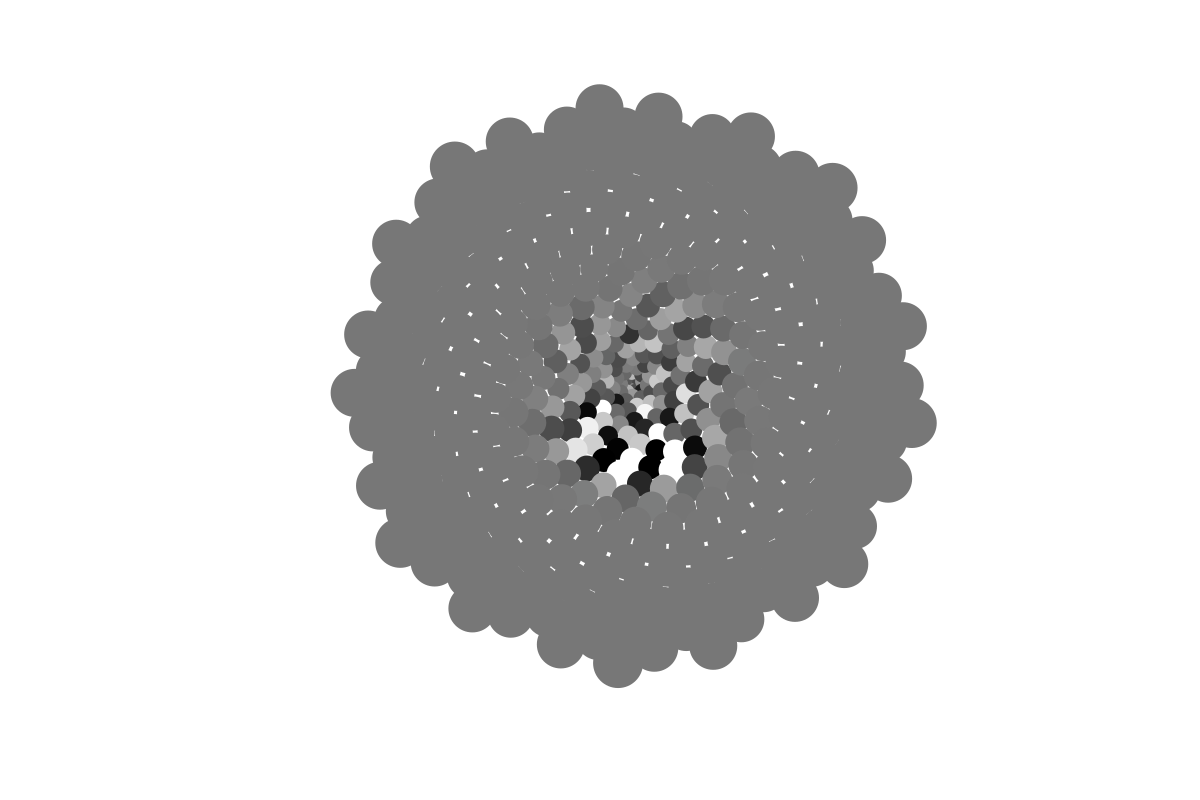}
      \caption{$\bpsi^{(6)}_{56,2}$}
    \end{subfigure}
    \begin{subfigure}{0.245\textwidth}
      \includegraphics[width = \textwidth]{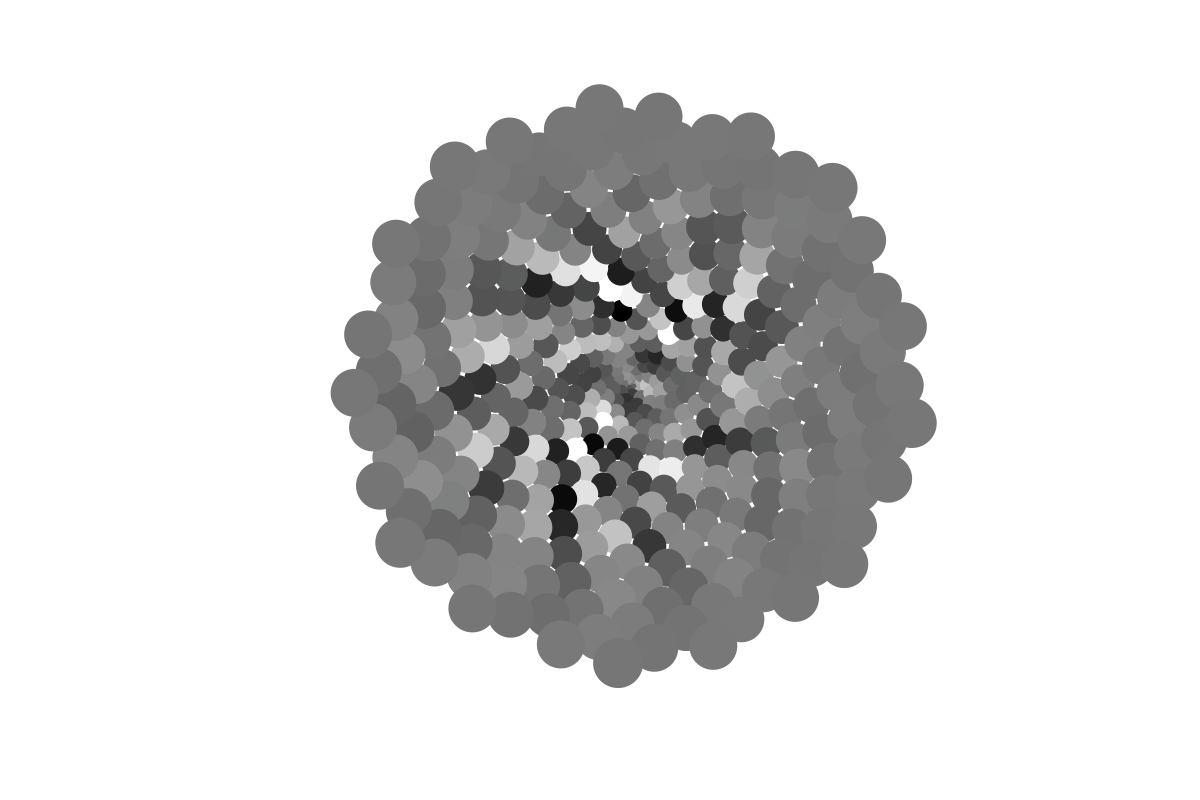}
      \caption{$\bpsi^{(7)}_{79,3}$}
    \end{subfigure}
    \begin{subfigure}{0.245\textwidth}
      \includegraphics[width = \textwidth]{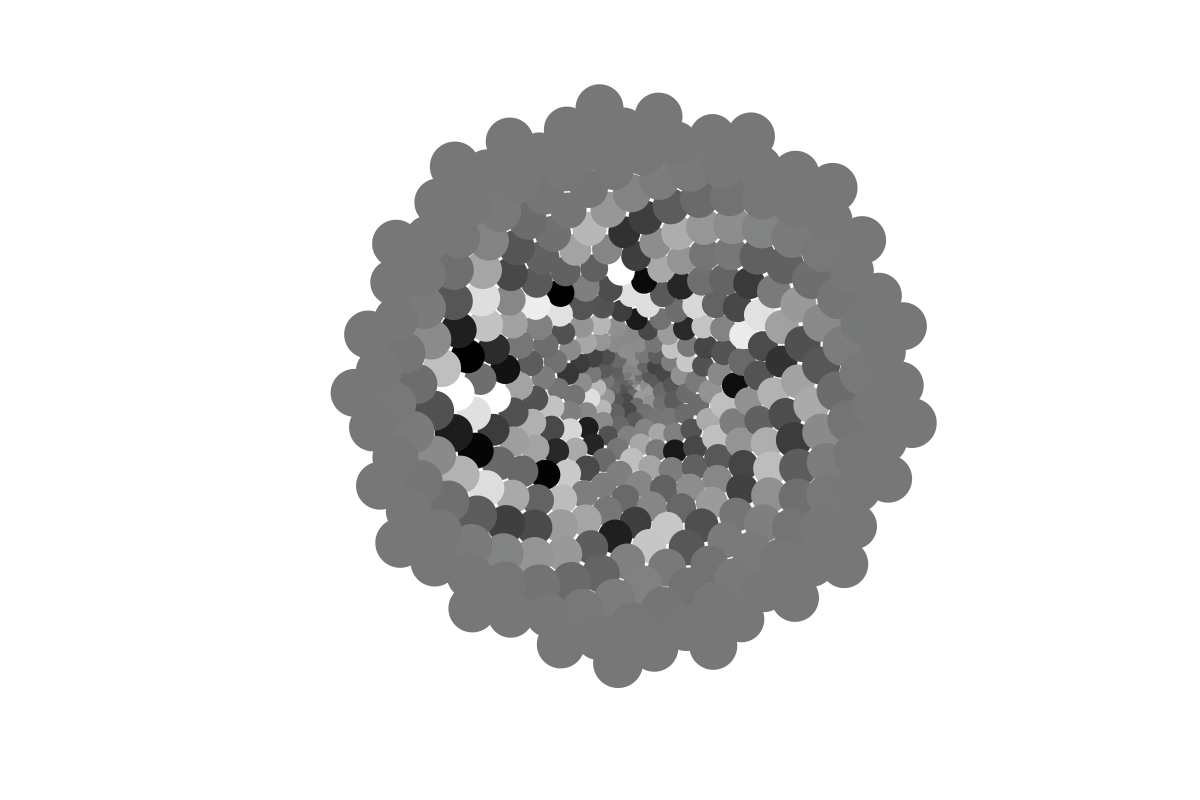}
      \caption{$\bpsi^{(11)}_{266,0} \equiv \bphi_{270}$}
    \end{subfigure}
    \\
    \begin{subfigure}{0.245\textwidth}
      \includegraphics[width = \textwidth]{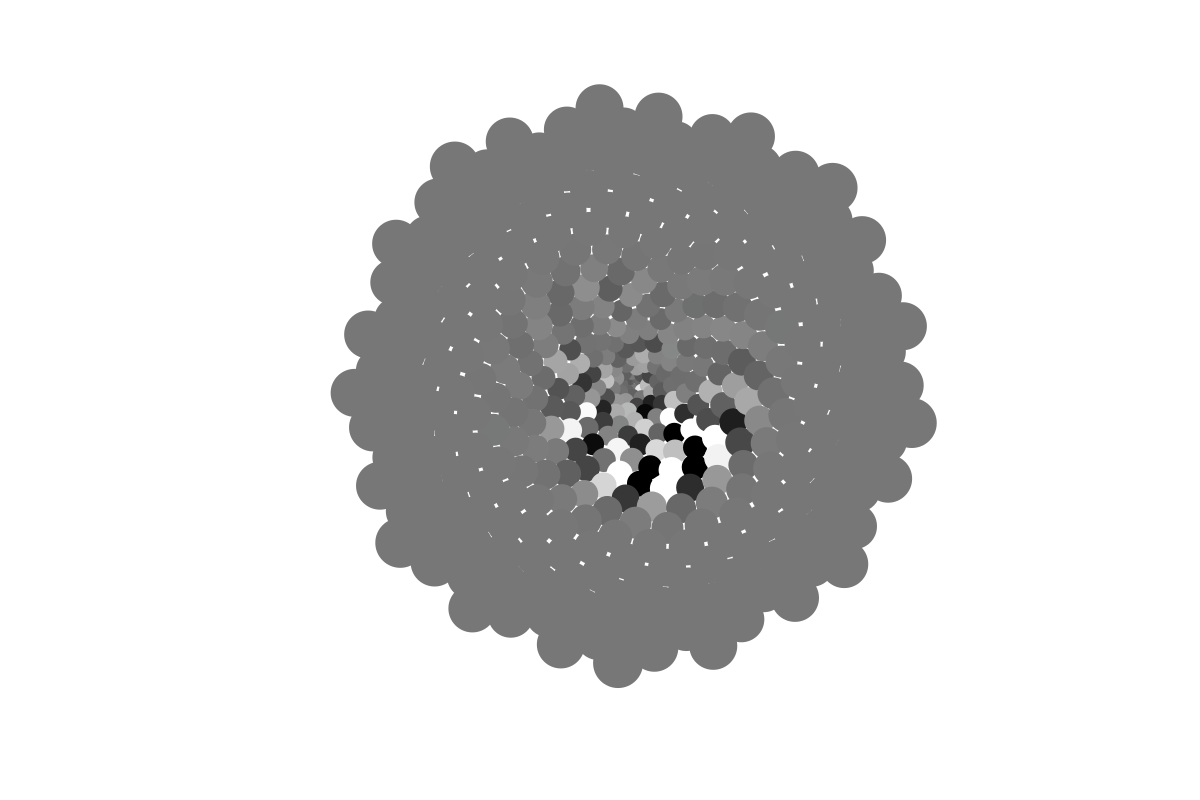}
      \caption{$\bpsi^{(5)}_{27,4}$}
    \end{subfigure}
    \begin{subfigure}{0.245\textwidth}
      \includegraphics[width = \textwidth]{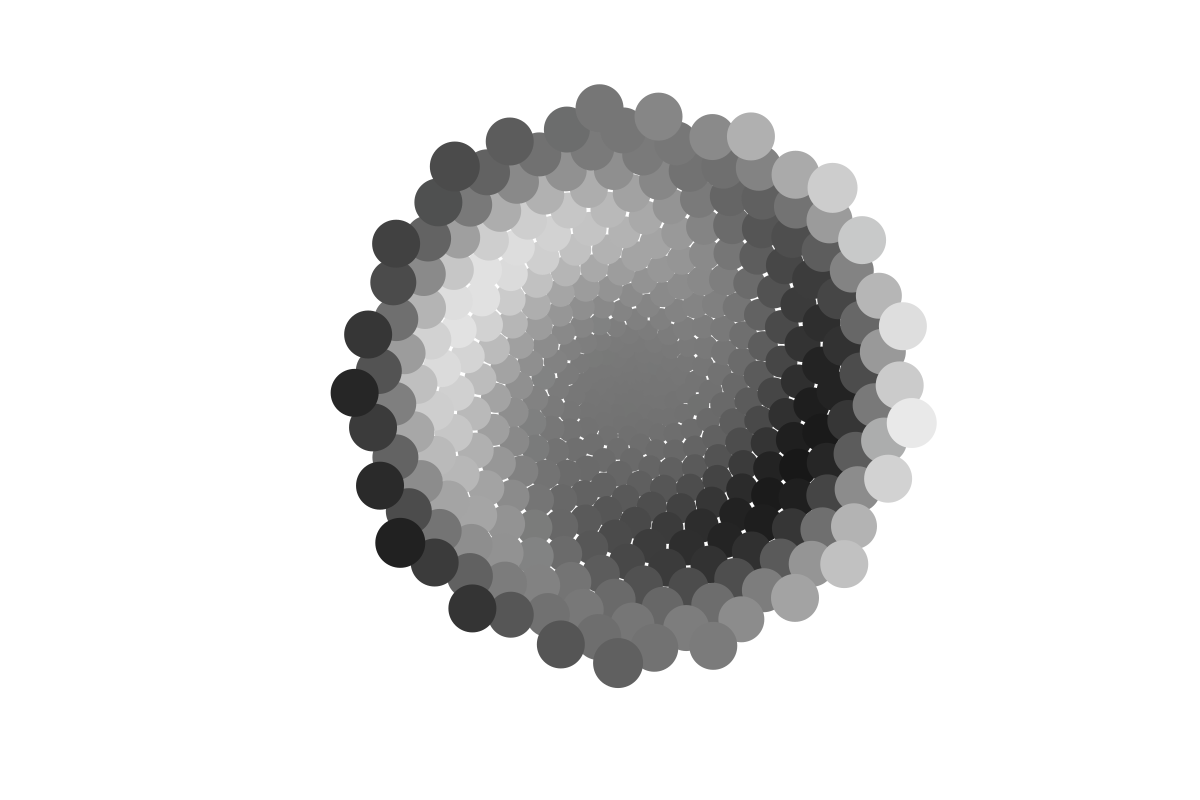}
      \caption{$\bpsi^{(11)}_{9,0} \equiv \bphi_6$}
    \end{subfigure}
    \begin{subfigure}{0.245\textwidth}
      \includegraphics[width = \textwidth]{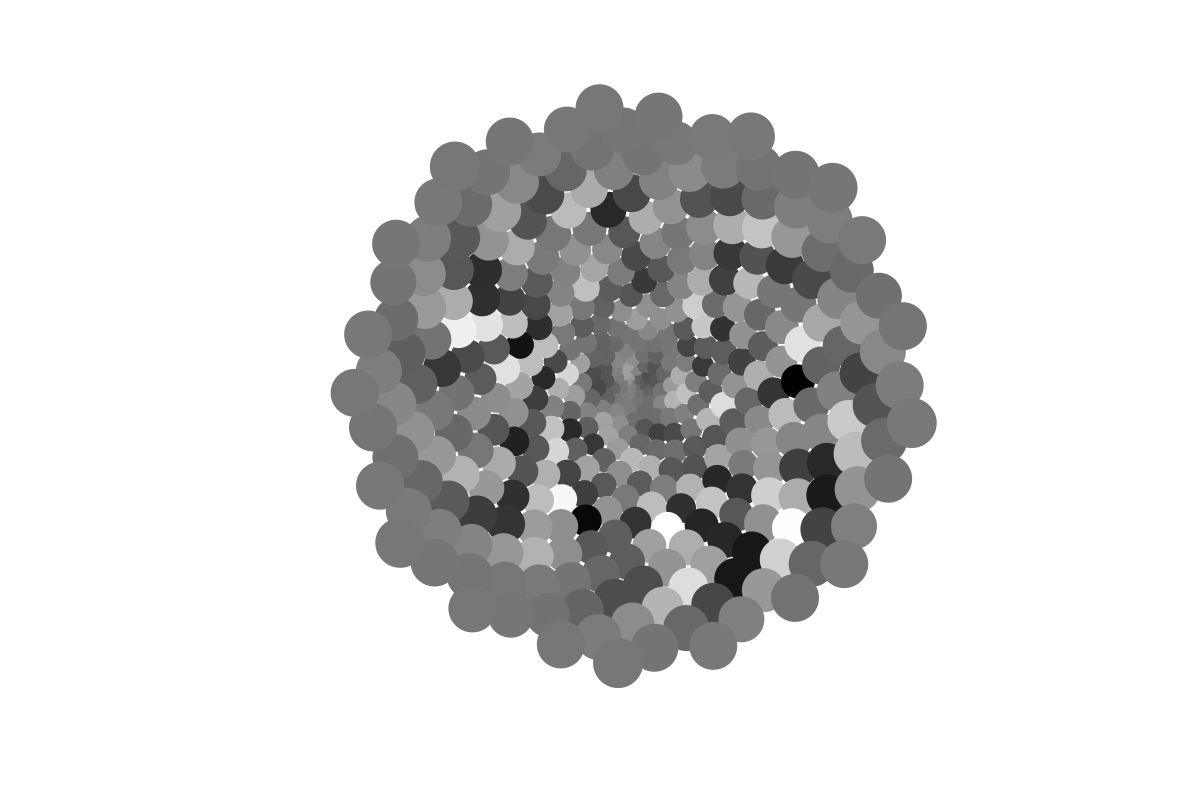}
      \caption{$\bpsi^{(11)}_{253,0} \equiv \bphi_{237}$}
    \end{subfigure}
    \begin{subfigure}{0.245\textwidth}
      \includegraphics[width = \textwidth]{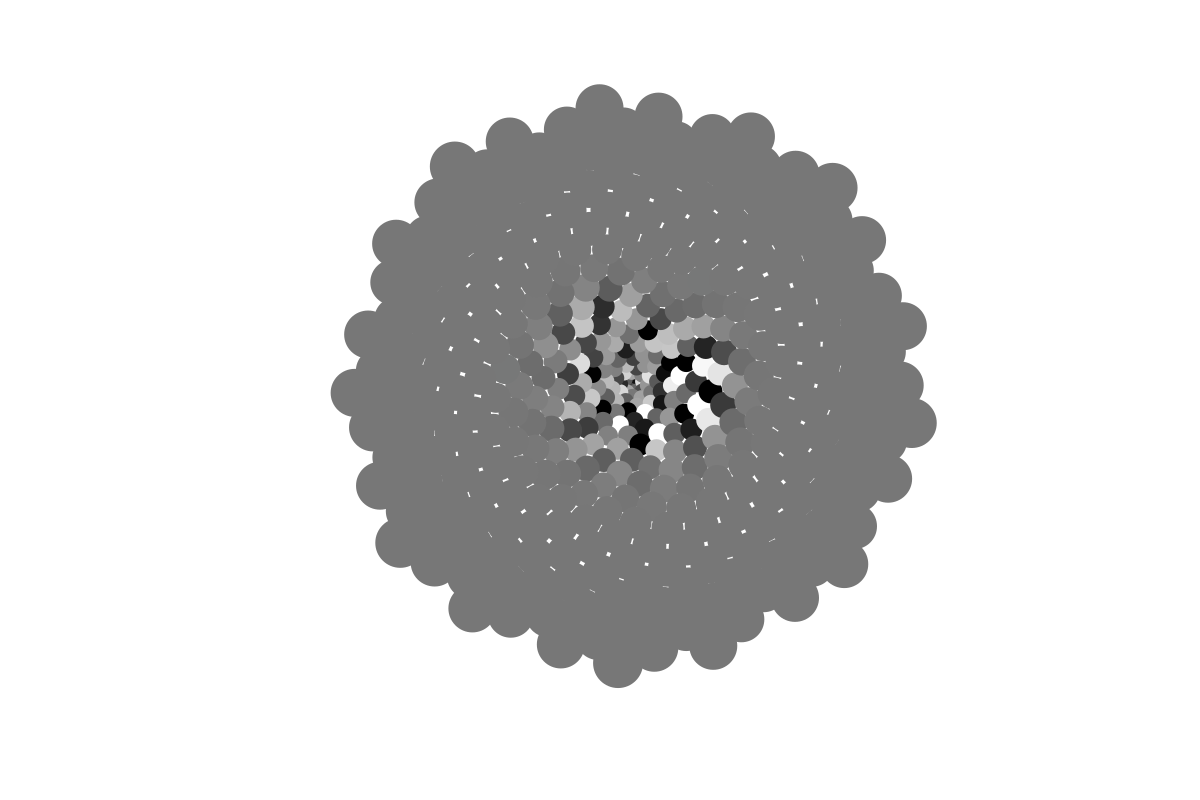}
      \caption{$\bpsi^{(8)}_{206,0}$}
    \end{subfigure}
    \\
    \begin{subfigure}{0.245\textwidth}
      \includegraphics[width = \textwidth]{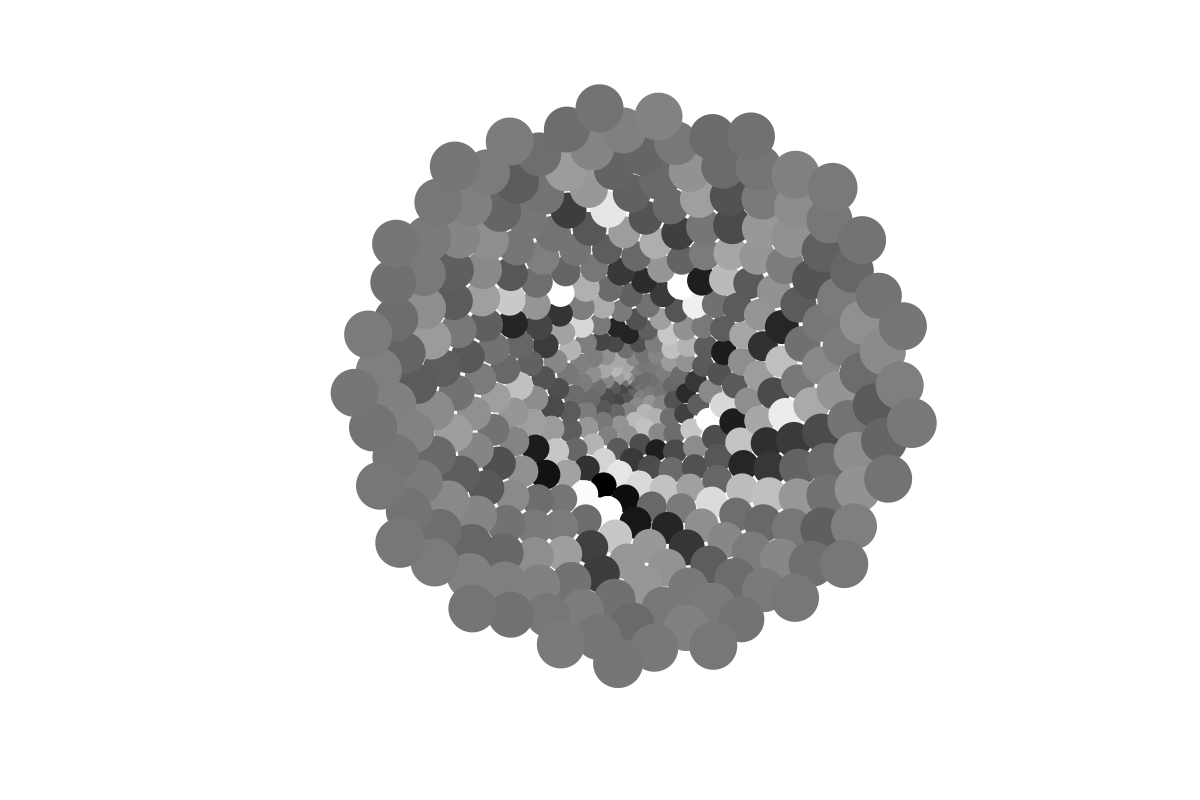}
      \caption{$\bpsi^{(7)}_{74,2}$}
    \end{subfigure}
    \begin{subfigure}{0.245\textwidth}
      \includegraphics[width = \textwidth]{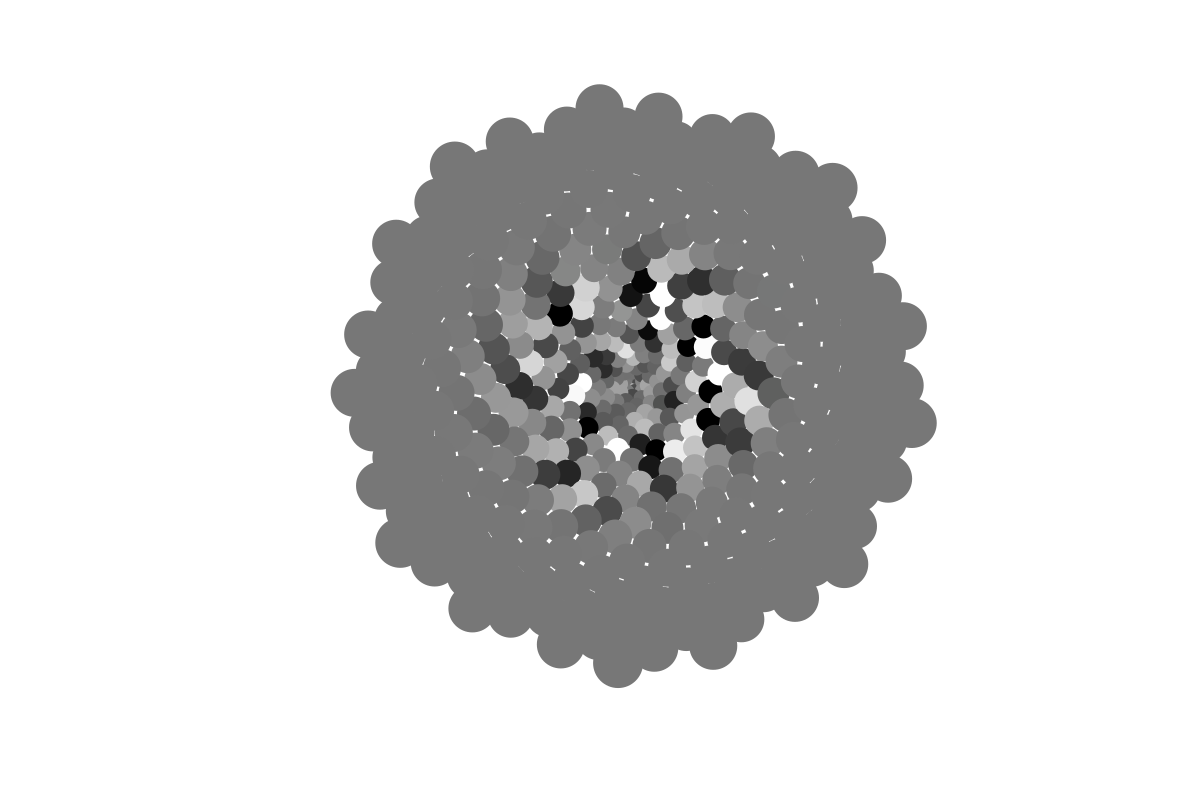}
      \caption{$\bpsi^{(8)}_{171,0}$}
    \end{subfigure}
    \begin{subfigure}{0.245\textwidth}
      \includegraphics[width = \textwidth]{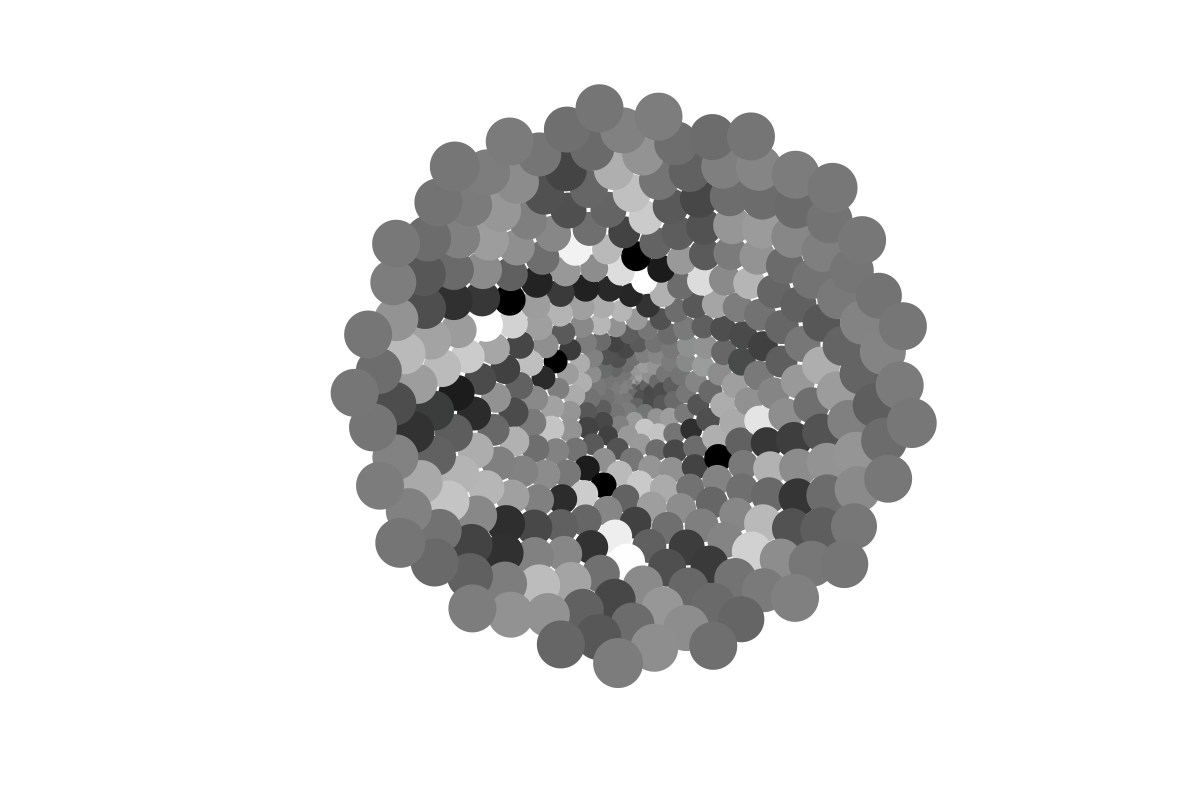}
      \caption{$\bpsi^{(7)}_{64,2}$}
    \end{subfigure}
    \begin{subfigure}{0.245\textwidth}
      \includegraphics[width = \textwidth]{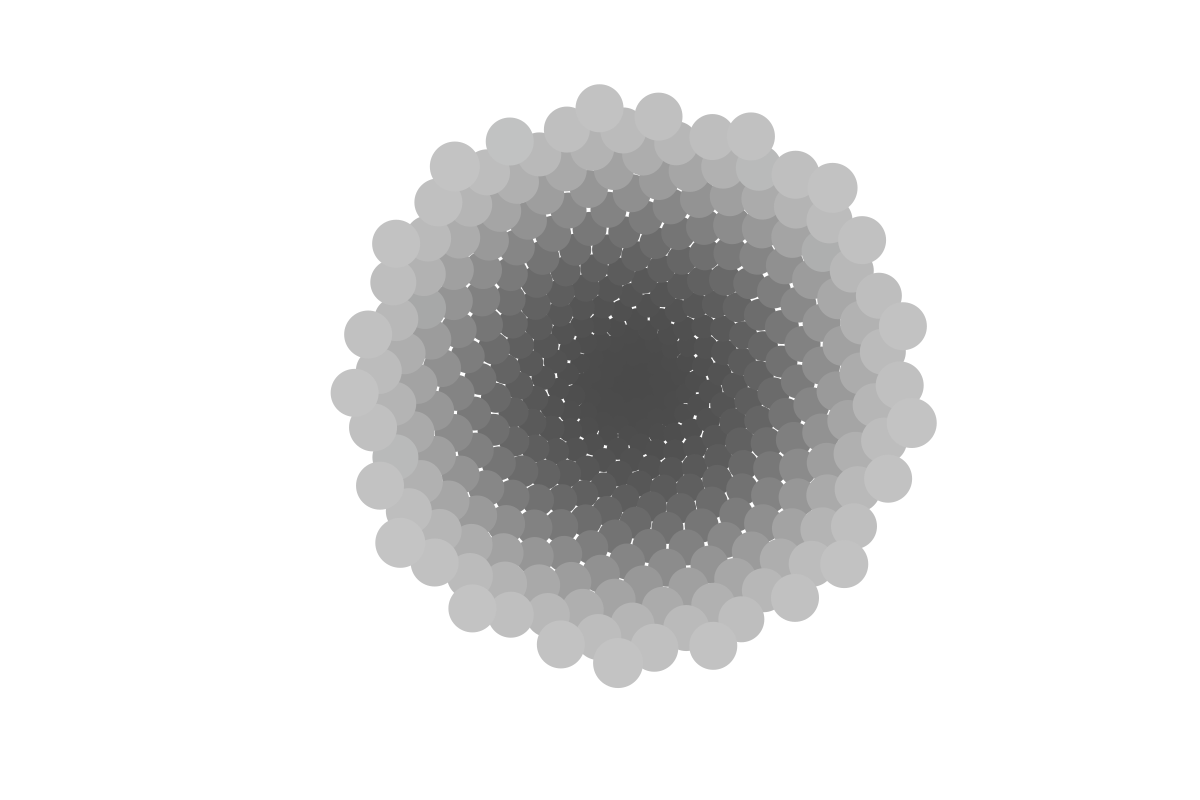}
      \caption{$\bpsi^{(11)}_{1,0} \equiv \bphi_1$}
    \end{subfigure}
    \caption{Sixteen most significant VM-NGWP best basis vectors (the DC vector not shown)
      for Barbara's pants.
      The basis vector amplitudes within $(-0.15, 0.15)$
        are mapped to the grayscale colormap.
      }
    \label{fig:barb-pants-top16}
  \end{center}
\end{figure}

Figure~\ref{fig:barb-pants-top16} shows the 16 most significant VM-NGWP best
basis vectors for approximating Barbara's pants signal.
We note that the majority of these basis vectors are of high-frequency nature
than those for the eye signal shown in Fig.~\ref{fig:barb-eye-top16},
which reflect the oscillating anisotropic patterns of her pants. 
The basis vectors $\bpsi^{(6)}_{3,1}$ (Fig.~\ref{fig:barb-pants-top16}a),
$\bpsi^{(11)}_{9,0}$ (Fig.~\ref{fig:barb-pants-top16}j), and
$\bpsi^{(11)}_{1,0}$ (Fig.~\ref{fig:barb-pants-top16}p) take care of shading
in this region while the other basis vectors extract oscillatory patterns
of various scales.
We also note that four among these top 16 best basis vectors are
the global graph Laplacian eigenvectors; see Fig.~\ref{fig:barb-pants-top16}h, j, k, p.

\subsection{Toronto Street Network}
\begin{figure}
  \begin{subfigure}{0.49\textwidth}
    \centering\includegraphics[width=\textwidth]{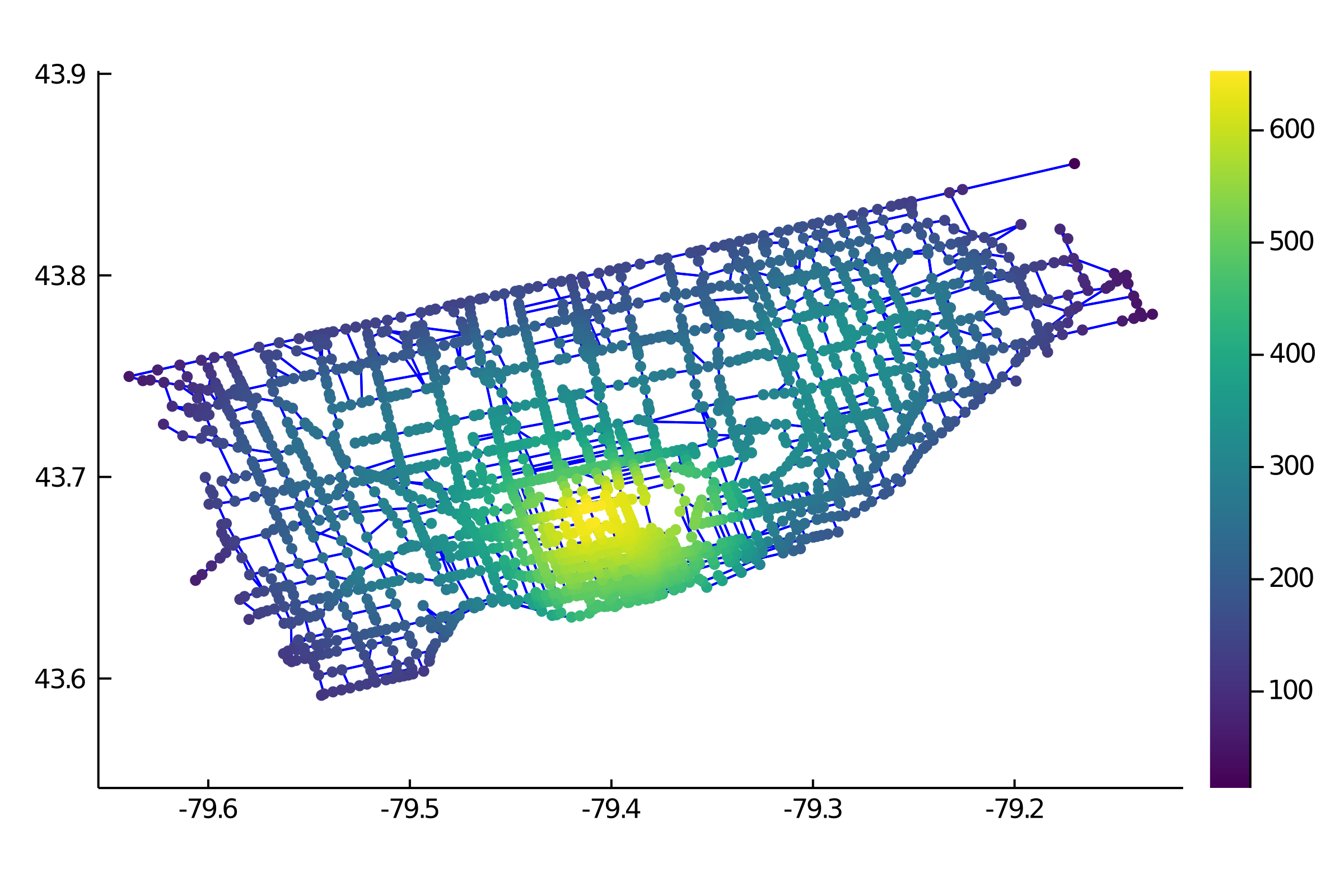}
    \caption{A smooth spatial distribution of the street intersections}
    \label{fig:toronto-fdensity-graph}
  \end{subfigure}
  \begin{subfigure}{0.49\textwidth}
    \centering\includegraphics[width=\textwidth]{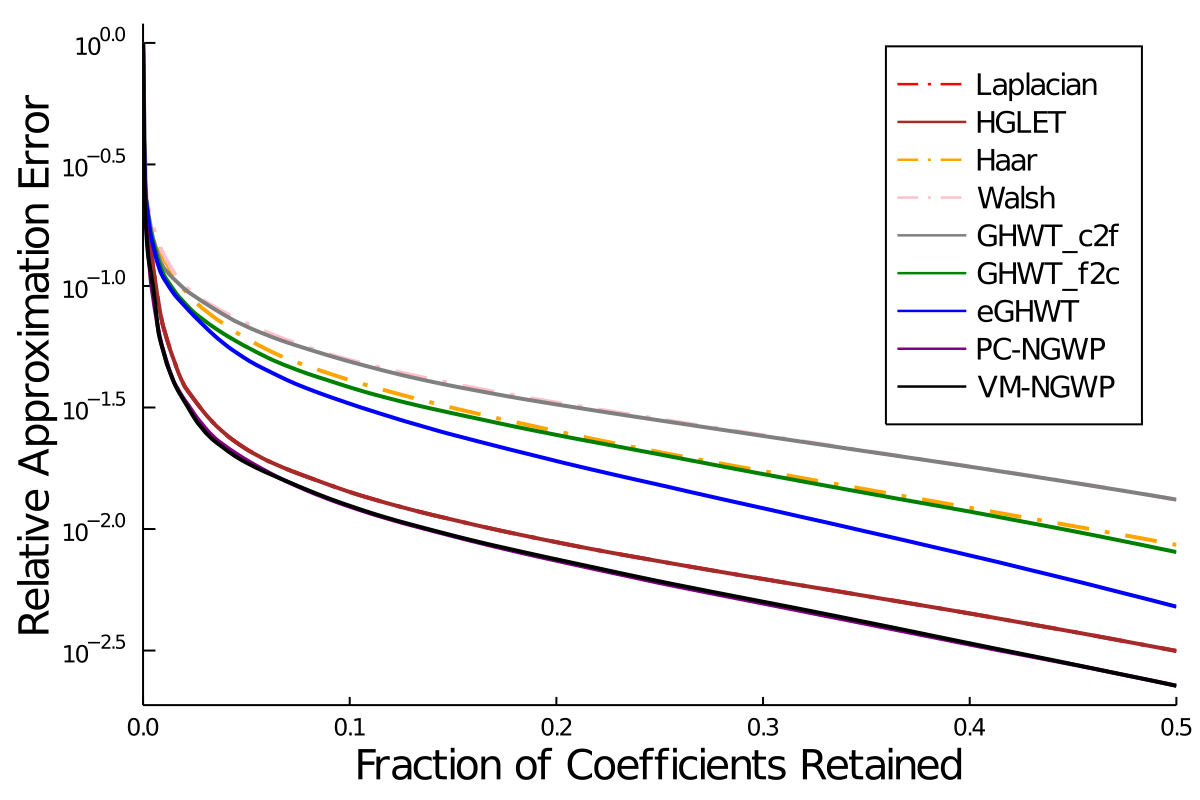}
    \caption{Approximation performance of various methods}
    \label{fig:toronto-fdensity-approx}
  \end{subfigure}
  \caption{A graph signal representing the smooth spatial distribution of
    the street intersections on the Toronto street network (a).
    The horizontal and vertical axes of this plot represent the longitude and
    latitude geo-coordinates of this area, respectively.
    The results of our approximation experiments (b)}
  \label{fig:toronto-fdensity}
\end{figure}

We obtained the street network data of the City of Toronto from its
open data portal\footnote{URL: \url{https://open.toronto.ca/dataset/traffic-signal-vehicle-and-pedestrian-volumes}}.
Using the street names and intersection coordinates included in the dataset,
we construct the graph representing the street network there with $N = 2275$
nodes and $M = 3381$ edges. Fig.~\ref{fig:toronto-fdensity-graph} displays
this graph. As before, each edge weight was set as the reciprocal of the
Euclidean distance between the endpoints of that edge.

We analyze two graph signals on this street network:
1) spatial distribution of the street intersections and 2) pedestrian volume
measured at each intersection.
The first graph signal was constructed by counting the number of the nodes within
the disk of radius $4.7$ km centered at each node.
In other words, this is a smooth version of histogram of the
distribution of street intersections computed with the overlapping circular bins
of equal size.
The longest edge length measured in the Euclidean distance among all these
$3381$ edges was chosen as this radius of this disk, which is located
at the northeast corner of this graph as one can easily see in 
Fig.~\ref{fig:toronto-fdensity-graph}.
The second graph signal is the most recent 8 peak-hour pedestrian volume counts
collected at intersections (i.e., nodes in this graph) where there are traffic
signals.
The dataset was collected between the hours of 7:30 am and 6:00 pm, over the
period of 03/22/2004--02/28/2018. 
From Fig.~\ref{fig:toronto-fdensity-approx}, we observe that qualitative
behaviors of these error curves are relatively similar to those of
Barbara's eye signal shown in Fig.~\ref{fig:barb-eye-approx}.
More precisely,
1) NGWP best bases outperformed all the others and the difference between
the VM-NGWP and the PC-NGWP is negligible;
2) the HGLET best basis chose the global graph Laplacian eigenbasis, which
worked quite well following the NGWP best bases; 
3) those bases based on the Haar-Walsh wavelet packet dictionaries
did not perform well.

\begin{figure}
  \begin{center}
    \begin{subfigure}{0.245\textwidth}
      \includegraphics[width = \textwidth]{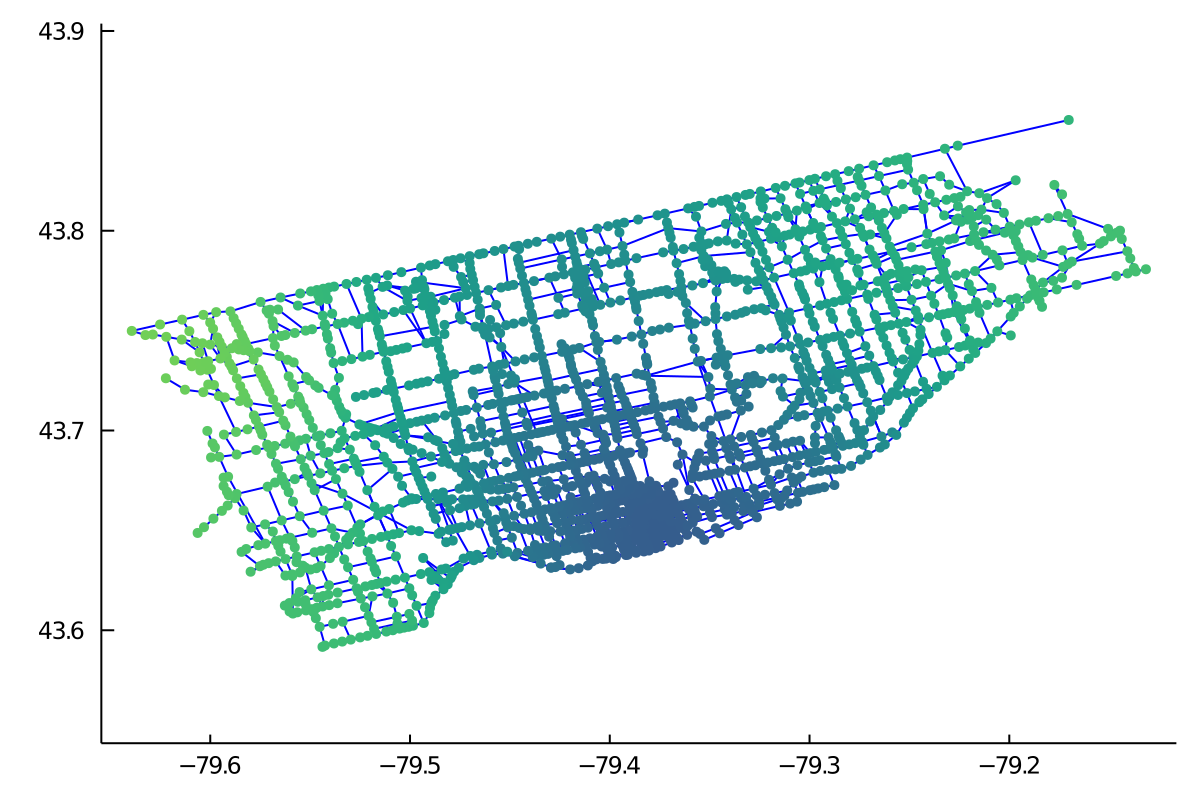}
      \caption{$\bpsi^{(43)}_{2,0} \equiv \bphi_2$}
    \end{subfigure}
    \begin{subfigure}{0.245\textwidth}
      \includegraphics[width = \textwidth]{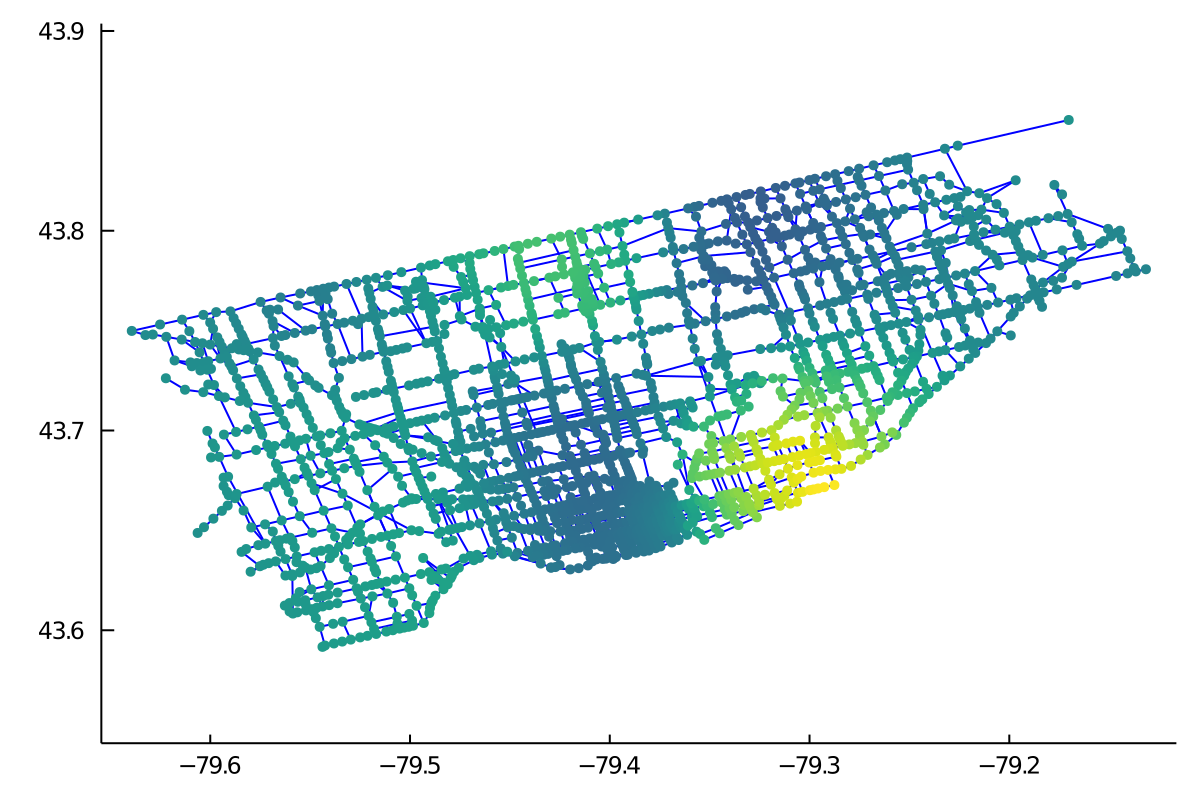}
      \caption{$\bpsi^{(6)}_{2,0}$}
    \end{subfigure}
    \begin{subfigure}{0.245\textwidth}
      \includegraphics[width = \textwidth]{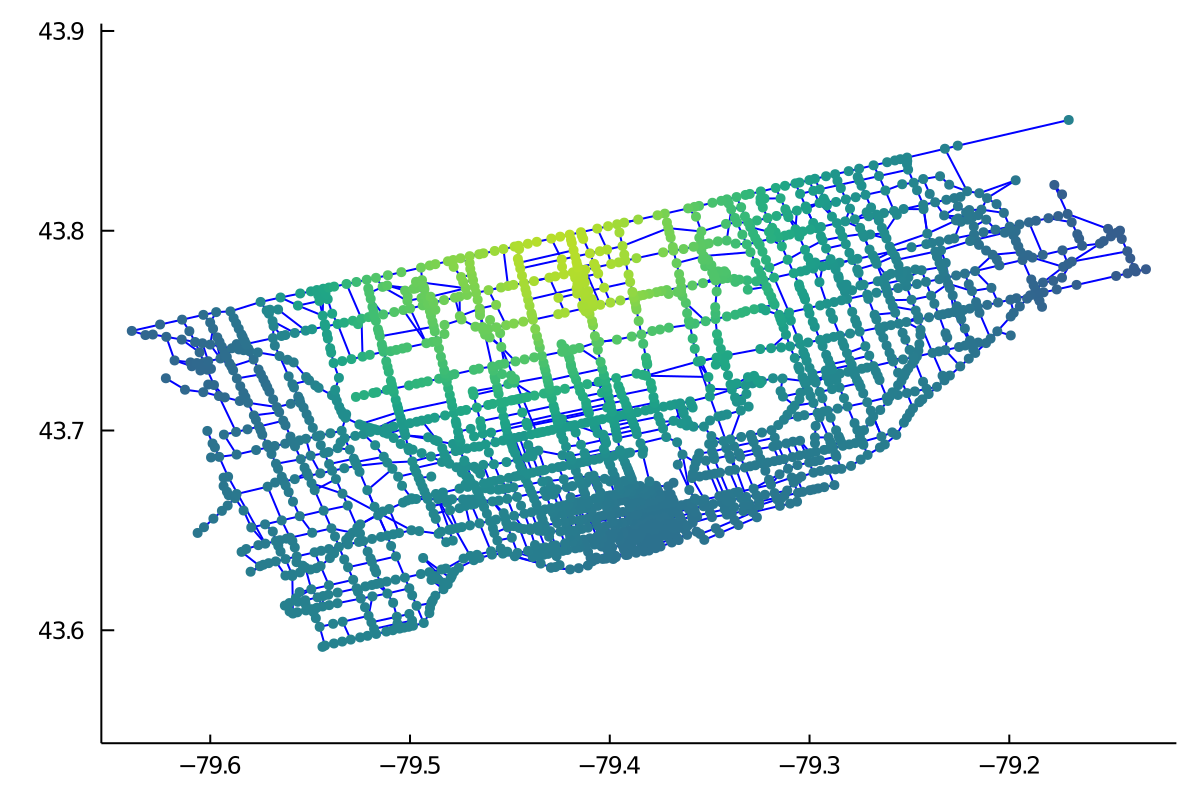}
      \caption{$\bpsi^{(6)}_{1,0}$}
    \end{subfigure}
    \begin{subfigure}{0.245\textwidth}
      \includegraphics[width = \textwidth]{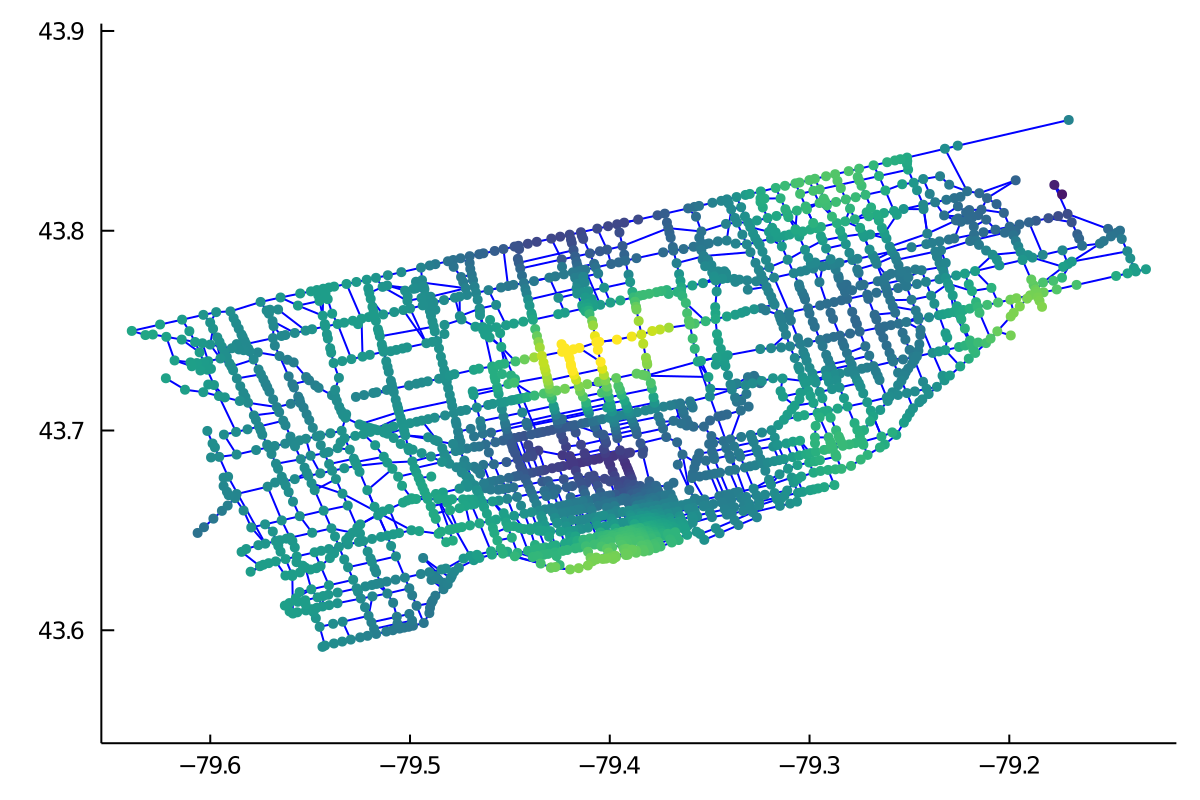}
      \caption{$\bpsi^{(6)}_{5,2}$}
    \end{subfigure}
    \\
    \begin{subfigure}{0.245\textwidth}
      \includegraphics[width = \textwidth]{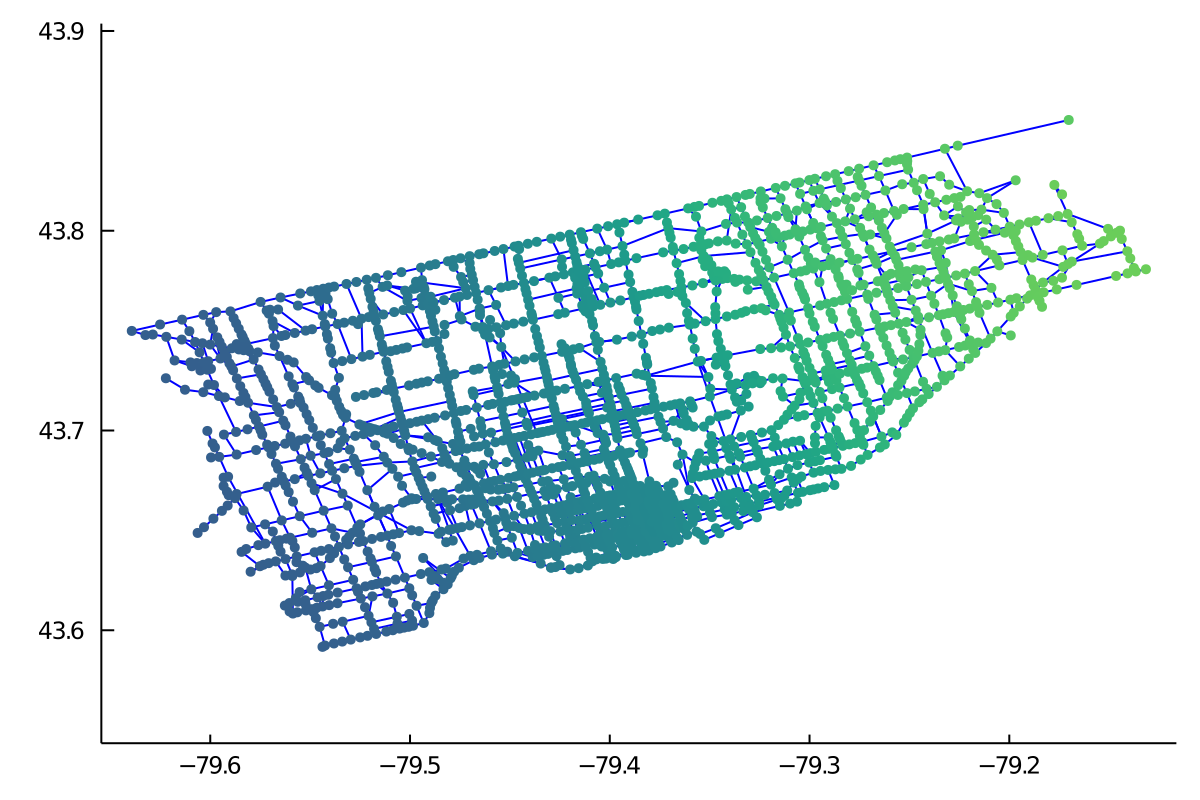}
      \caption{$\bpsi^{(43)}_{1,0} \equiv \bphi_1$}
    \end{subfigure}
    \begin{subfigure}{0.245\textwidth}
      \includegraphics[width = \textwidth]{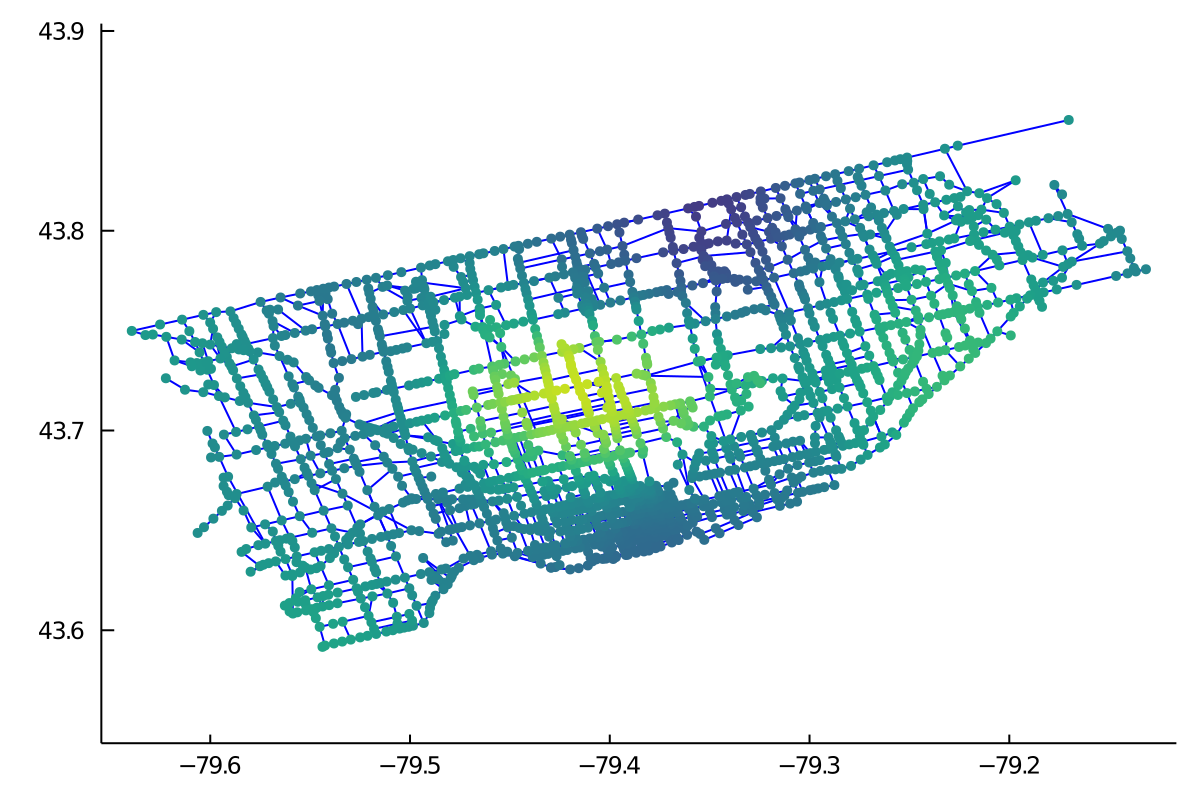}
      \caption{$\bpsi^{(6)}_{2,4}$}
    \end{subfigure}
    \begin{subfigure}{0.245\textwidth}
      \includegraphics[width = \textwidth]{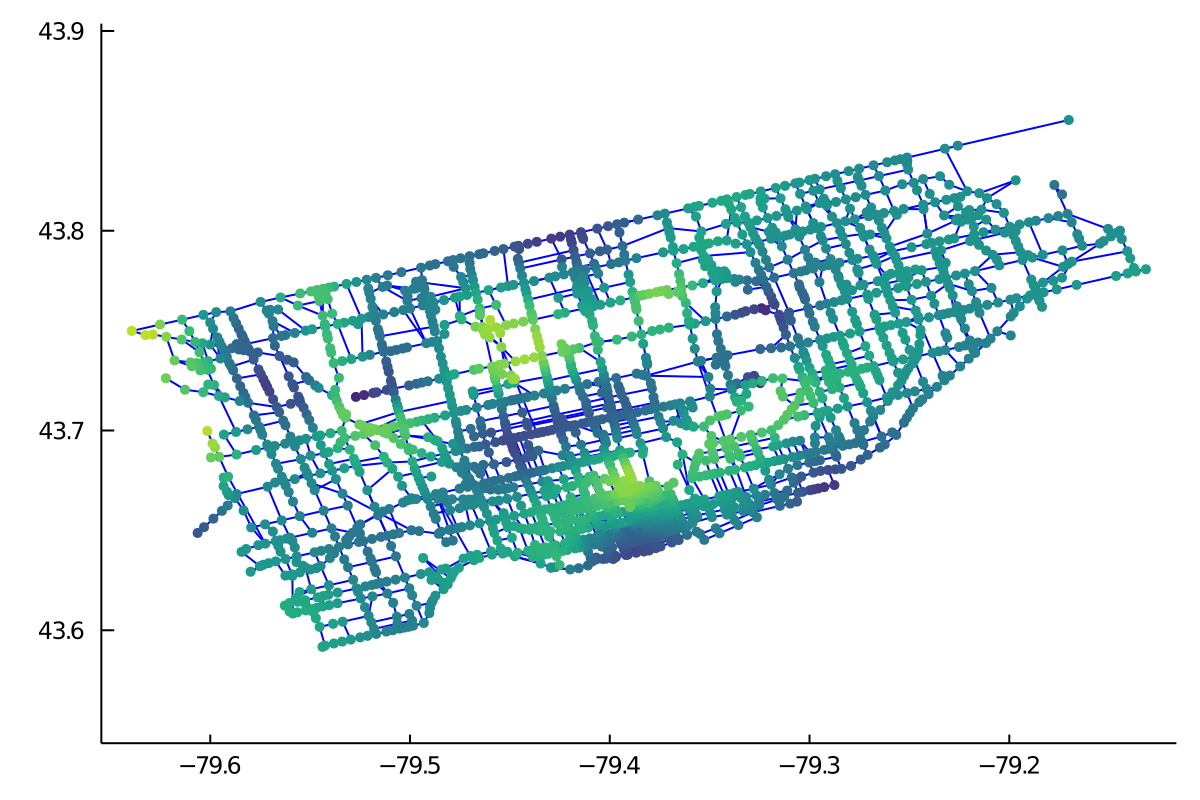}
      \caption{$\bpsi^{(8)}_{24,0}$}
    \end{subfigure}
    \begin{subfigure}{0.245\textwidth}
      \includegraphics[width = \textwidth]{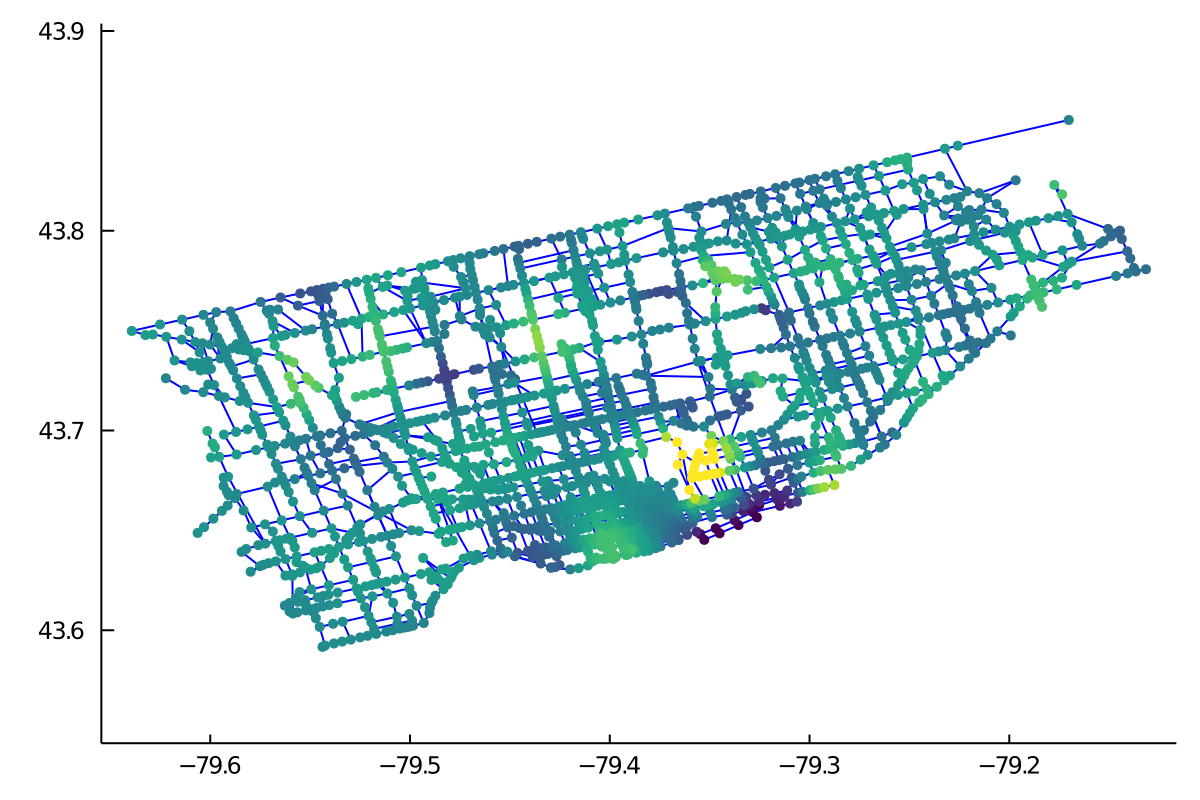}
      \caption{$\bpsi^{(5)}_{4,4}$}
    \end{subfigure}
    \\
    \begin{subfigure}{0.245\textwidth}
      \includegraphics[width = \textwidth]{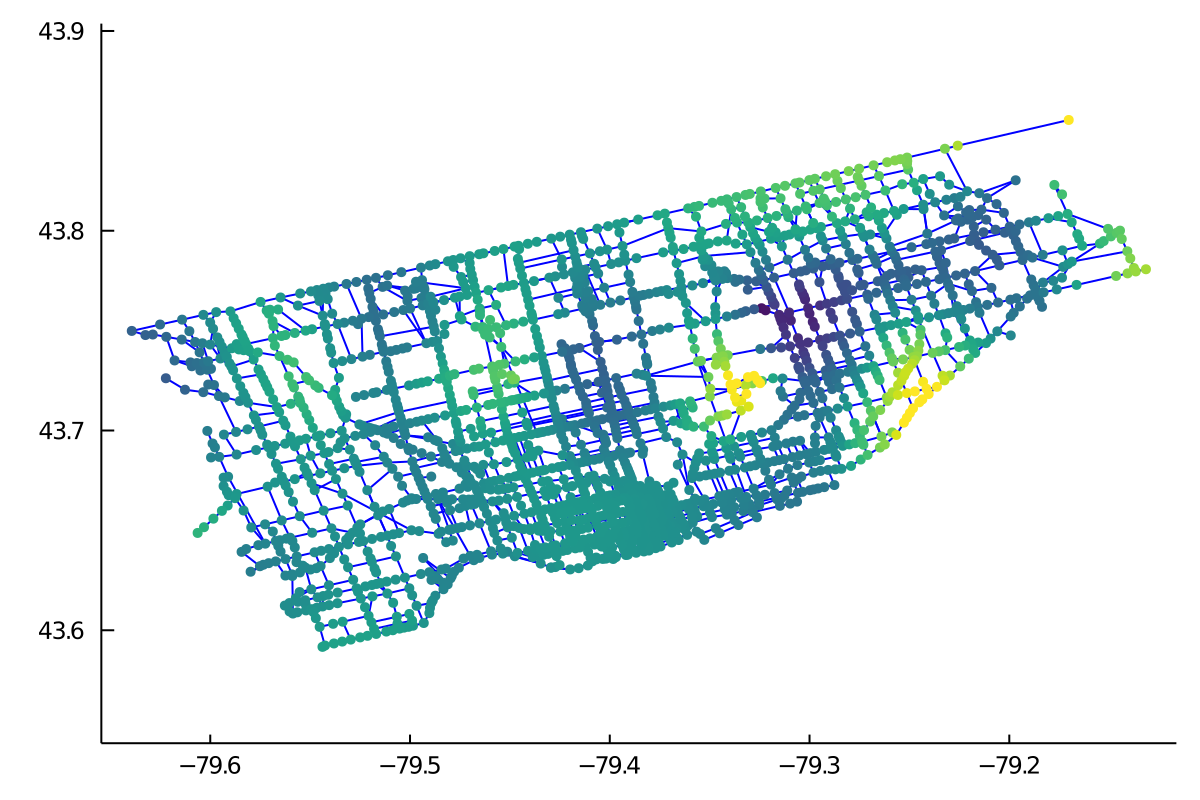}
      \caption{$\bpsi^{(6)}_{5,3}$}
    \end{subfigure}
    \begin{subfigure}{0.245\textwidth}
      \includegraphics[width = \textwidth]{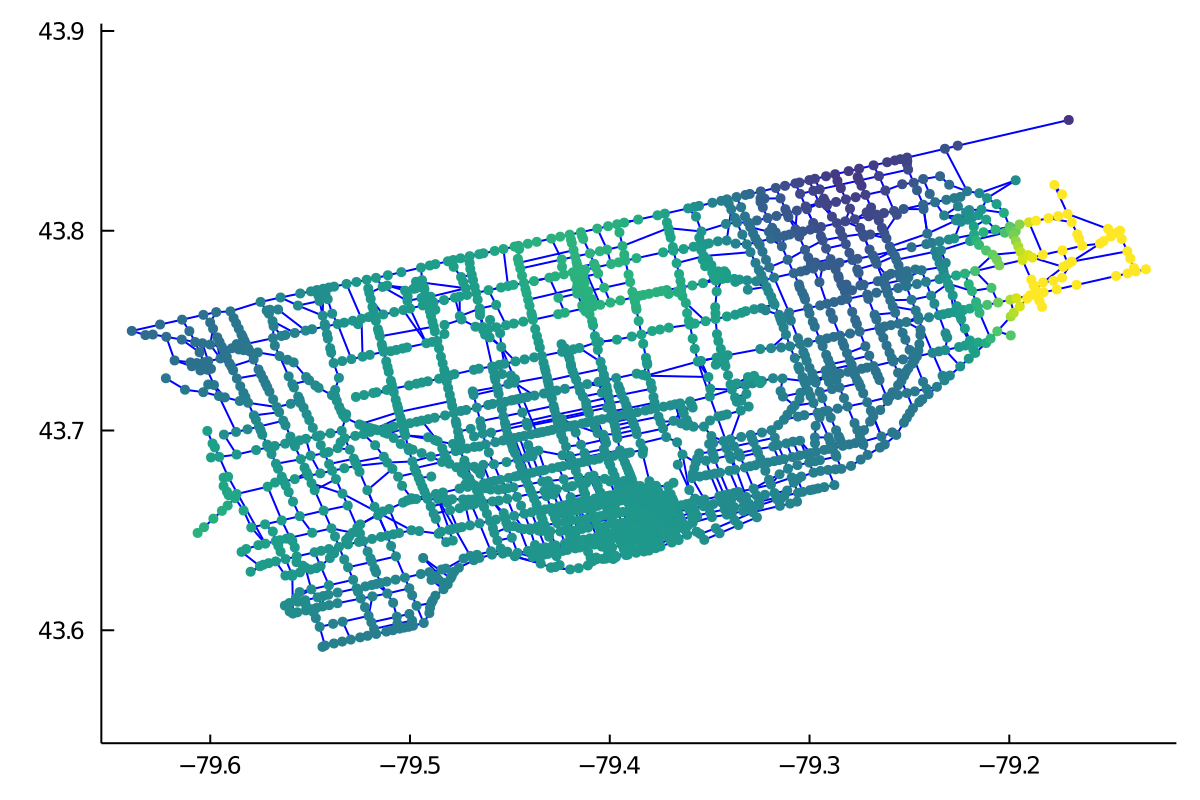}
      \caption{$\bpsi^{(6)}_{2,2}$}
    \end{subfigure}
    \begin{subfigure}{0.245\textwidth}
      \includegraphics[width = \textwidth]{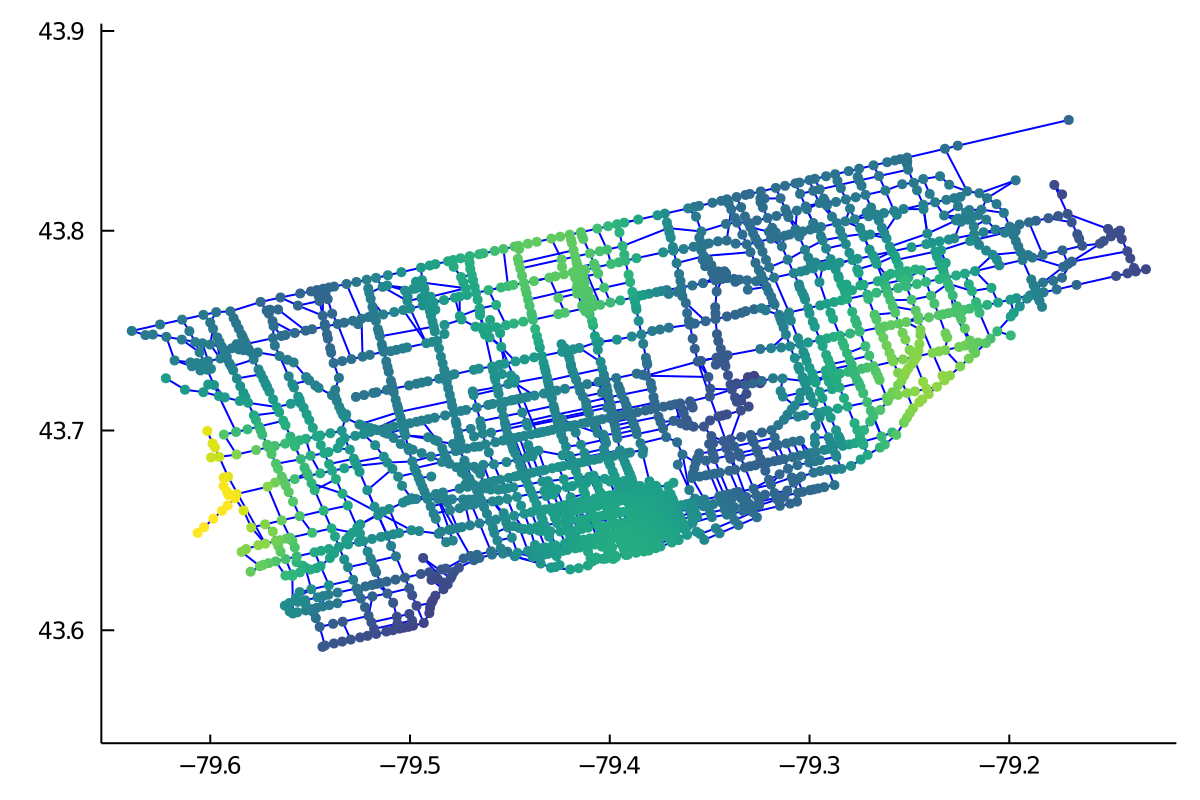}
      \caption{$\bpsi^{(43)}_{12,0} \equiv \bphi_{10}$}
    \end{subfigure}
    \begin{subfigure}{0.245\textwidth}
      \includegraphics[width = \textwidth]{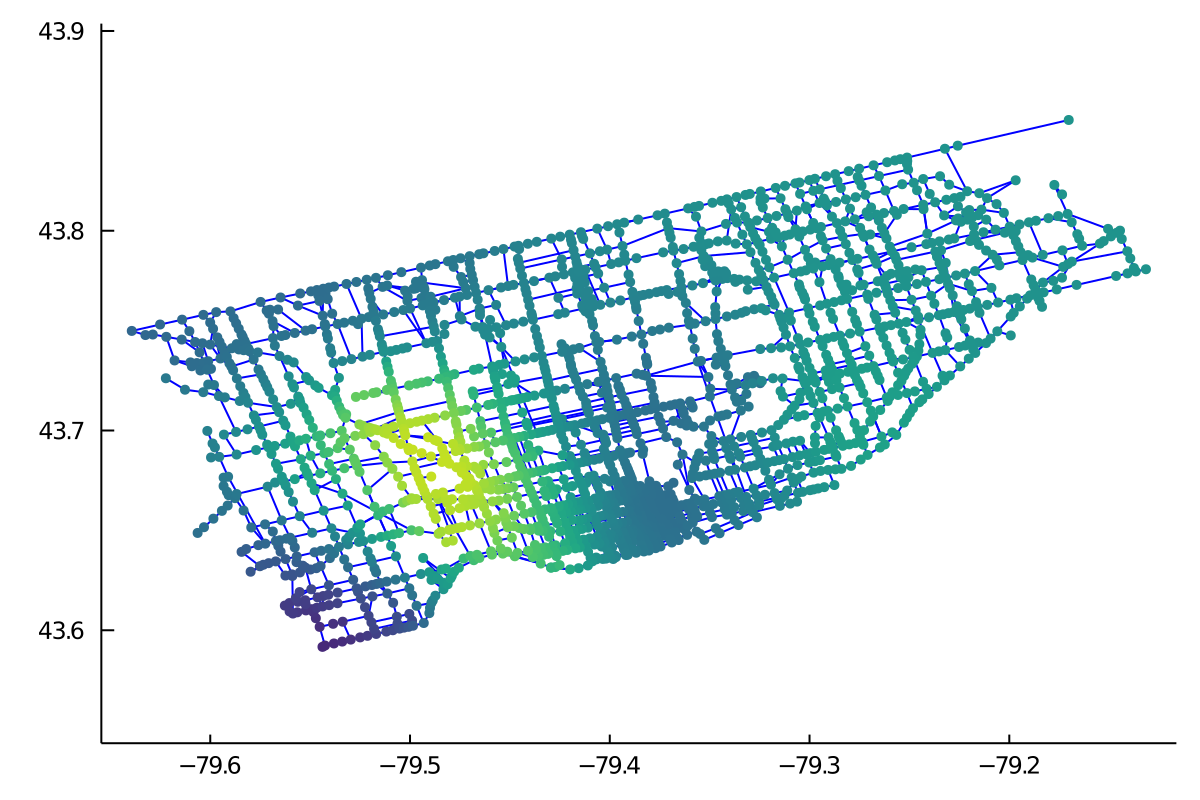}
      \caption{$\bpsi^{(6)}_{2,3}$}
    \end{subfigure}
    \\
    \begin{subfigure}{0.245\textwidth}
      \includegraphics[width = \textwidth]{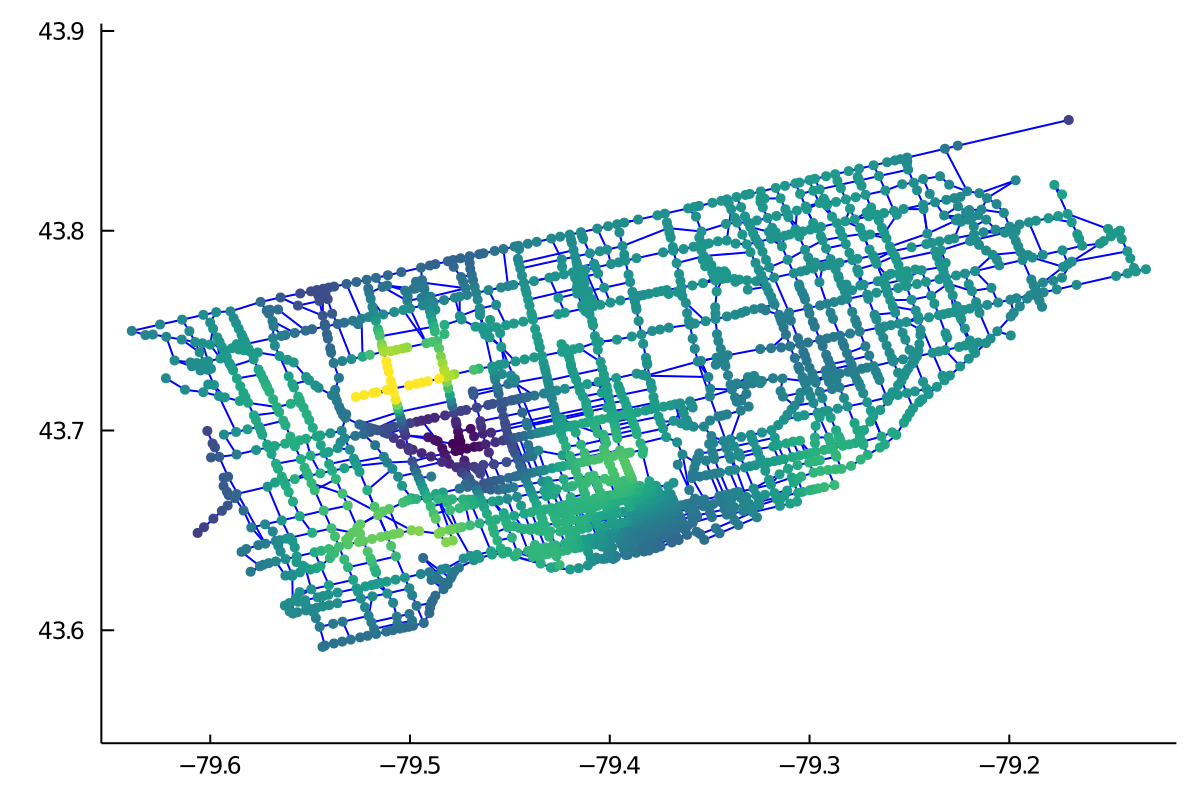}
      \caption{$\bpsi^{(6)}_{5,5}$}
    \end{subfigure}
    \begin{subfigure}{0.245\textwidth}
      \includegraphics[width = \textwidth]{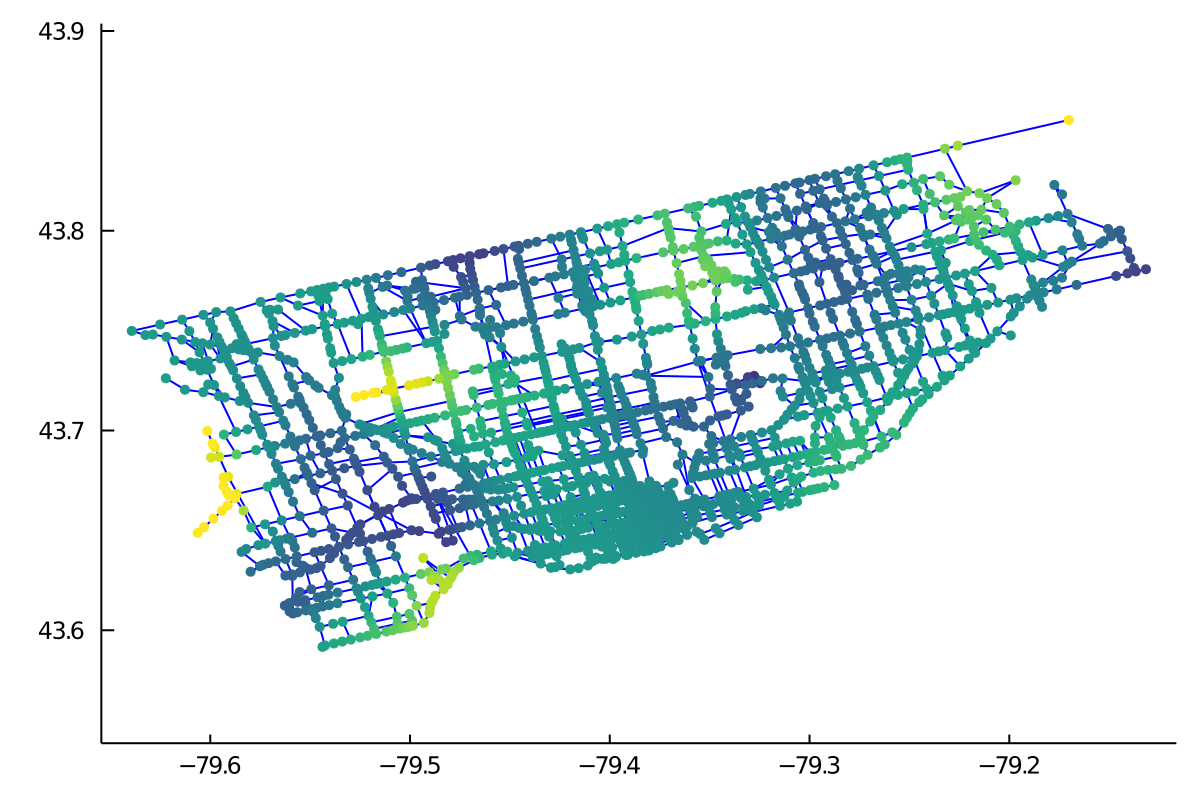}
      \caption{$\bpsi^{(43)}_{18,0} \equiv \bphi_{18}$}
    \end{subfigure}
    \begin{subfigure}{0.245\textwidth}
      \includegraphics[width = \textwidth]{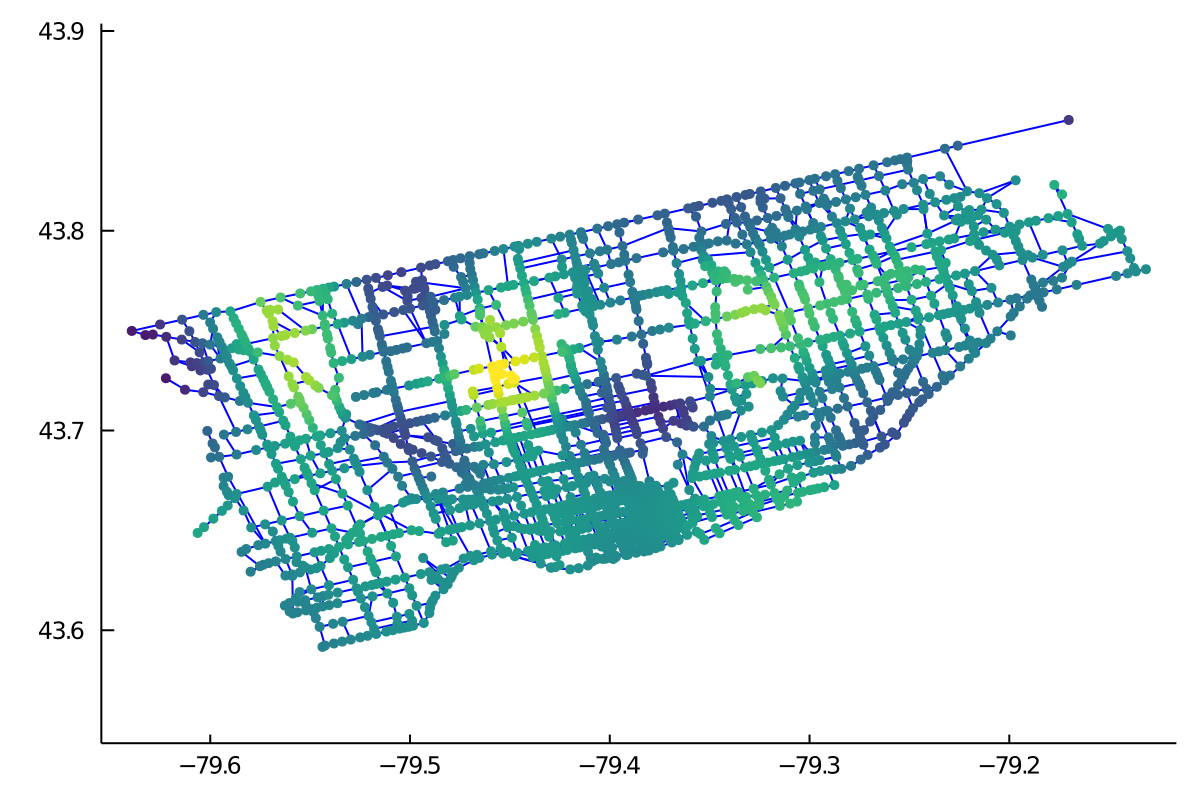}
      \caption{$\bpsi^{(6)}_{5,1}$}
    \end{subfigure}
    \begin{subfigure}{0.245\textwidth}
      \includegraphics[width = \textwidth]{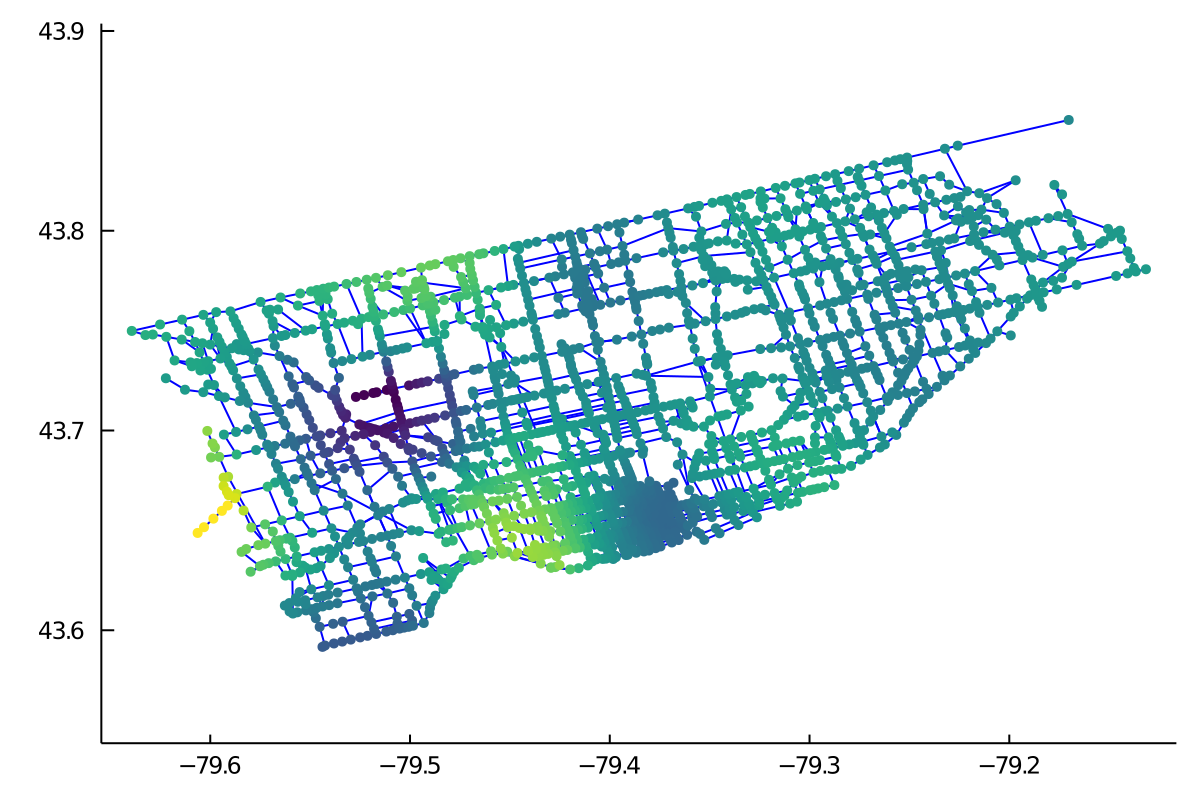}
      \caption{$\bpsi^{(7)}_{8,0}$}
    \end{subfigure}
    \caption{Sixteen most significant VM-NGWP best basis vectors (the DC vector not shown)
      for street intersection density data on the Toronto street map.
      The basis vector amplitudes within $(-0.075, 0.075)$
        are mapped to the viridis colormap.
      }
    \label{fig:toronto-fdensity-top16}
  \end{center}
\end{figure}

In order to examine what kind of basis vectors were chosen to approximate
this smooth histogram of street intersections, we display the most important
16 VM-NGWP best basis vectors in Fig.~\ref{fig:toronto-fdensity-top16}.
We note that these top basis vectors exhibit different spatial scales.
The basis vectors with levels $j=5$ and $j=6$ are relatively localized
to specific regions of Toronto.
For example, $\bpsi^{(5)}_{4,4}$ (Fig.~\ref{fig:toronto-fdensity-top16}h)
tries to differentiate the eastern neighbor of the dense downtown region
along the north-south direction while $\bpsi^{(6)}_{2,0}$ (Fig.~\ref{fig:toronto-fdensity-top16}b) tries to do the same along the east-west direction.
$\bpsi^{(6)}_{2,2}$ (Fig.~\ref{fig:toronto-fdensity-top16}j) tries to
differentiate the intersection density around the northeast region of Toronto.
On the other hand, there are coarse scale basis vectors with $j=\jmax=43$,
which are in fact the global Laplacian eigenvectors, i.e.,
Fig.~\ref{fig:toronto-fdensity-top16}a, e, k, n. It is not surprising that these
coarse scale basis vectors were selected as a part of the VM-NGWP best basis
considering that the global graph Laplacian eigenbasis performed quite well on
this graph signal as shown in Fig.~\ref{fig:toronto-fdensity-approx}.

\begin{figure}
  \begin{subfigure}{0.49\textwidth}
    \centering\includegraphics[width=\textwidth]{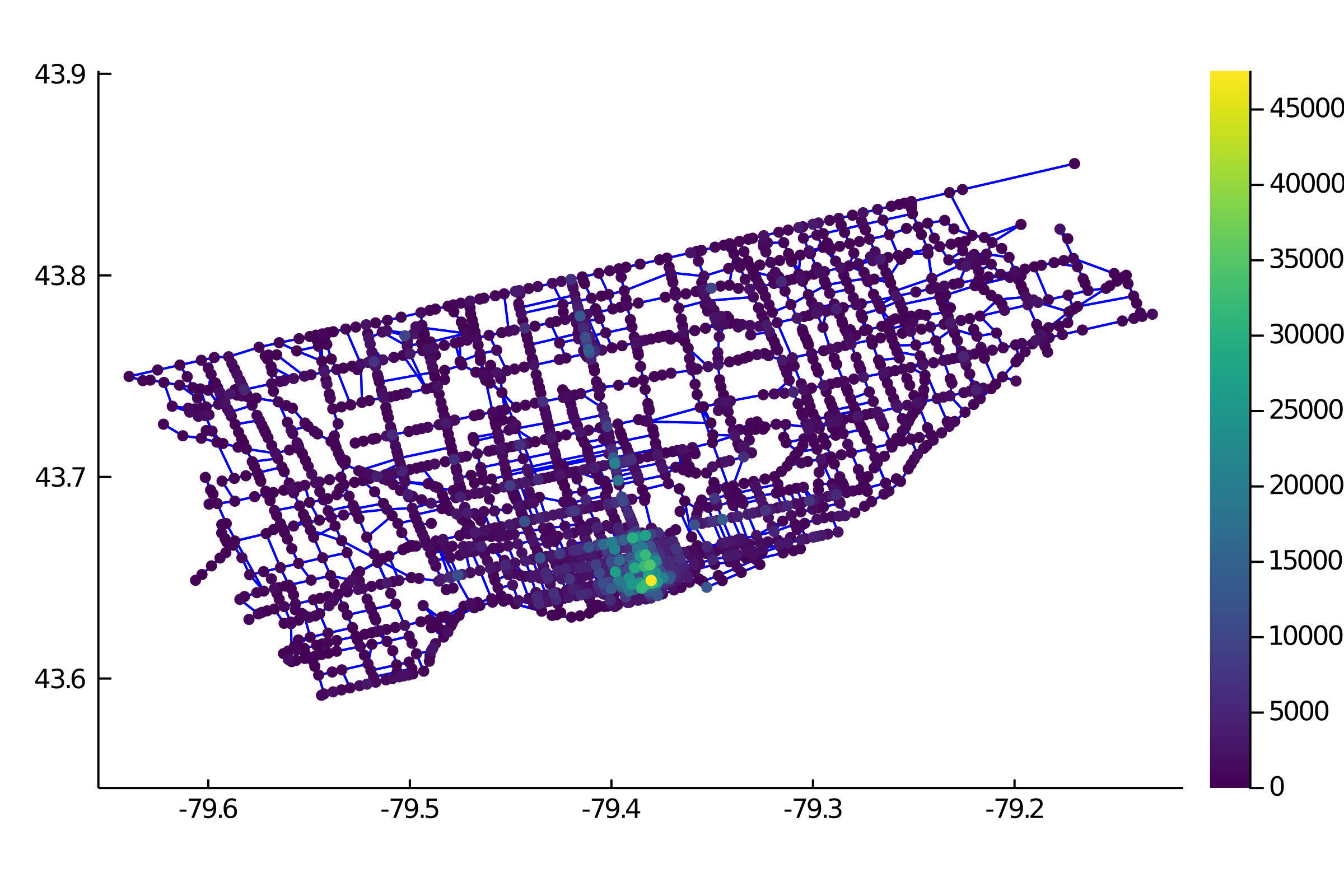}
    \caption{Pedestrian volume data measured at the street intersections}
    \label{fig:toronto-fp-graph}
  \end{subfigure}
  \begin{subfigure}{0.49\textwidth}
    \centering\includegraphics[width=\textwidth]{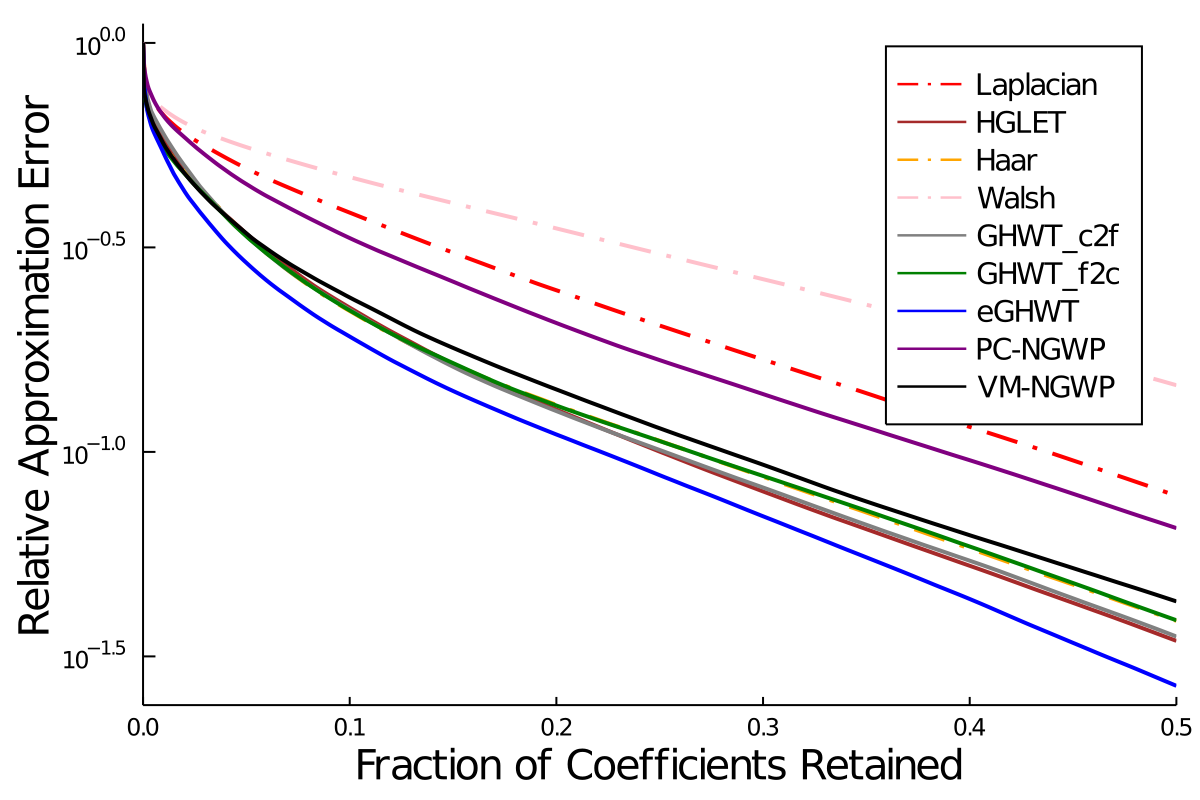}
    \caption{Approximation performance of various methods}
    \label{fig:toronto-fp-approx}
  \end{subfigure}
  \caption{The pedestrian volume graph signal on the Toronto street network (a);
    the results of our approximation experiments (b)}
  \label{fig:toronto-fp}
\end{figure}

Now, let us analyze the pedestrian volume data measured at the street
intersections as shown in Fig.~\ref{fig:toronto-fp-graph}, which is highly localized around the specific part of the downtown region (the dense
  region in the lower middle section) of the street graph.
Figure~\ref{fig:toronto-fp-approx} shows the approximation errors
of various methods.
From Fig.~\ref{fig:toronto-fp-approx}, we observe the following:
1) the eGHWT best basis clearly outperformed all the other methods;
2) the HGLET best basis and the GHWT c2f best basis followed the eGHWT best
basis;
3) the graph Haar basis and the GHWT f2c best basis were next best performers;
4) The VM-NGWP best basis followed those five while the PC-NGWP best basis was
the distant seventh performer; and
5) the global bases such as the graph Laplacian eigenbasis and the graph Walsh
based were the worst performers.
Considering the non-smooth and highly localized nature of the input signal,
it is not surprising that the global bases did not perform well and that
the non-smooth local bases (the eGHWT, the GHWT best bases, the graph Haar
basis) and the basis vectors whose supports strictly follow the partition
pattern of the primal graph (the HGLET best basis) had an edge over the
NGWP best bases that contain smooth basis vectors whose supports may spread
among neighboring regions.

\begin{figure}
  \begin{center}
    \begin{subfigure}{0.245\textwidth}
      \includegraphics[width = \textwidth]{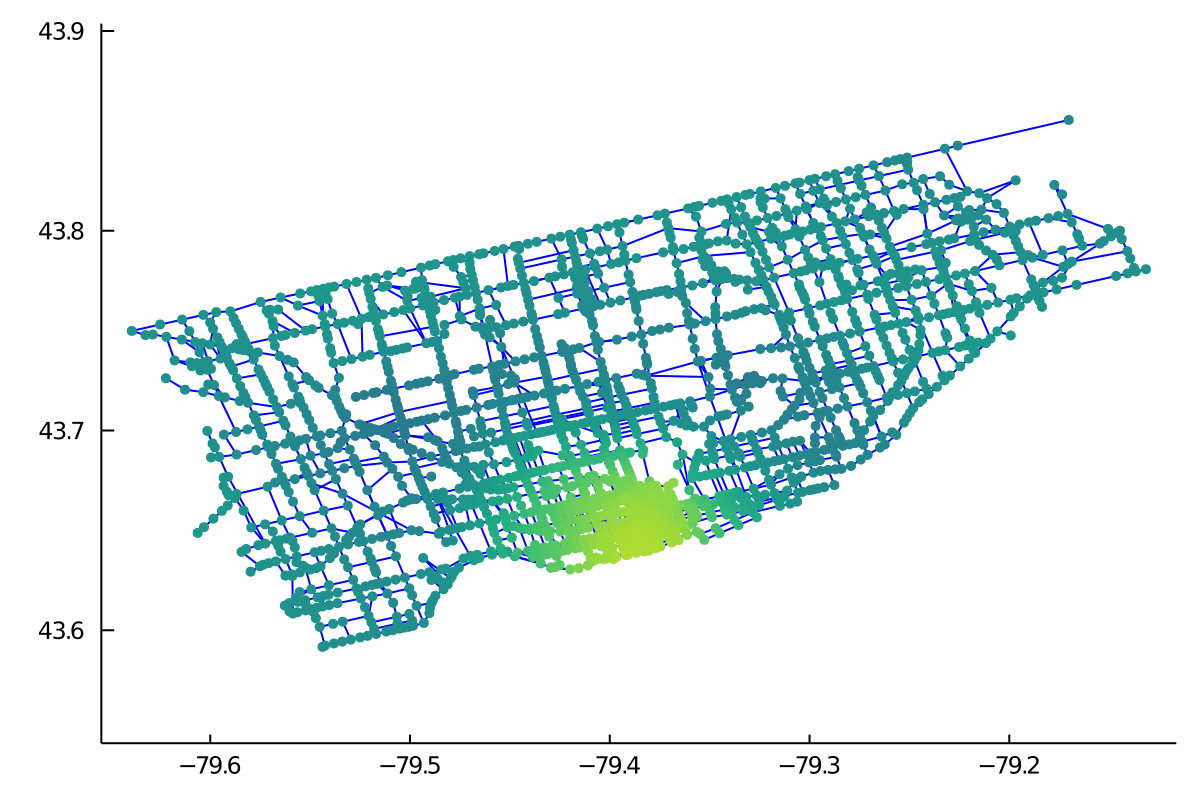}
      \caption{$\bpsi^{(4)}_{0,2}$}
    \end{subfigure}
    \begin{subfigure}{0.245\textwidth}
      \includegraphics[width = \textwidth]{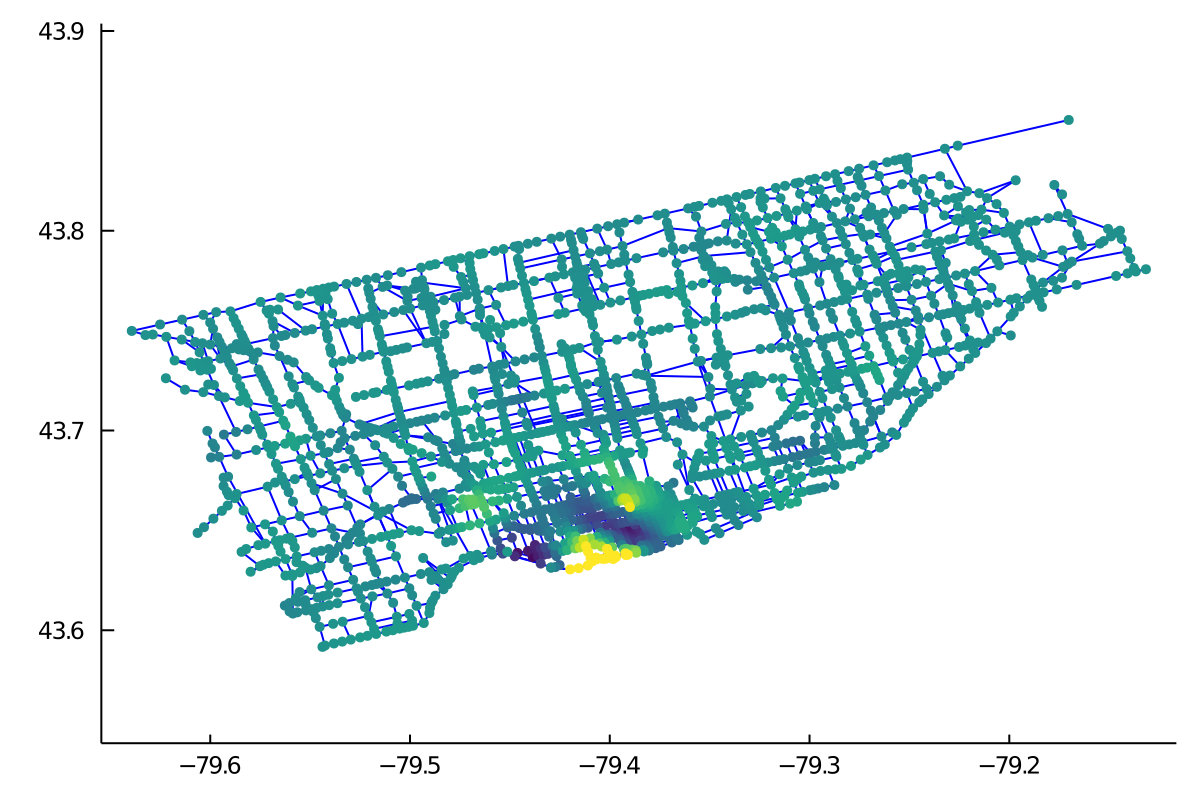}
      \caption{$\bpsi^{(3)}_{1,22}$}
    \end{subfigure}
    \begin{subfigure}{0.245\textwidth}
      \includegraphics[width = \textwidth]{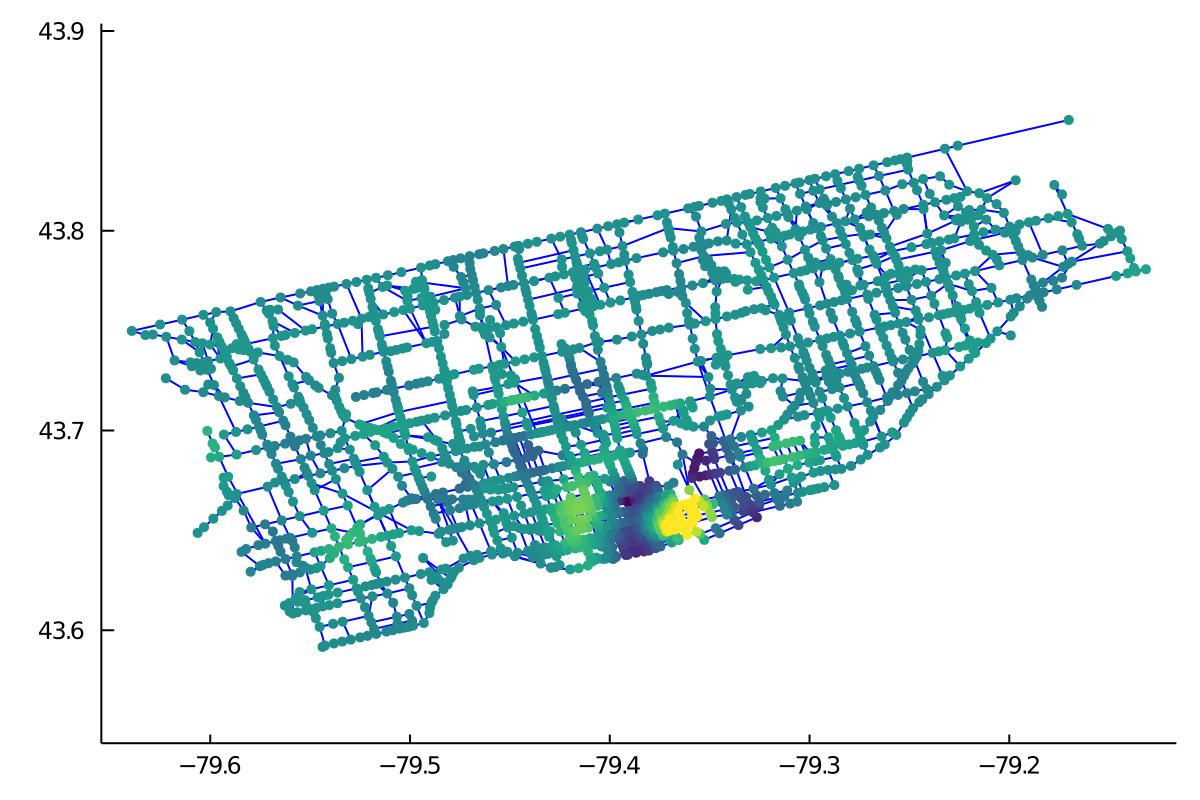}
      \caption{$\bpsi^{(3)}_{1,75}$}
    \end{subfigure}
    \begin{subfigure}{0.245\textwidth}
      \includegraphics[width = \textwidth]{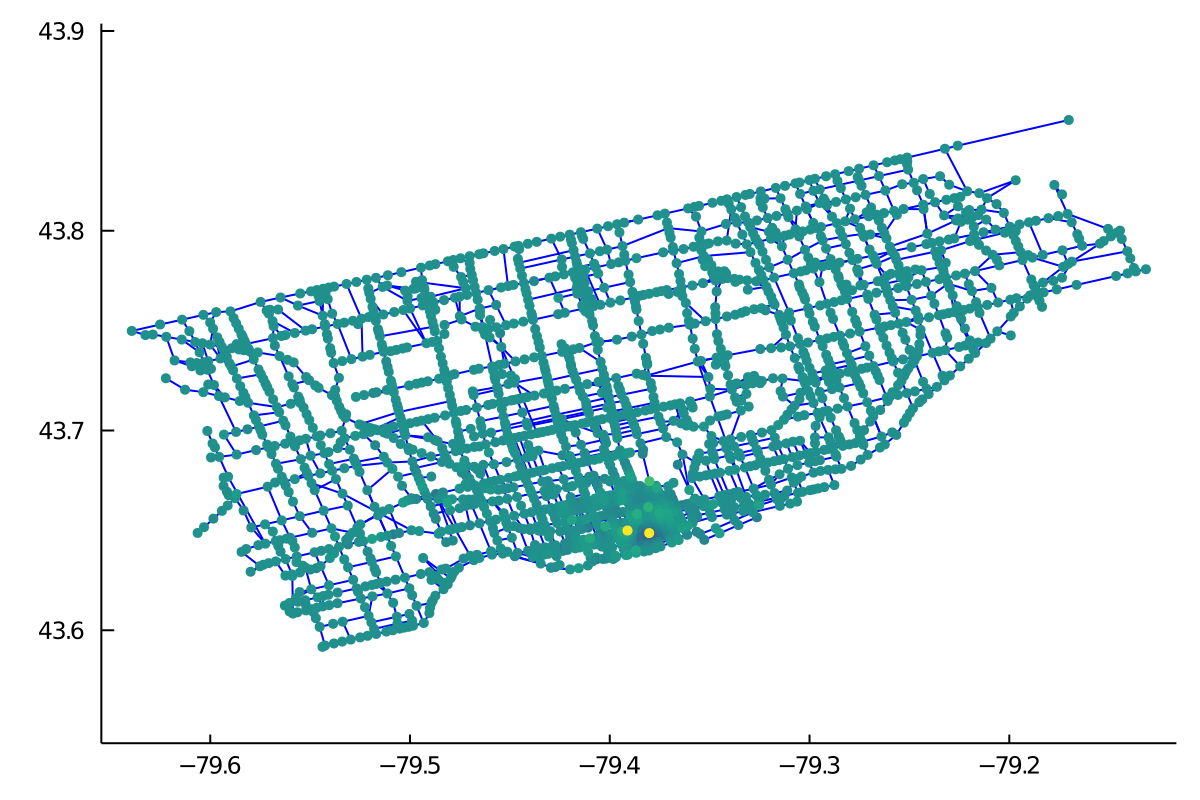}
      \caption{$\bpsi^{(1)}_{1,1245}$}
    \end{subfigure}
    \\
    \begin{subfigure}{0.245\textwidth}
      \includegraphics[width = \textwidth]{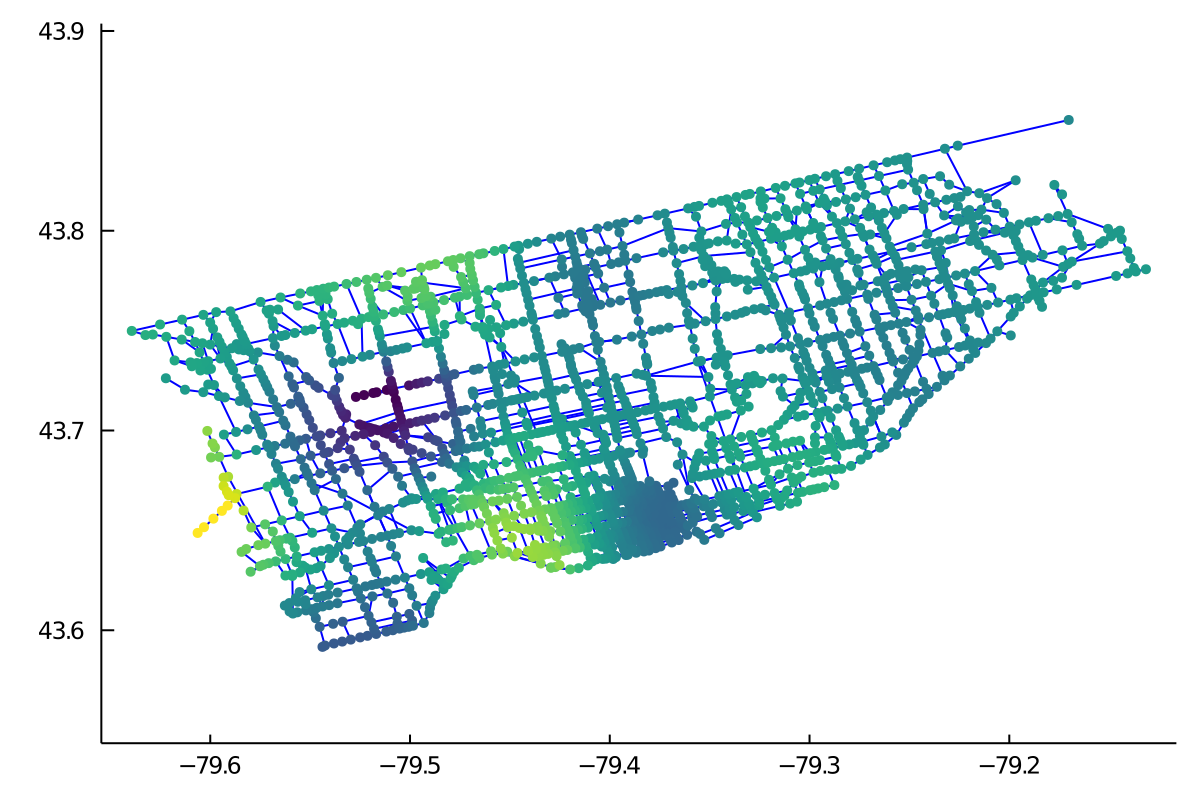}
      \caption{$\bpsi^{(7)}_{8,0}$}
    \end{subfigure}
    \begin{subfigure}{0.245\textwidth}
      \includegraphics[width = \textwidth]{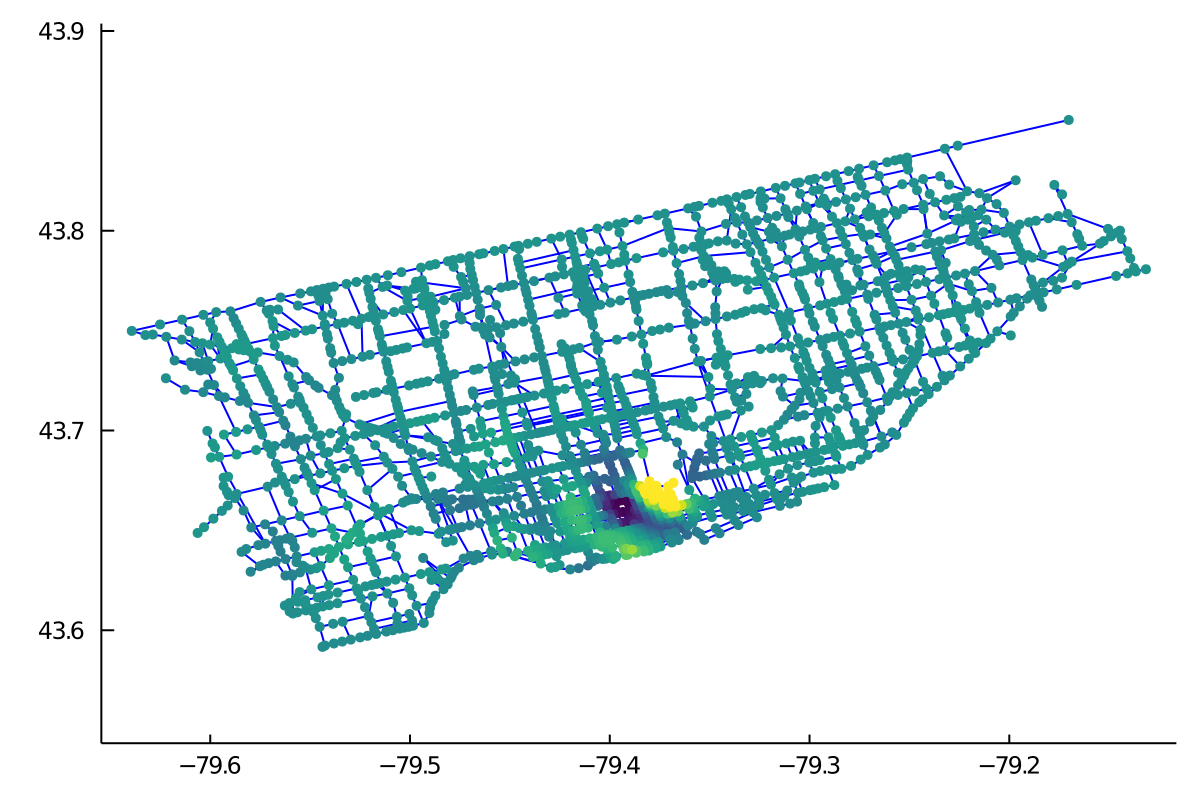}
      \caption{$\bpsi^{(3)}_{1,87}$}
    \end{subfigure}
    \begin{subfigure}{0.245\textwidth}
      \includegraphics[width = \textwidth]{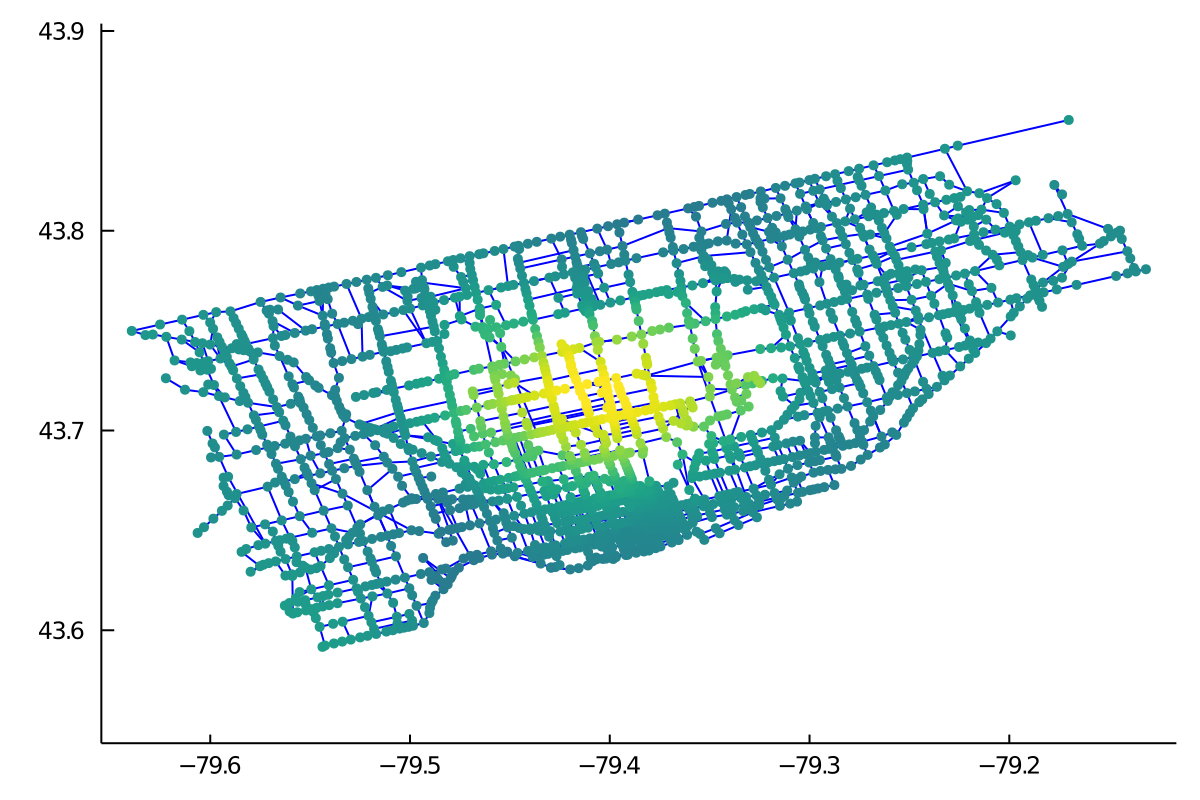}
      \caption{$\bpsi^{(4)}_{0,11}$}
    \end{subfigure}
    \begin{subfigure}{0.245\textwidth}
      \includegraphics[width = \textwidth]{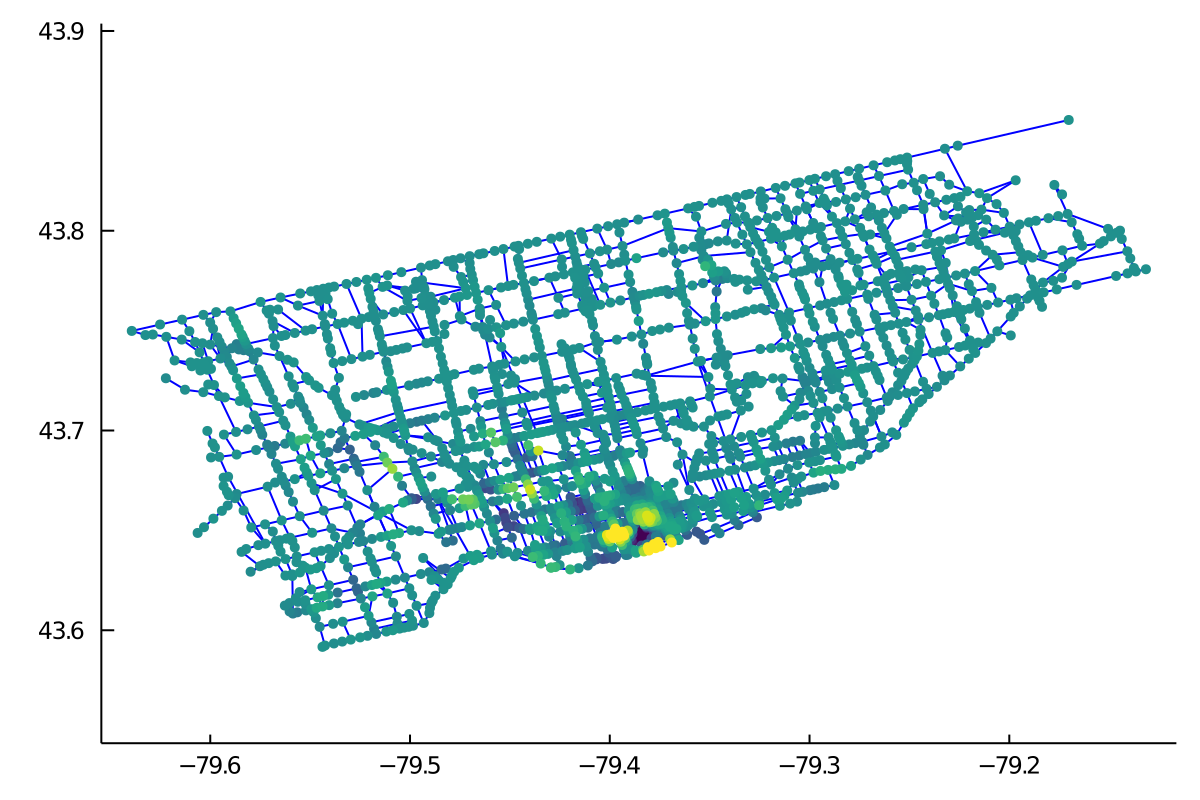}
      \caption{$\bpsi^{(2)}_{1,136}$}
    \end{subfigure}
    \\
    \begin{subfigure}{0.245\textwidth}
      \includegraphics[width = \textwidth]{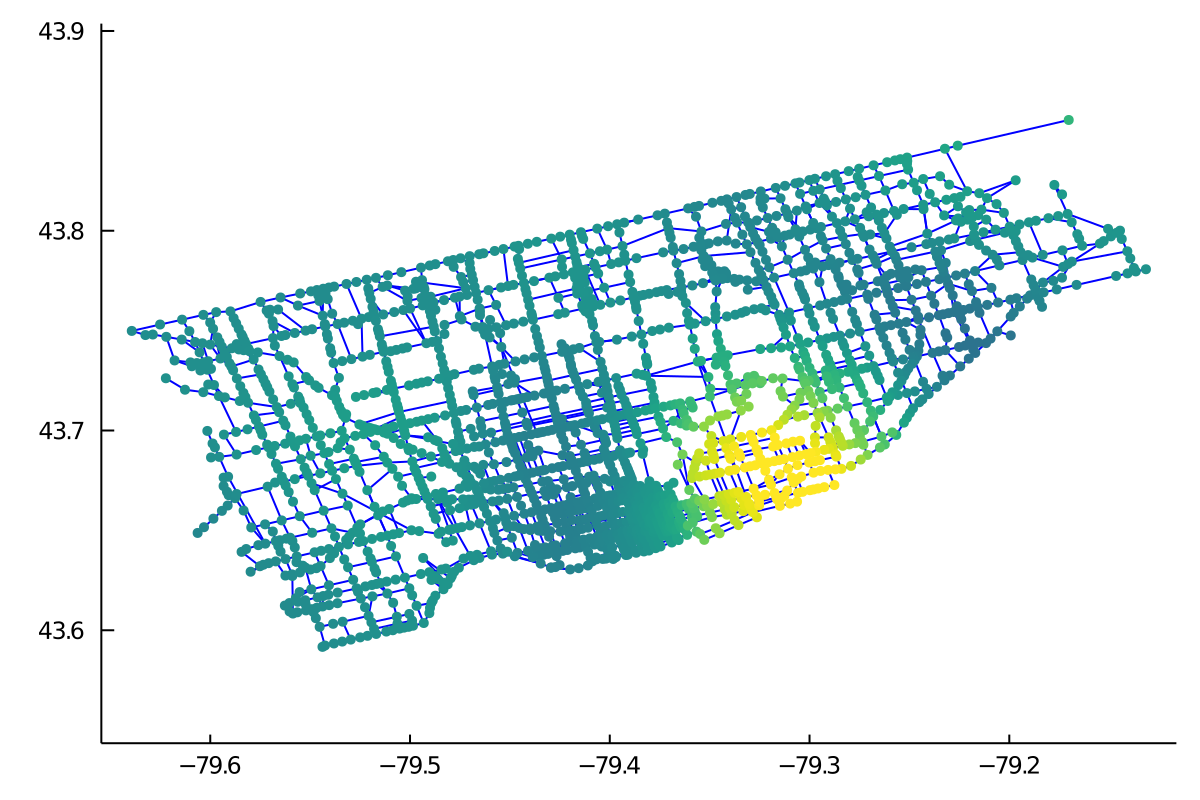}
      \caption{$\bpsi^{(4)}_{0,5}$}
    \end{subfigure}
    \begin{subfigure}{0.245\textwidth}
      \includegraphics[width = \textwidth]{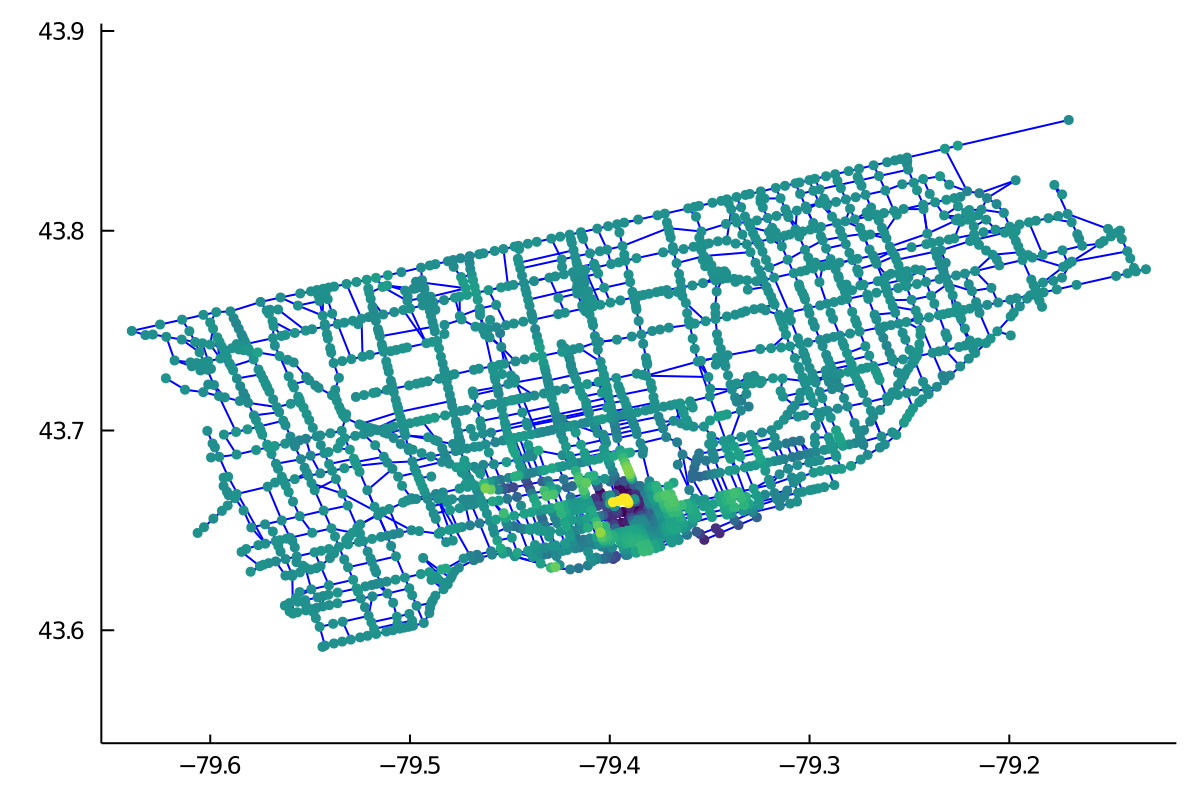}
      \caption{$\bpsi^{(2)}_{1,16}$}
    \end{subfigure}
    \begin{subfigure}{0.245\textwidth}
      \includegraphics[width = \textwidth]{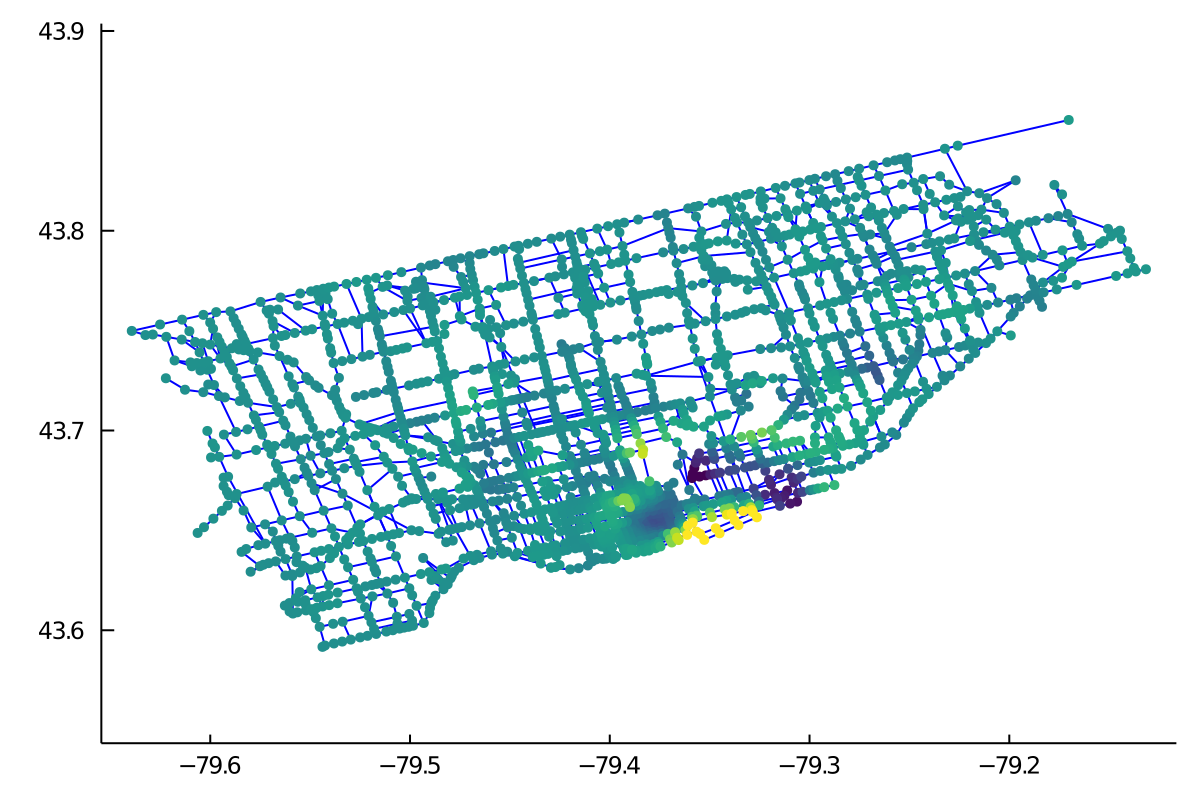}
      \caption{$\bpsi^{(3)}_{1,98}$}
    \end{subfigure}
    \begin{subfigure}{0.245\textwidth}
      \includegraphics[width = \textwidth]{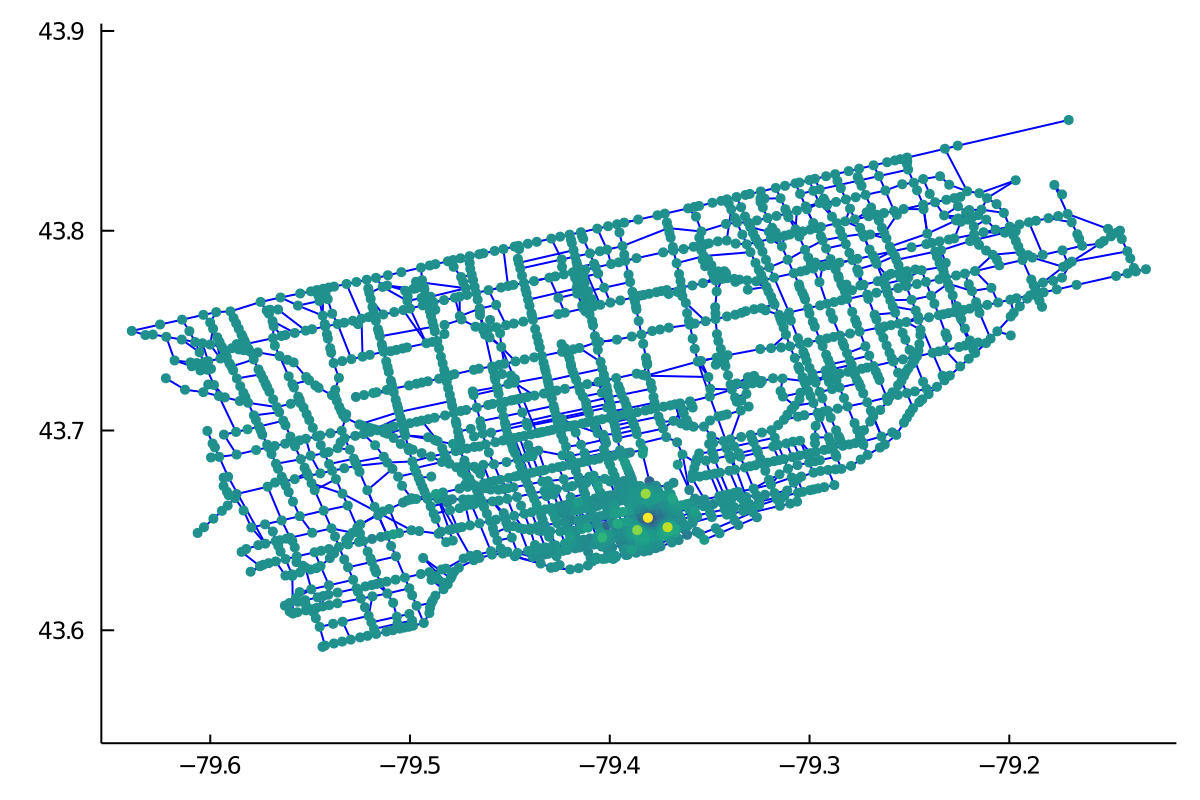}
      \caption{$\bpsi^{(1)}_{1,1660}$}
    \end{subfigure}
    \\
    \begin{subfigure}{0.245\textwidth}
      \includegraphics[width = \textwidth]{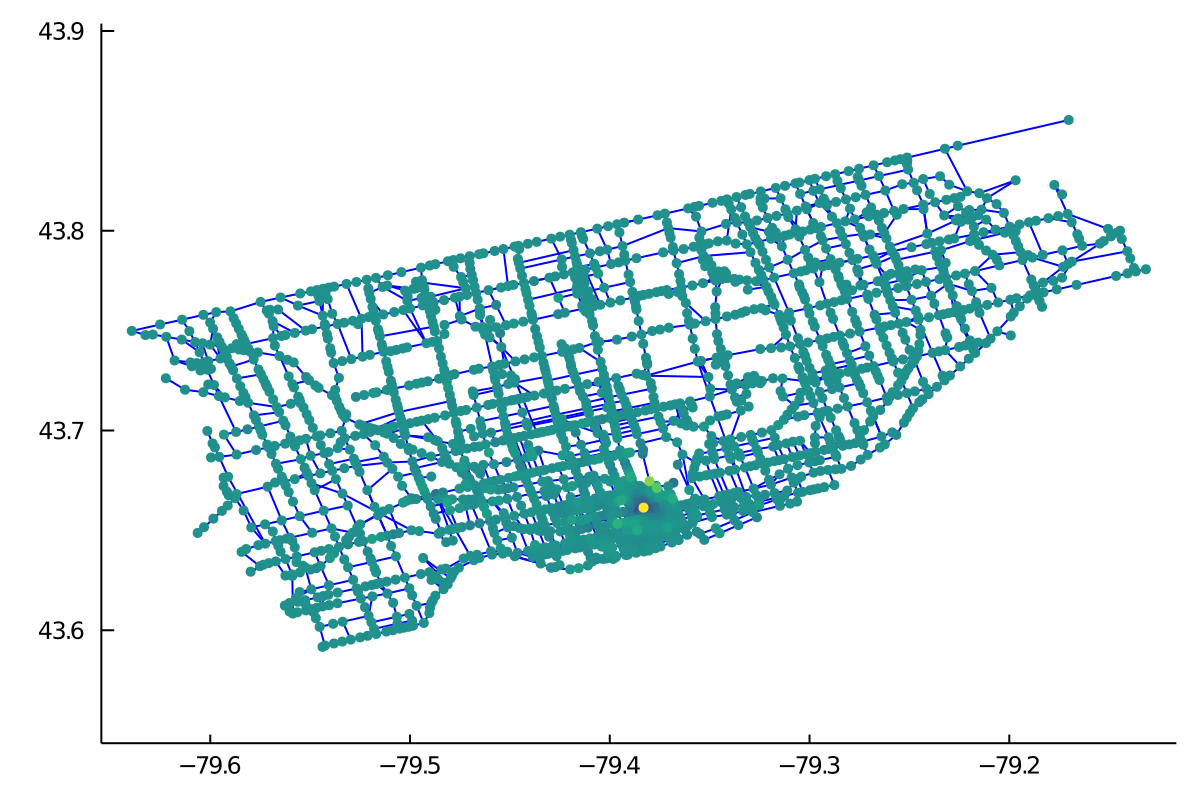}
      \caption{$\bpsi^{(1)}_{1,1585}$}
    \end{subfigure}
    \begin{subfigure}{0.245\textwidth}
      \includegraphics[width = \textwidth]{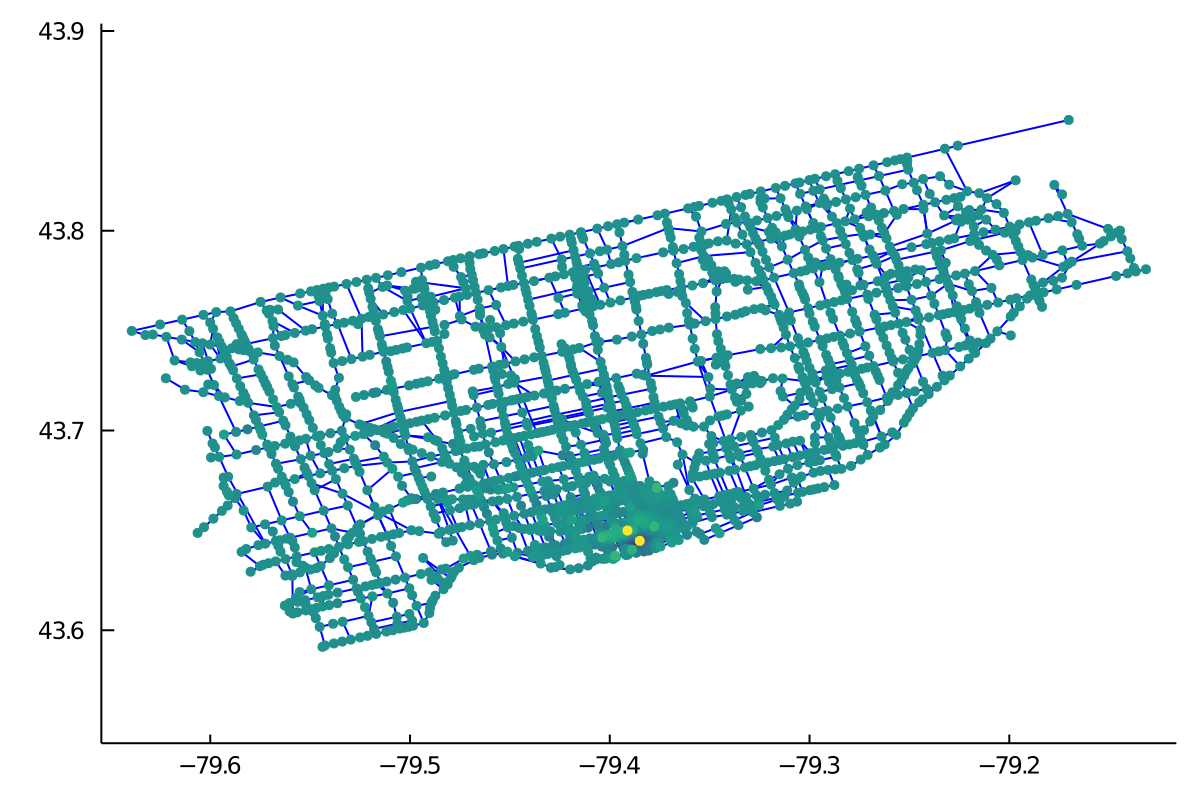}
      \caption{$\bpsi^{(1)}_{1,1161}$}
    \end{subfigure}
    \begin{subfigure}{0.245\textwidth}
      \includegraphics[width = \textwidth]{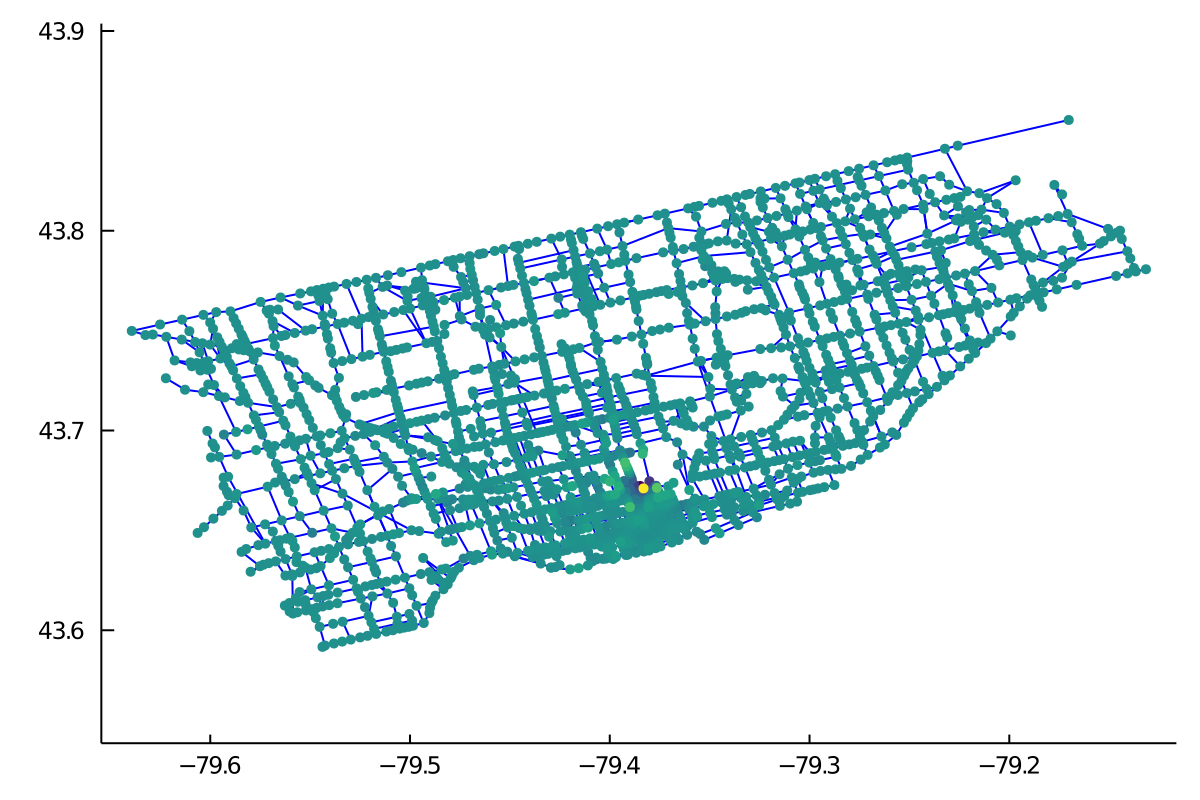}
      \caption{$\bpsi^{(1)}_{1,1487}$}
    \end{subfigure}
    \begin{subfigure}{0.245\textwidth}
      \includegraphics[width = \textwidth]{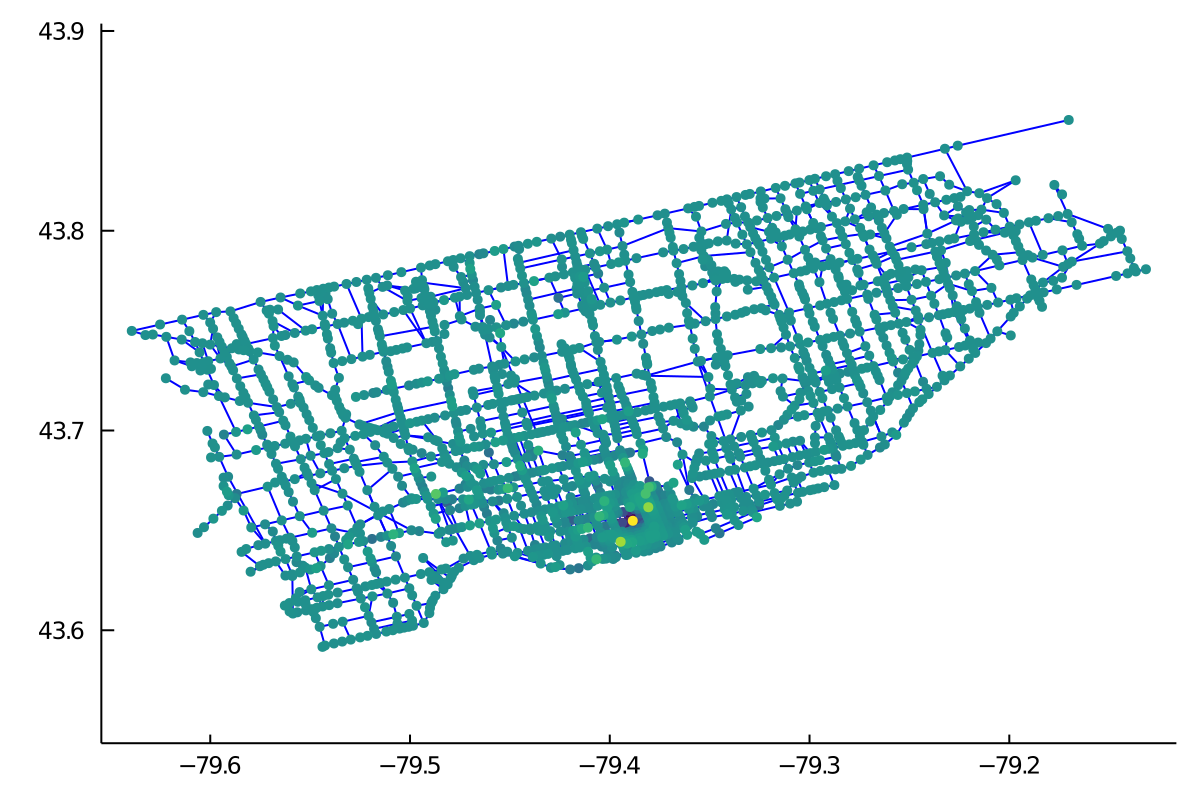}
      \caption{$\bpsi^{(1)}_{1,1612}$}
    \end{subfigure}
    \caption{Sixteen most significant VM-NGWP best basis vectors
      for pedestrian volume data on the Toronto street map.
      The basis vector amplitudes within $(-0.075, 0.075)$
        are mapped to the viridis colormap.}
    \label{fig:toronto-fp-top16}
  \end{center}
\end{figure}

\begin{figure}
  \begin{center}
    \begin{subfigure}{0.245\textwidth}
      \includegraphics[width = \textwidth]{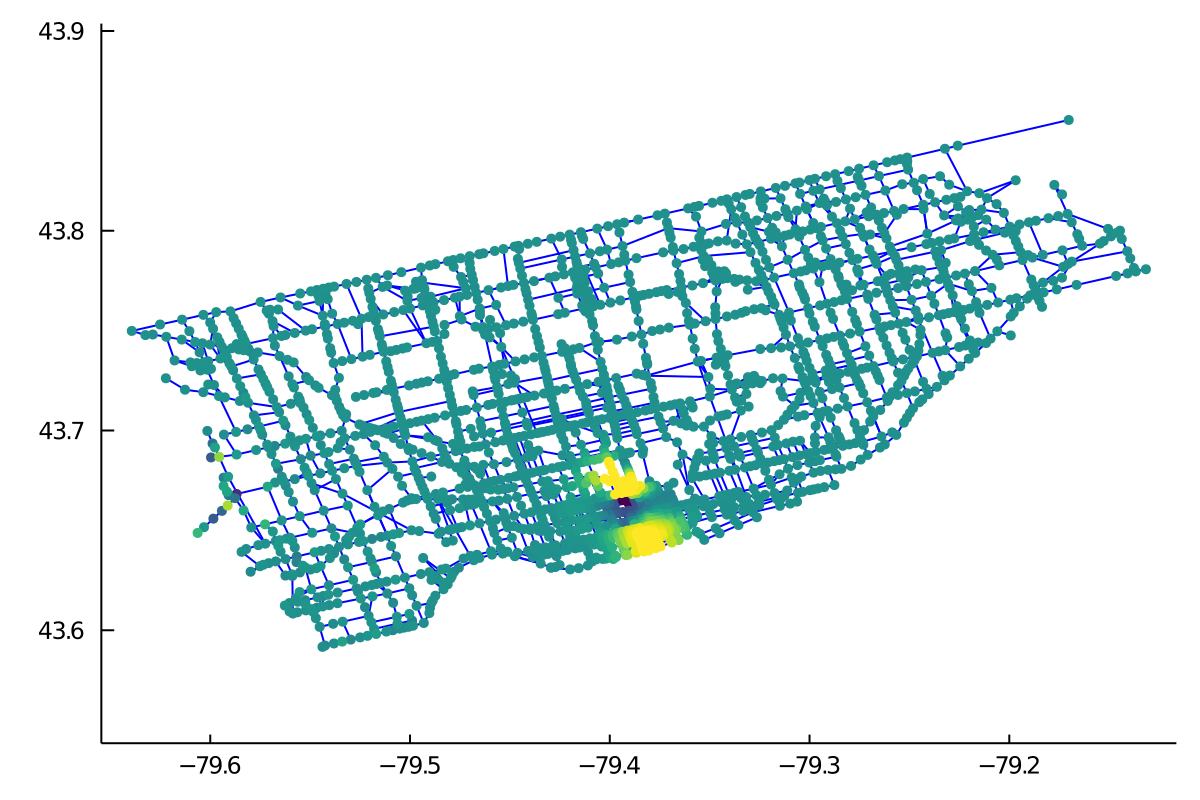}
      \caption{$\bpsi^{(1)}_{0,587}$}
    \end{subfigure}
    \begin{subfigure}{0.245\textwidth}
      \includegraphics[width = \textwidth]{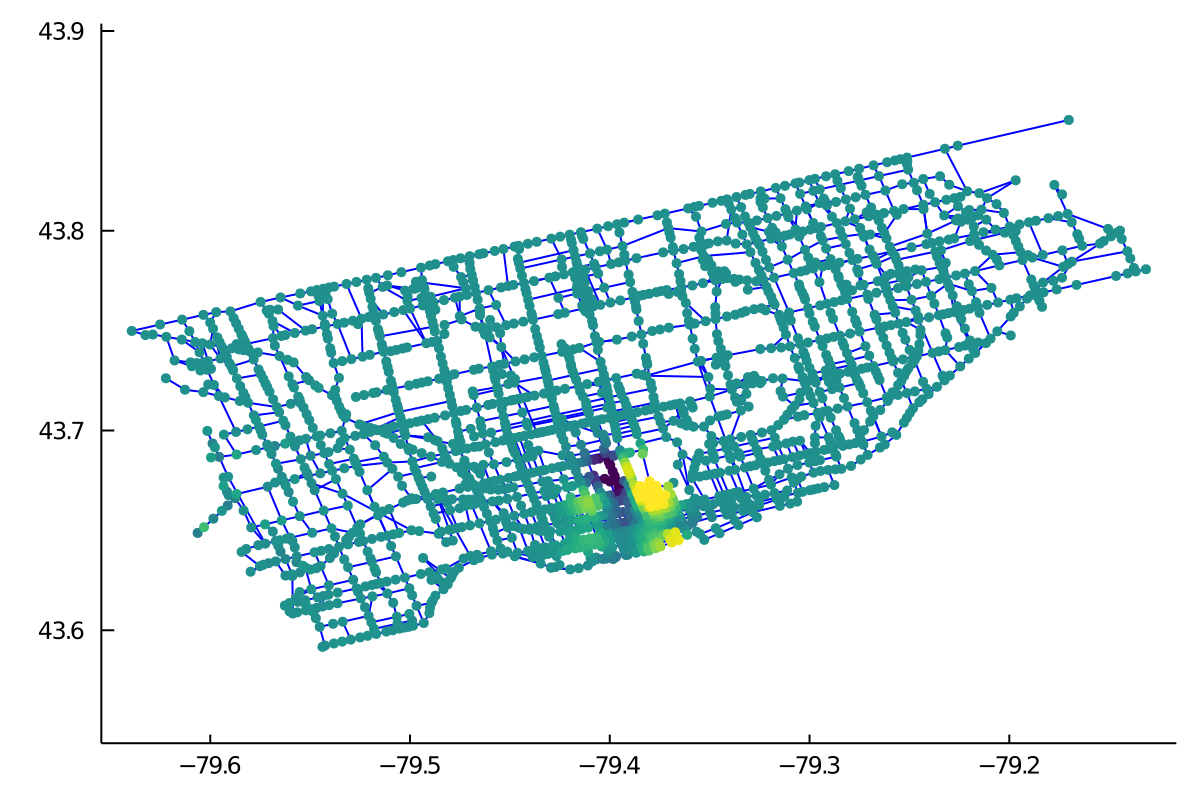}
      \caption{$\bpsi^{(1)}_{0,580}$}
    \end{subfigure}
    \begin{subfigure}{0.245\textwidth}
      \includegraphics[width = \textwidth]{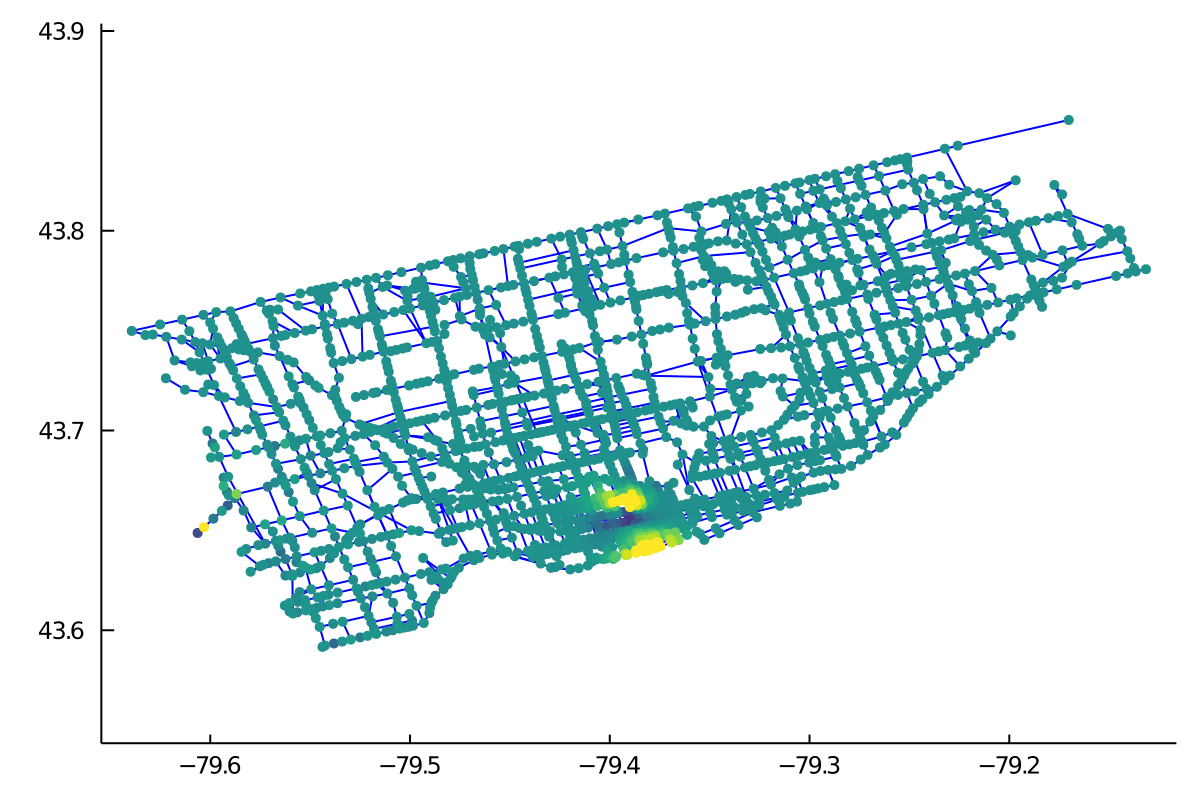}
      \caption{$\bpsi^{(1)}_{0,588}$}
    \end{subfigure}
    \begin{subfigure}{0.245\textwidth}
      \includegraphics[width = \textwidth]{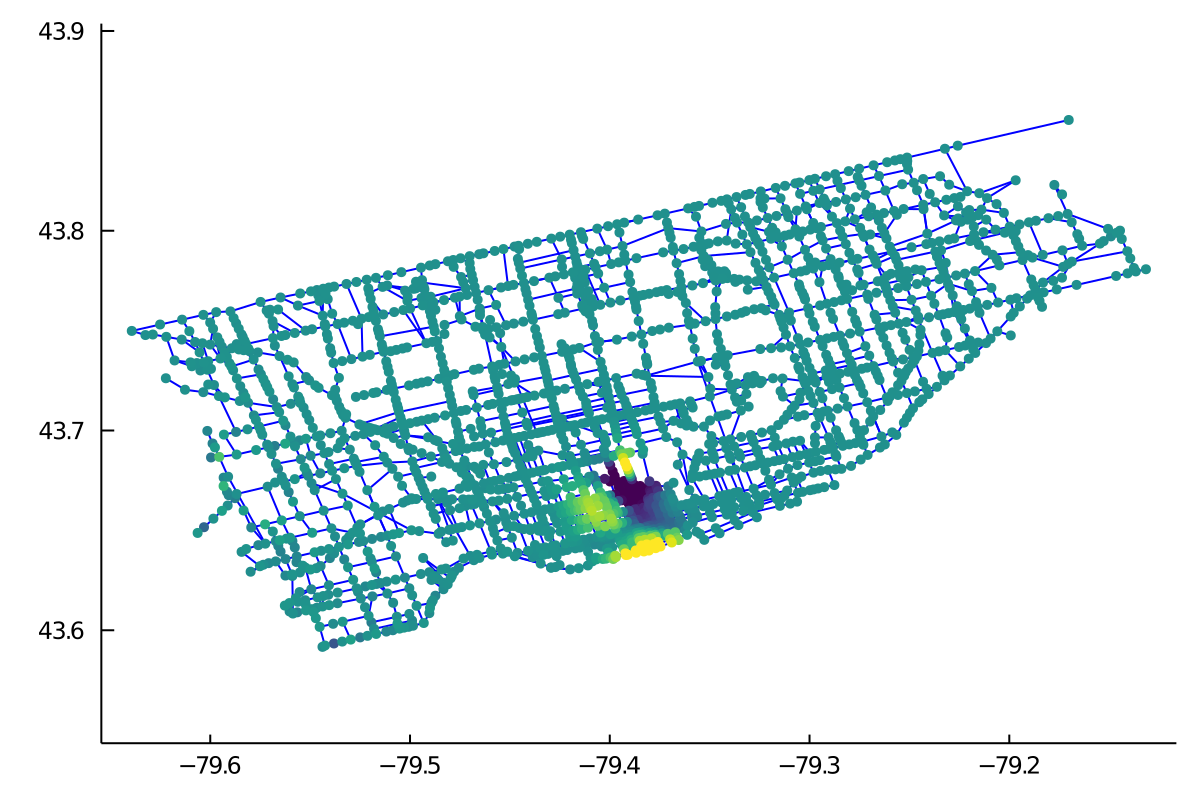}
      \caption{$\bpsi^{(1)}_{0,584}$}
    \end{subfigure}
    \\
    \begin{subfigure}{0.245\textwidth}
      \includegraphics[width = \textwidth]{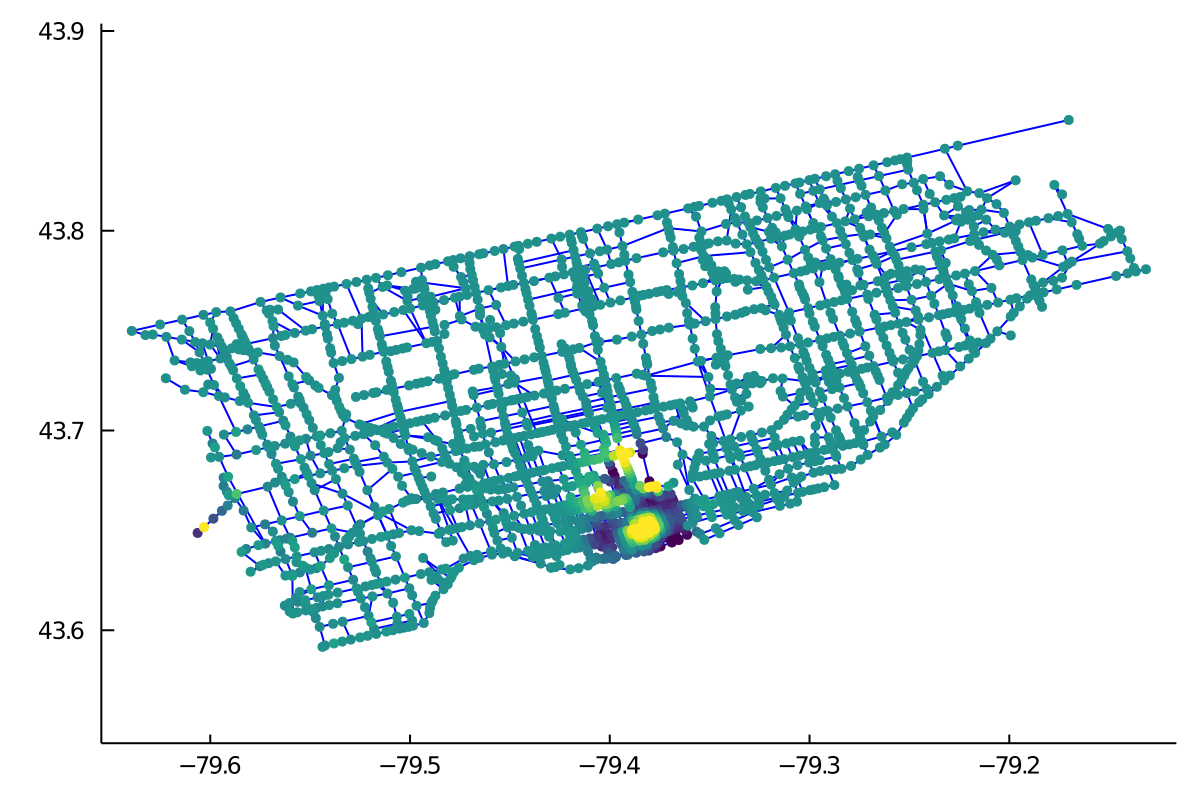}
      \caption{$\bpsi^{(1)}_{0,583}$}
    \end{subfigure}
    \begin{subfigure}{0.245\textwidth}
      \includegraphics[width = \textwidth]{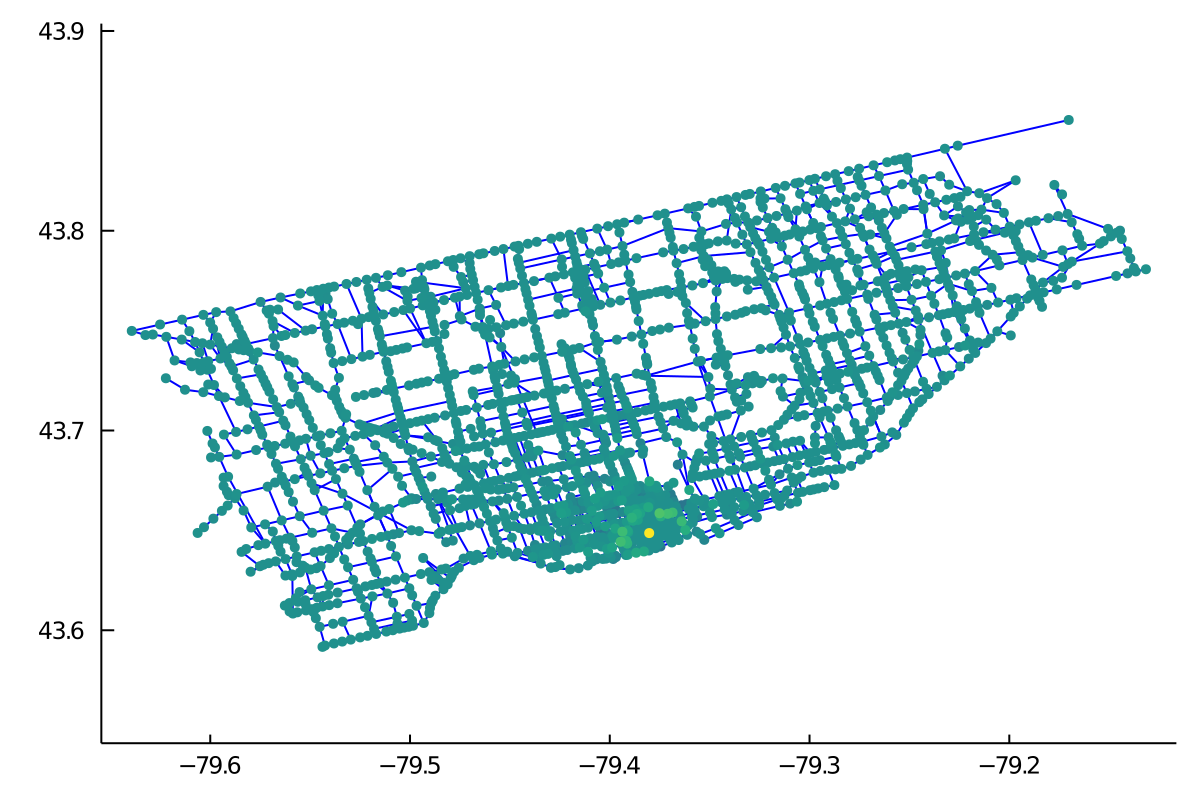}
      \caption{$\bpsi^{(1)}_{1,1188}$}
    \end{subfigure}
    \begin{subfigure}{0.245\textwidth}
      \includegraphics[width = \textwidth]{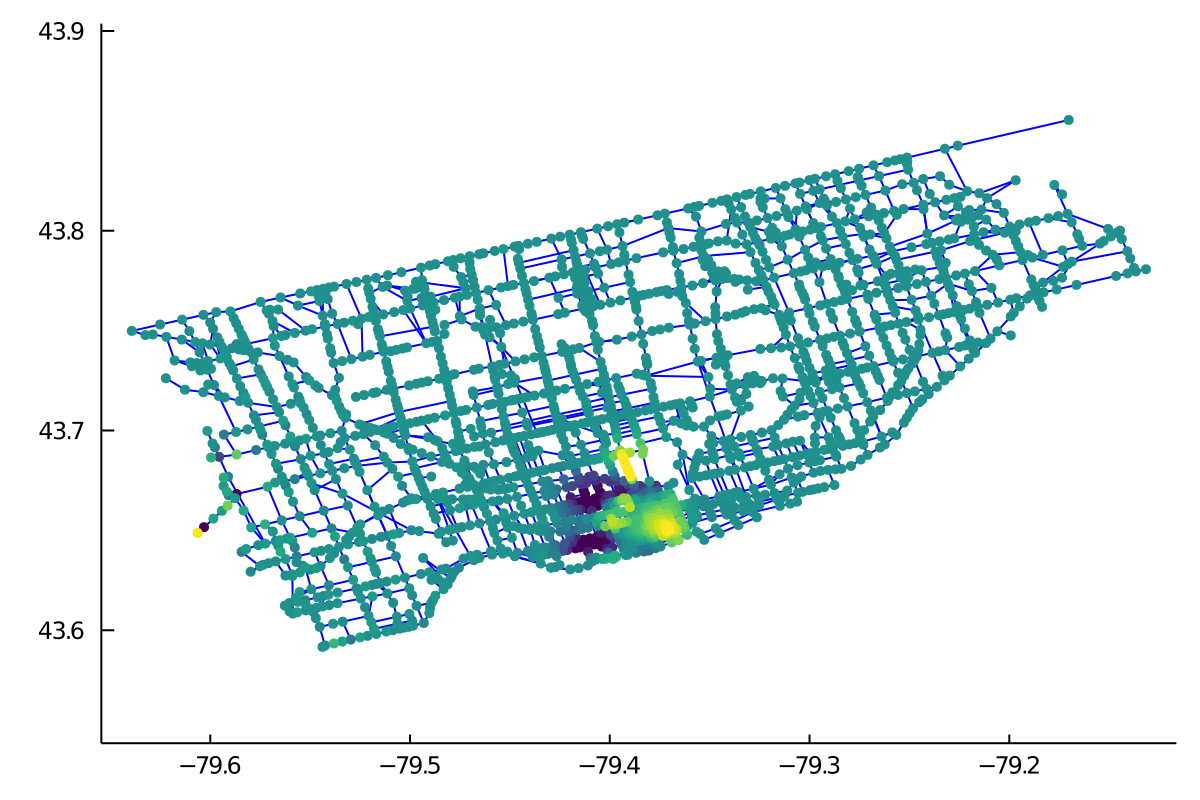}
      \caption{$\bpsi^{(1)}_{0,581}$}
    \end{subfigure}
    \begin{subfigure}{0.245\textwidth}
      \includegraphics[width = \textwidth]{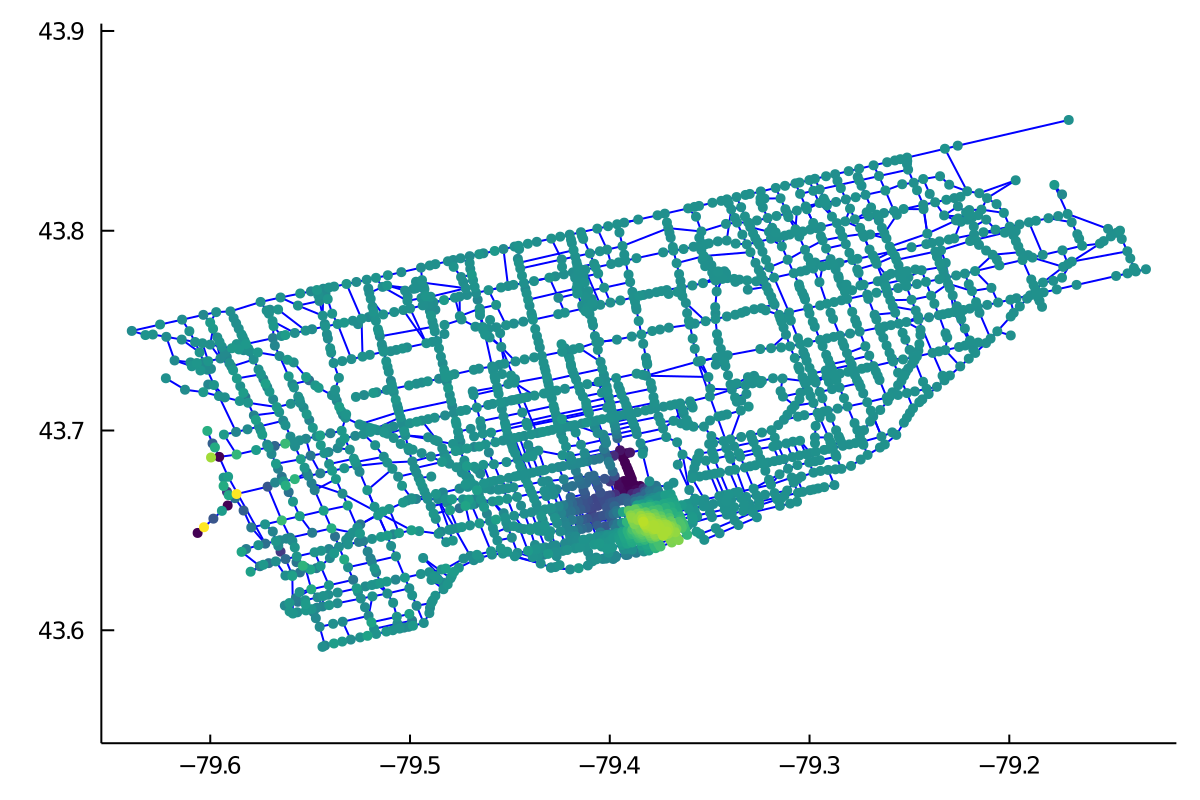}
      \caption{$\bpsi^{(1)}_{0,585}$}
    \end{subfigure}
    \\
    \begin{subfigure}{0.245\textwidth}
      \includegraphics[width = \textwidth]{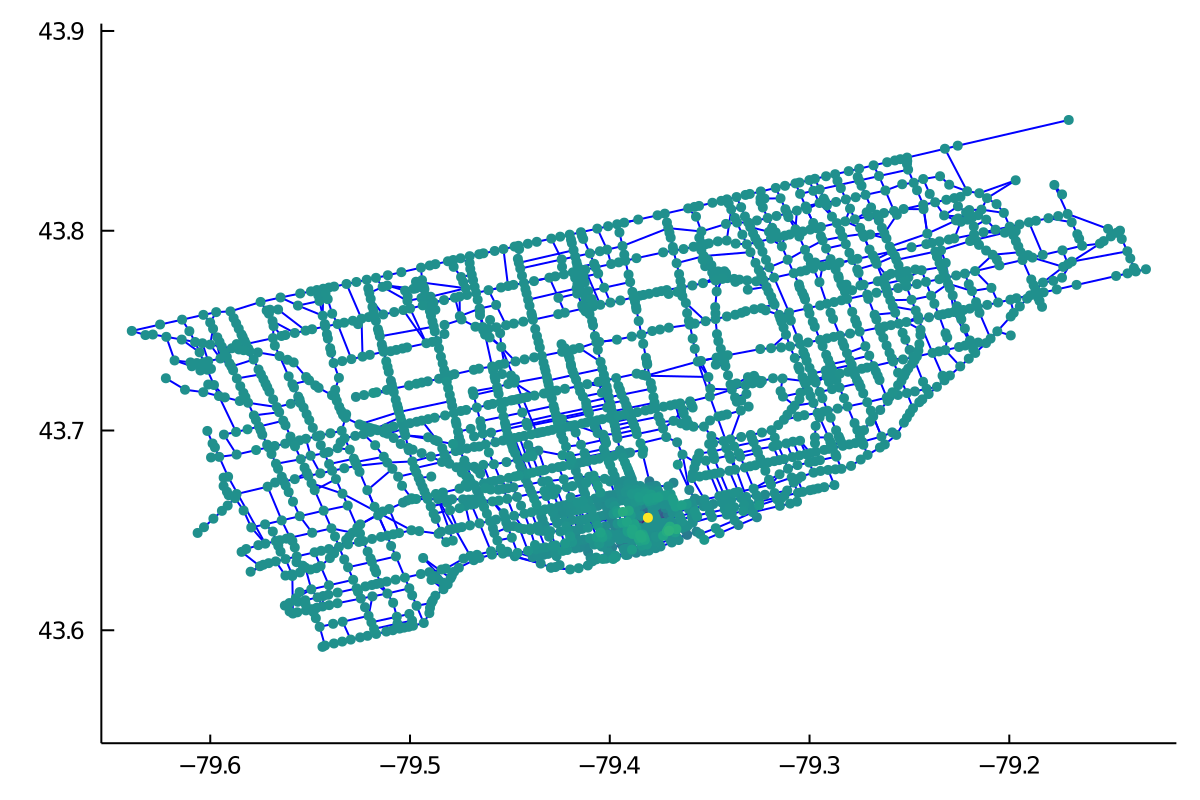}
      \caption{$\bpsi^{(1)}_{1,906}$}
    \end{subfigure}
    \begin{subfigure}{0.245\textwidth}
      \includegraphics[width = \textwidth]{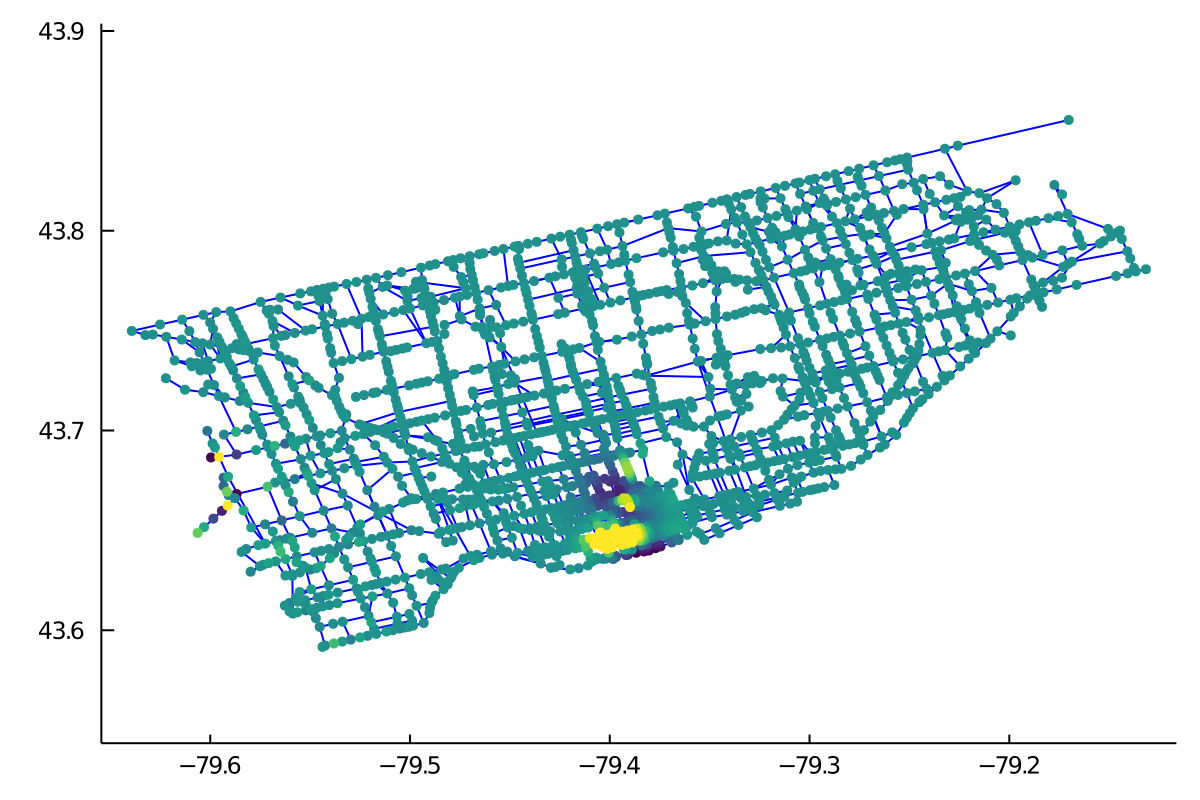}
      \caption{$\bpsi^{(1)}_{0,582}$}
    \end{subfigure}
    \begin{subfigure}{0.245\textwidth}
      \includegraphics[width = \textwidth]{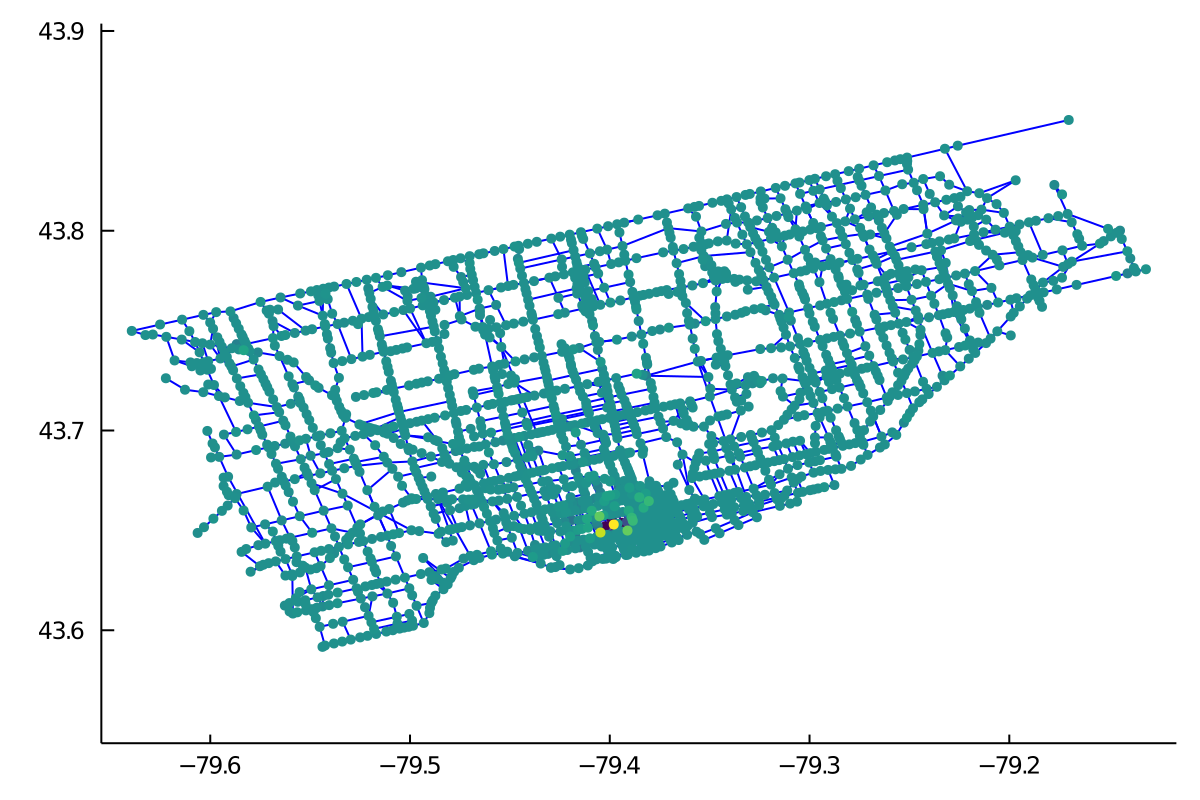}
      \caption{$\bpsi^{(1)}_{1,1421}$}
    \end{subfigure}
    \begin{subfigure}{0.245\textwidth}
      \includegraphics[width = \textwidth]{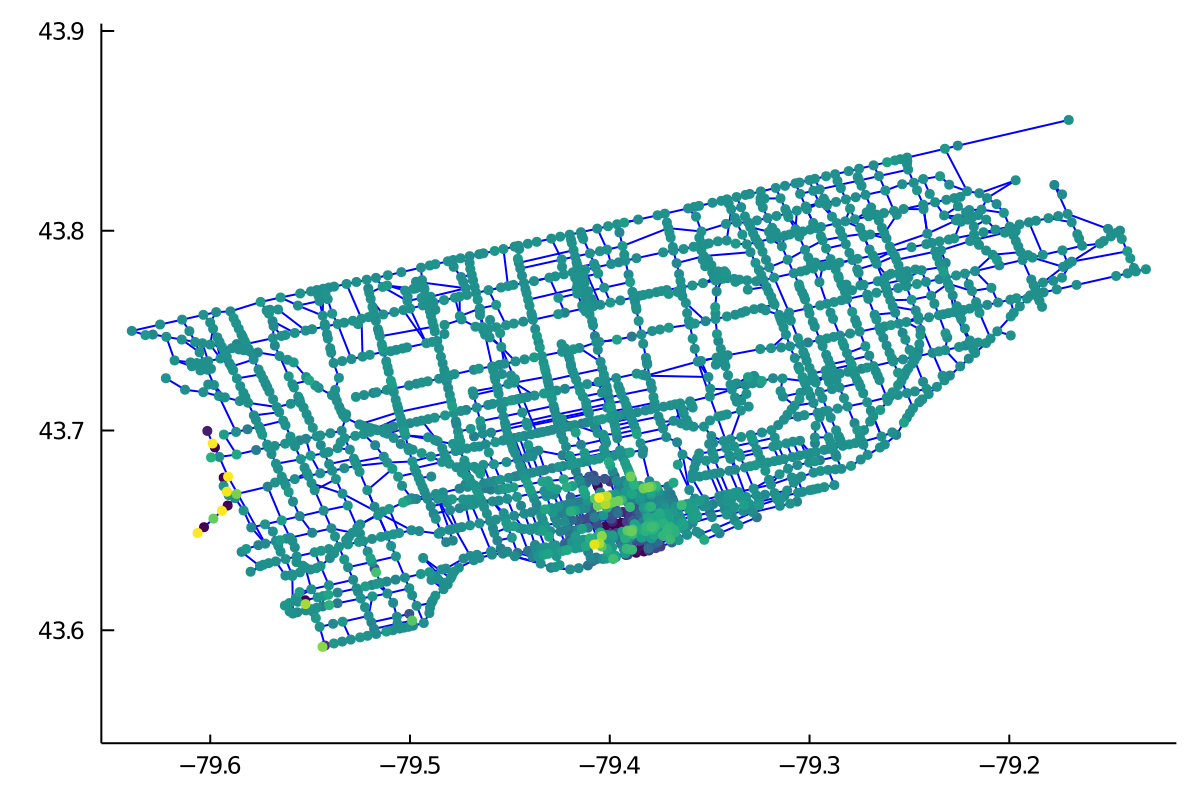}
      \caption{$\bpsi^{(1)}_{1,1685}$}
    \end{subfigure}
    \\
    \begin{subfigure}{0.245\textwidth}
      \includegraphics[width = \textwidth]{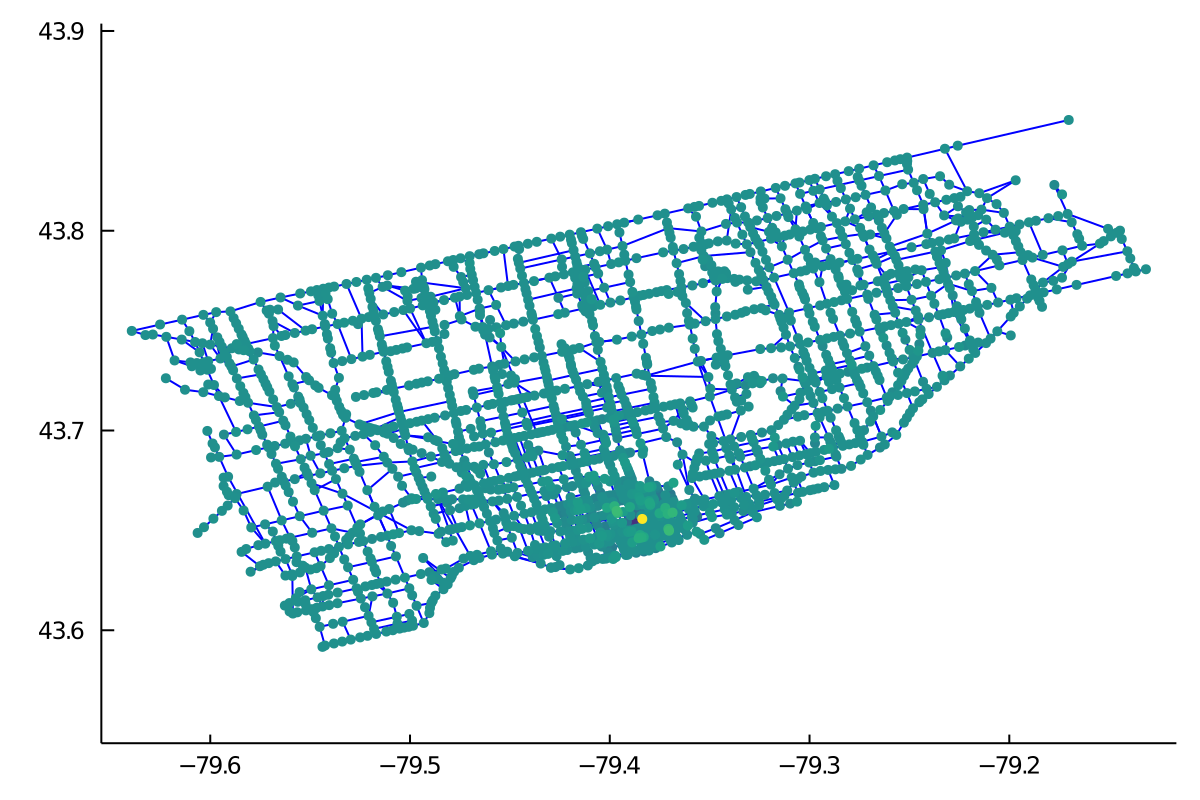}
      \caption{$\bpsi^{(1)}_{1,1159}$}
    \end{subfigure}
    \begin{subfigure}{0.245\textwidth}
      \includegraphics[width = \textwidth]{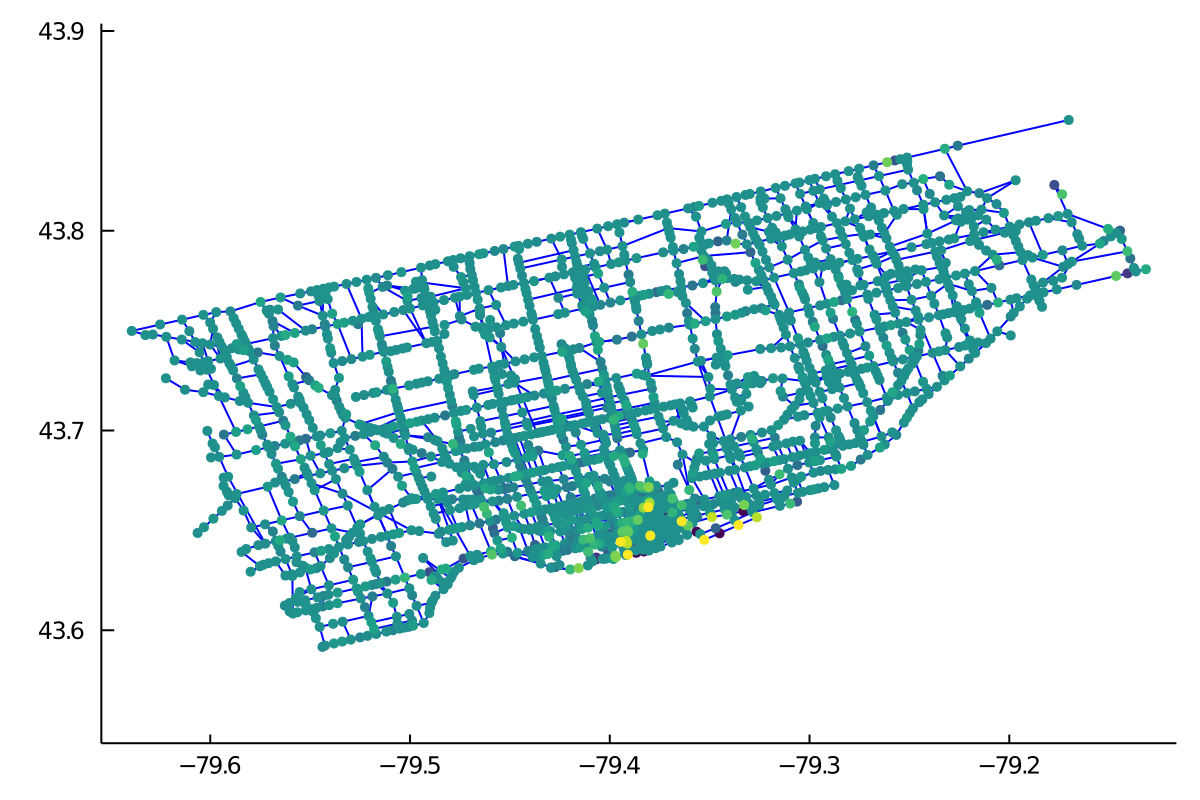}
      \caption{$\bpsi^{(1)}_{1,1214}$}
    \end{subfigure}
    \begin{subfigure}{0.245\textwidth}
      \includegraphics[width = \textwidth]{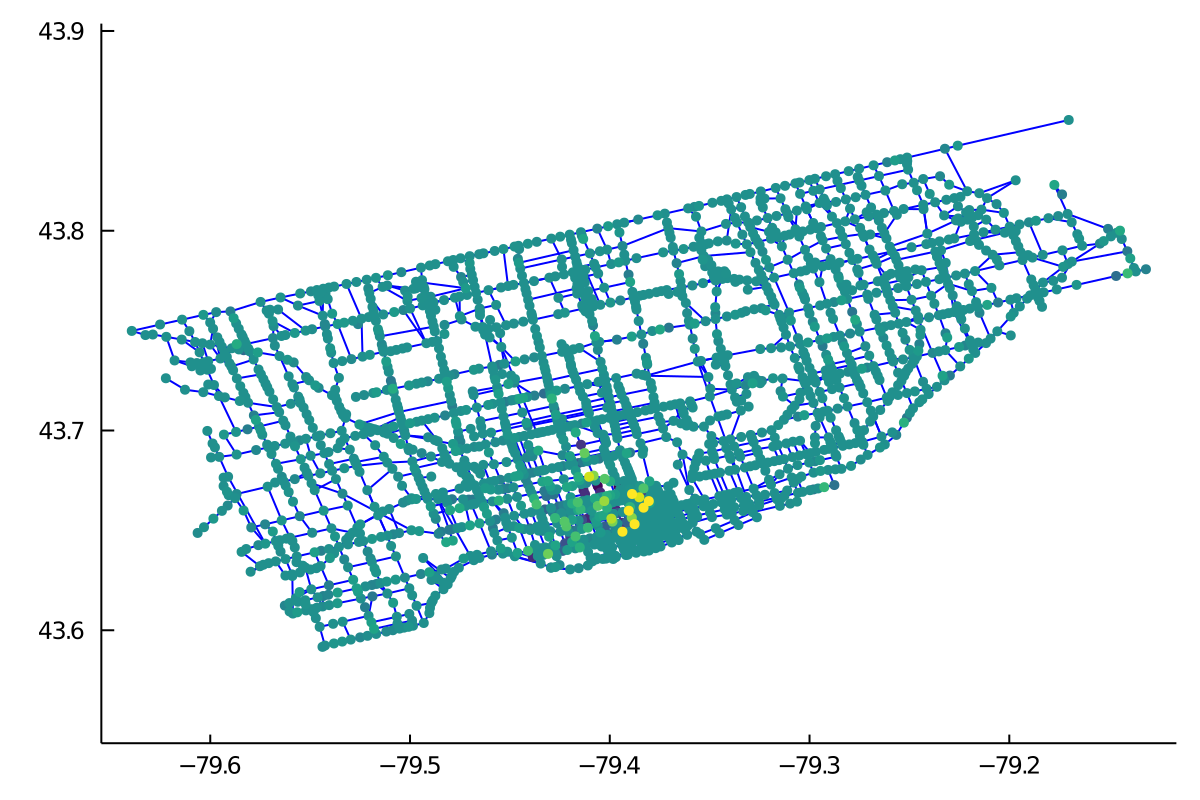}
      \caption{$\bpsi^{(1)}_{1,1419}$}
    \end{subfigure}
    \begin{subfigure}{0.245\textwidth}
      \includegraphics[width = \textwidth]{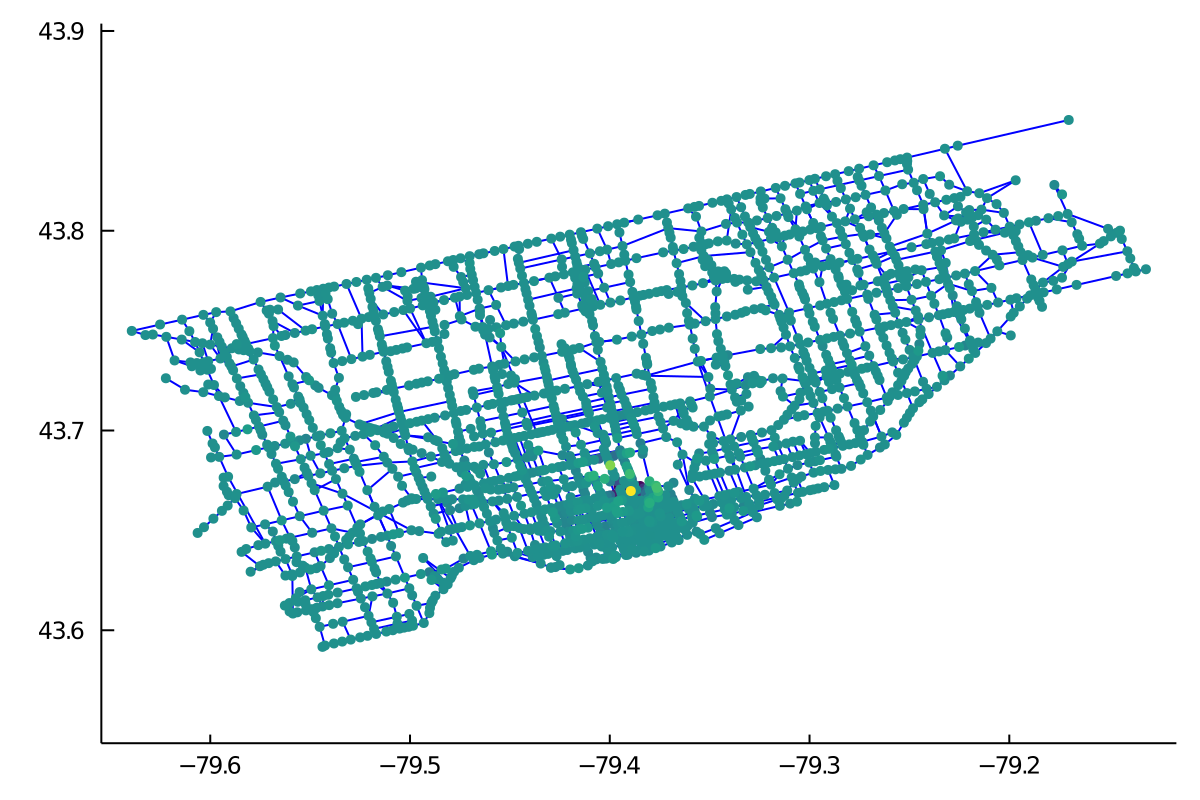}
      \caption{$\bpsi^{(1)}_{1,1203}$}
    \end{subfigure}
    \caption{Sixteen most significant PC-NGWP best basis vectors
      for pedestrian volume data on the Toronto street map.
      The basis vector amplitudes within $(-0.075, 0.075)$
      are mapped to the viridis colormap.}
    \label{fig:toronto-fp-top16-pc}
  \end{center}
\end{figure}

In order to examine the performance difference between the VM-NGWP best basis
and the PC-NGWP best basis, we display their 16 most significant basis vectors
in Figs.~\ref{fig:toronto-fp-top16} and \ref{fig:toronto-fp-top16-pc},
respectively.
We note that the top VM-NGWP best basis vectors exhibit local to intermediate
spatial scales.
The basis vectors with $j=1$ (Fig.~\ref{fig:toronto-fp-top16}d, l, m, n, o, p)
are highly localized at certain nodes within the dense downtown region 
while the basis vectors with $j=2, 3, 4$ (Fig.~\ref{fig:toronto-fp-top16}b, c, f,
g, h, i, j, k) try to characterize the pedestrian volume within the downtown
region and its neighbors as oriented edge detectors.
The top basis vector $\bpsi^{(4)}_{0,2}$ in Fig.~\ref{fig:toronto-fp-top16}a
works as a local averaging operator around the downtown region.
  On the other hand, the top PC-NGWP best basis vectors are more localized
than those of the VM-NGWP best basis vectors. As one can see from
Fig.~\ref{fig:toronto-fp-top16-pc}, there are neither medium nor coarse scale
basis vectors in these top 16 basis vectors. The reason behind these performance
difference between the VM-NGWP and the PC-NGWP is the following.
The VM-NGWP best basis for this graph signal turned out to be 
``almost'' the graph Shannon wavelet basis with the deepest level $J=4$, i.e.,
the basis for the union of the following subspaces:
$V^{\star (4)}_{0}$, $V^{\star (7)}_{8}$, $V^{\star (43)}_{18} (=\{\bphi_{18}\})$,
$V^{\star (43)}_{19} (=\{\bphi_{19}\})$, $V^{\star (6)}_{5}$, $V^{\star (5)}_{3}$,
$V^{\star (3)}_{1}$, $V^{\star (2)}_{1}$, and $V^{\star (1)}_{1}$.
Note that
$V^{\star (7)}_{8} \cup V^{\star (43)}_{18} \cup V^{\star (43)}_{19} \cup V^{\star (6)}_{5} \cup V^{\star (5)}_{3} = V^{\star(4)}_{1}$, hence this is ``almost'' the graph
Shannon wavelet basis with $J=4$, which is the basis for the union
$V^{\star(4)}_{0} \cup V^{\star(4)}_{1} \cup V^{\star (3)}_{1} \cup V^{\star (2)}_{1} \cup V^{\star (1)}_{1}$. Hence, this VM-NGWP best basis should behave similarly to that
graph Shannon wavelet basis (except the mother wavelet vectors at level $j=4$).
In particular, it does not contain oscillatory basis vectors of large scale that
are not really necessary to approximate this highly localized pedestrian volume
data.
On the other hand, the PC-NGWP best basis turned out to be the graph Shannon
wavelet basis with $J=1$, i.e., the basis for the subspaces $V^{\star(1)}_0$ and
$V^{\star(1)}_1$. 
Since the pedestrian volume data is quite non-smooth and localized,
$\bdelta$-like basis vectors with scale $j=1$ in the PC-NGWP dictionary tend to
generate sparser coefficients, i.e., having a small number of large magnitude
coefficients with many negligible ones. Therefore, the best basis algorithm with
$\ell^1$-norm ends up favoring those fine scale basis vectors in the PC-NGWP
best basis for this graph signal.

%% file: discussion.tex

\section{Discussion}
\label{sec:discussion}

In this article, we proposed two ways to construct graph wavelet packet
dictionaries that fully utilize the natural dual domain of an input graph:
the VM-NGWP and the PC-NGWP dictionaries. Then, using two different graph
signals on each of the two different graphs, we compared their performance in
approximating a given graph signal against our previous multiscale graph basis
dictionaries, such as the HGLET, GHWT, and eGHWT dictionaries,
which include the graph Haar and the graph Walsh bases. 
Our proposed dictionaries outperformed the others on locally
smooth graph signals, and performed reasonably well for a graph signal
sampled on an image containing oriented anisotropic texture patterns.
On the other hand, our new dictionaries were beaten by the eGHWT on the
non-smooth and localized graph signal. One of the potential reasons for such
a behavior is the fact that our dictionaries are a direct generalization of
the ``Shannon'' wavelet packet dictionaries, i.e., their ``frequency'' domain
support is localized and well-controlled while the ``time'' domain support
is not compact. In order to improve the performance of our dictionaries
for such non-smooth localized graph signals, we need to bipartition $G^\star$
recursively but \emph{smoothly with overlaps}, which may lead to a graph version
of the \emph{Meyer wavelet packet dictionary}~\cite[Sect.~7.2.2, 8.4.2]{MALLAT-BOOK3},
whose basis vectors are more localized in the ``time'' domain than those of
the Shannon wavelet packet dictionary. 
In fact, it is interesting to investigate such smooth partitioning with
overlaps not only on $G^\star$ but also on $G$ itself since it may lead to
the graph version of the local cosine basis dictionary~\cite{LCT},
\cite[Sect.~8.4.3]{MALLAT-BOOK3}.

We also note that the VM-NGWP dictionary performed generally better than
the PC-NGWP dictionary for the graph signals we have examined. One of the
possible reasons is the use of the explicit localization procedure, i.e.,
the varimax rotation, in the former; the latter allows one to try to ``pinpoint''
a particular primal node where the basis vector should concentrate,
but the MGSLp procedure unfortunately shuffles and slightly delocalizes
the basis vectors after orthogonalization.
On the other hand, the difference in their computational costs is just a
constant in $O(N^3)$ operations. Hence, it is important to investigate how to
reduce the computational complexity in both cases. One such possibility is to
use only the first $N_0$ graph Laplacian eigenvectors with $N_0 \ll N$.
Clearly, one cannot represent a given graph signal precisely with $N_0$
eigenvectors, but this scenario may be acceptable for certain applications
including graph signal clustering, classification, and regression.
Of course, it is of our interest to investigate whether we can come up
with faster versions of the varimax rotation algorithm and the
MGSLp algorithm, which forms one of our future research projects.

Finally, we would like to emphasize that the natural dual domain $G^\star$ can be
used in applications beyond dictionary construction. Such applications include:
graph cuts and spectral clustering \cite{NG-JORDAN-WEISS, SHI-MALIK} to move
beyond noted limitations to using the first few eigenvectors \cite{NADLER-GALUN};
graph visualization and embeddings \cite{BELKIN-NIYOGI} to represent embeddings
with lower distortion \cite{JONES-MAGGIONI-SCHUL} as was done
in \cite{KOHLI-CLONINGER-MISHNE}; and anomaly detection through spectral methods
\cite{EGILMEZ-ORTEGA, MISHNE-COHEN} by going beyond the first few eigenvectors
\cite{CHENG-MISHNE, CLONINGER-CZAJA}. It is interesting to investigate going
beyond the initial set of eigenvectors with small eigenvalues in a method
informed by $G^\star$, and the effect of $G^\star$ on such methods.

%% file: acknowledgments.tex
\section*{Acknowledgments}
This research was partially supported by the US National Science Foundation
grants DMS-1819222, DMS-1912747, CCF-1934568, DMS-2012266; the US Office of
Naval Research grant N00014-20-1-2381; and Russell Sage Foundation Grant 2196.
We also thank the three anonymous reviewers for their insightful comments
  and suggestions, which helped us to improve the presentation and
  quality of this article.

%% file: varimax.tex
\section{Varimax Rotation}
\label{app:varimax}
The algorithm of varimax rotation we adopted and used in this article is
the so-called Basic Singular Value (BSV) algorithm proposed by Jennrich \cite{JENNRICH}. 
The algorithm in Eq.~\eqref{eq:varimax} can be summarized as follows:
\begin{enumerate}
\item[(0)] Initialize an orthogonal rotation matrix $R$.
\item[(1)] Compute $\dd f/\dd R$, where $f$ is the objective function defined in Eq.~\eqref{eq:varimax}, and $\dd f/\dd R$ is the matrix of partial derivatives of $f$ at $R$.
\item[(2)] Find the singular value decomposition $U \Sigma V^*$ of $\dd f/\dd R$.
\item[(3)] Replace $R$ by $UV^*$ and go to (1) or stop.
\end{enumerate}
Algorithm~\ref{algo:VM-rotation} below describes the details.
Jennrich also showed that under certain general conditions, this algorithm
converges to a stationary point from any initial estimate.
The BSV algorithm seems to be the standard varimax rotation algorithm
available in many packages, e.g., MATLAB\,\textsuperscript{\footnotesize\textregistered}\footnote{MATLAB is a registered trademark of The MathWorks, Inc.}, R, etc.
\begin{algorithm}[h]
\DontPrintSemicolon 
\SetKwData{tol}{tol}
\SetKwData{maxit}{maxit}
\KwIn{Full column rank input matrix whose columns to be rotated
  $A \in \R^{N \times m}$ ($m \leq N$);
  maximum number of iteration steps
  \maxit(default value: $1000$); relative tolerance \tol(default value: 1e-12)}
\KwOut{Rotated matrix $B\in \R^{N \times m}$}
$B=A$ \tcp*{initialize the output matrix}
$S=0$ \tcp*{initialize the nuclear norm} 

\For{$i$ from $1$ to \maxit}{
  $S_0 = S$

  $[U, \Sigma, V] = \operatorname{svd}(A^\transp \cdot(N \cdot B^{\circ 3} - B \cdot \operatorname{diag}(B^\transp \cdot B)))$ \tcp*{$B^{\circ 3} \define B \circ B \circ B$\\ where $\circ$ is the Hadamard (entrywise) product}
    
  $T = U \cdot V^*$ \tcp*{update the orthogonal rotation matrix}
  
  $S = \operatorname{trace}(\Sigma)$
    
  $B = A \cdot T$ \tcp*{update the rotated matrix}
  
  \If{$|S - S_0|/S<$ \tol}{ 
    break  \tcp*{stop when $S$ does not change much}
  }
}
\Return{$B$}
\caption{The Varimax Rotation Algorithm}
\label{algo:VM-rotation}
\end{algorithm}

%% file: mgslp.tex
\section{Modified Gram-Schmidt with \texorpdfstring{$\ell^p$}{lp} Pivoting Orthogonalization}
\label{app:mgslp}
We have implemented a simplified version of the modified Gram-Schmidt with mixed
$\ell^2$-$\ell^p$ ($0 < p < 2$) pivoting algorithm described and used
in~\cite{CoifmanRonaldR2006Dw}.  Algorithm~\ref{algo:MGS-Lp-pivoting} below
describes the details.
Our version skips the the step of computing the largest $\ell^2$ norm and
picking the parameter $\lambda$ (a notation used in~\cite{CoifmanRonaldR2006Dw})
to increase the numerical stability.
Instead, we directly set up a tolerance parameter, $\tol$, for robustness.
On the other hand, we keep the $\ell^p$ ($0 < p < 2$) pivoting procedure in
MGS (i.e., always perform the orthogonalization process of the vector with
minimum $\ell^p$-norm in the candidate pool), which nicely preserves the
sparsity of the obtained wavelet-like vectors after the orthogonalization
process. The MGSLp algorithm is summarized as follows.
\begin{algorithm}[h]
\DontPrintSemicolon 
\SetKwData{tol}{tol}
\KwIn{List of unit vectors $v = [v_1, \ldots, v_m] \in \R^{N \times m}$;
norm parameter $0 < p < 2$ (default value: $1$);
error tolerance \tol (default value: 1e-12)} 
\KwOut{List of orthonormal vectors $q = [q_1, \ldots, q_r] \in \R^{N \times r}$
where $r=\rank(v)$}
$q = \emptyset$ \tcp*{initialize the output list}
$w = [\|v_1\|_p, \ldots, \|v_m\|_p]$

\For{$i$ from $1$ to $m$}{
  $k = i-1+\operatorname{findmin}(w)$ \tcp*{find the minimum $\ell^p$-norm index}
    
  swap($v_i$, $v_k$) \tcp*{pivoting}
    
  \If{$\|v_i\|_2 < \tol$}{ 
    break  \tcp*{check linear dependency}
  }
    
  $\tilde{v} = v_i / \| v_i \|_2$ 
    
  $w = \emptyset$ \tcp*{re-initialize the $\ell^p$-norm vector}

  \For{$j$ from $i+1$ to $m$}{
    $v_j = v_j - (\tilde{v}^\transp \cdot v_j)\tilde{v}$
    
    $w$ = append($w$, $\|v_j\|_p$)
  }
    
  $q$ = append($q$, $\tilde{v}$)
}

\Return{$q$}
\caption{Modified Gram-Schmidt Orthogonalization with $\ell^p$ pivoting}
\label{algo:MGS-Lp-pivoting}
\end{algorithm}